%% file: making_talk_cheap.tex
\documentclass[11pt,dvipsnames]{article}
\input{\detokenize{packagesJMP.tex}}
\graphicspath{{graphs/}}

\makeatletter
\newcommand\notsotiny{\@setfontsize\notsotiny{6.5}{7.5}}
\newcommand\notsosmall{\@setfontsize\notsosmall{8}{9}}
\newcommand\notsosmallbutbigger{\@setfontsize\notsosmallbutbigger{9}{10}}
\newcommand\notsosmallbutevenbigger{\@setfontsize\notsosmallbutevenbigger{10}{11}}
\makeatother

\title{\vspace*{-1.5cm}\bfseries Making Talk Cheap: \\ Generative AI and Labor Market Signaling\thanks{We are indebted to Adam Kapor, Jakub Kastl, and Alessandro Lizzeri for their invaluable guidance and support. We are also grateful to Elena Aguilar, Nick Buchholz, Jacob Dorn, Michael Droste, Allison Green, Jeff Gortmaker, Kate Ho, Simon Jäger, Quan Le, Thomas Le Barbanchon, Rachel Moore, Stephen Redding, Jordan Richmond, Ben Scuderi, Wilbur Townsend, Carolyn Tsao, So Hye Yoon, and Yuci Zhou for helpful comments and discussions. We thank Matt Barrie and the entire team at \texttt{Freelancer.com} for their openness and willingness to collaborate with academic researchers. We are especially grateful to Andrew Bateman and Will Bosler, without whom this research agenda would not have been possible. We additionally thank Torsha Chakraverty for great research assistance. This research benefited from financial support from the International Economics Section (IES) at Princeton University and the William S. Dietrich II Economic Theory Center, whose support is gratefully acknowledged. We also thank the Research Computing at ITC, Dartmouth College, for their technical support in implementing and scaling up our LLM-based customization measures for freelancers' job application proposals. All errors are our own.}}
\author{%
  \Large Ana\"{i}s Galdin\thanks{Dartmouth College, Tuck School of Business. Email: anais.galdin@dartmouth.edu}%
  \hfill
  \Large Jesse Silbert\thanks{Princeton University, Department of Economics. Email: jesseas@princeton.edu}\\[0.4ex]
  {Dartmouth College, Tuck \hspace{3.6cm} Princeton University}\\[0.4ex]
  \makebox[\textwidth][r]{\makebox[0.48\textwidth][c]{\plum{Job Market Paper}}}%
}
\date{\vspace{0.1cm}\large November 11, 2025 \\ \href{https://jesse-silbert.github.io/website/silbert_jmp.pdf}{\color{Aquamarine4}{\textbf{[Click for the latest version]}}} }

\begin{document}
{\small
\maketitle
}

\begin{abstract}
    \begin{spacing}{1.0}
\footnotesize 

Large language models (LLMs) like ChatGPT have significantly lowered the cost of producing written content. This paper studies how LLMs, through lowering writing costs, disrupt markets that traditionally relied on writing as a costly signal of quality (e.g., job applications, college essays). Using data from \texttt{Freelancer.com}, a major digital labor platform, we explore the effects of LLMs' disruption of labor market signaling on equilibrium market outcomes. We develop a novel LLM-based measure to quantify the extent to which  an application is tailored to a given job posting. Taking the measure to the data, we find that employers had a high willingness to pay for workers with more customized applications in the period before LLMs were introduced, but not after. To isolate and quantify the effect of LLMs' disruption of signaling on equilibrium outcomes, we develop and estimate a structural model of labor market signaling, in which workers invest costly effort to produce noisy signals that predict their ability in equilibrium. We use the estimated model to simulate a counterfactual equilibrium in which LLMs render written applications useless in signaling workers' ability. Without costly signaling, employers are less able to identify high-ability workers, causing the market to become significantly less meritocratic: compared to the pre-LLM equilibrium, workers in the top quintile of the ability distribution are hired 19\% less often, workers in the bottom quintile are hired 14\% more often.
\end{spacing}
\end{abstract}

\newpage
\begin{spacing}{1.5}
\section{Introduction}\label{introduction}

This paper studies the equilibrium effects of large language models (LLMs) like ChatGPT disrupting the signaling value of written communication on hiring patterns and labor market efficiency. To measure the signals workers send to employers, we develop a novel LLM-based approach to quantify how tailored an application is to a given job posting. Using data from \texttt{Freelancer.com}, a major digital labor platform, we provide new descriptive evidence that prior to the mass adoption of LLMs, workers used the customization of their applications to signal their ability, but that after LLMs, these signals became far less informative. To quantify the equilibrium effect of LLMs' disruption of signaling, we build and estimate a structural model of the pre-LLM market that embeds signaling in the style of \citet{spence1973} into a model of labor supply and demand. Using the estimated model, we explore a counterfactual equilibrium in which LLMs reduce writing costs to zero in the pre-LLM market, thereby eliminating labor market signaling.

Written communication has long been used as a tool to facilitate matching in many markets, ranging from workers writing cover letters to employers to prospective students writing application essays to college admission committees. Because writing requires time and effort, the act of writing itself can send a signal (\cite{spence1973}). For instance, a worker interested in and well-suited to a specific job can send a customized cover letter instead of a generic one, thereby credibly signaling her value to the employer.

In recent years, however, the advent of generative artificial intelligence (AI) in the form of LLMs has dramatically lowered the cost of writing. LLMs can produce polished, human-sounding text in seconds at virtually no cost (\cite{dathathri2024scalable}). Thus, this technology may threaten the signaling value of writing in the labor market. If all workers can cheaply produce highly customized, expertly written applications with the aid of an LLM, then employers may no longer be able to use written applications as credible signals of whom to hire. Given this potential threat to labor market signaling, it is important to empirically evaluate both the extent to which writing served as a costly signal before the mass adoption of LLMs and the extent to which LLMs now diminish the signaling value of writing and, in turn, impact matching in labor markets.

Our empirical context is the market for coding jobs on \texttt{Freelancer.com}, a major digital labor platform (DLP). On this platform, employers, seeking to outsource one-off digital tasks (e.g., website design), post jobs to find and hire freelance workers to complete those tasks. Workers apply to jobs with a short, written application---a ``proposal''---as well as an asking wage---a bid.

This setting provides two distinct advantages for studying how LLMs disrupt labor market signaling. First, we can measure both the signals workers send and the effort they exert when sending those signals using the platform's detailed text and click data on each application and job posting. In particular, we measure signals by quantifying the extent to which the text of each proposal is customized and relevant to the text of the corresponding job post, using an LLM to approximate human judgment at scale. Moreover, we measure signaling effort using click data to measure the time each worker spends on each application. Second, we can observe via click data which applications were written using an LLM-powered writing tool the platform introduced in April 2023 designed to automate proposal writing.

Leveraging this unique setting, we provide new descriptive evidence that LLMs have disrupted previously informative labor market signaling. We organize our descriptive evidence into three main findings. First, we show that before the mass adoption of LLMs, employers had a significantly higher willingness to pay for workers who sent more customized proposals. Estimating a reduced-form multinomial logit model of employer demand using our measure of signal, we find that, all else equal, workers with a one standard deviation higher signal have the same increased chance of being hired as workers with a \$26 lower bid.\footnote{Bids in our sample range from \$30 to \$250, with a standard deviation of about \$66.} Second, we provide evidence that before the adoption of LLMs, employers valued workers' signals because signals were predictive of workers' effort, which in turn predicted workers' ability to complete the posted job successfully. Third, we find, however, that after the mass adoption of LLMs, these patterns weaken significantly or disappear completely: employer willingness to pay for workers sending higher signals falls sharply, proposals written with the platform's native AI-writing tool exhibit a negative correlation between effort and signal, and signals no longer predict successful job completion conditional on being hired. Taken together, these findings suggest that workers’ job-specific proposals functioned as Spence-like signals of worker ability prior to the advent of generative AI, but that LLMs disrupted this signaling mechanism on which employers previously relied to make hiring decisions.

While these findings provide evidence of LLMs disrupting labor market signaling, they do not reveal changes to the distribution of hired workers' abilities, nor do they isolate the signaling channel from other potential effects of LLMs on labor supply and demand.\footnote{For example, LLMs may increase worker productivity, thereby affecting labor supply, or they may change the nature of tasks employers seek to outsource, thereby affecting labor demand.}

To quantify how LLMs impact equilibrium hiring through their disruption of signaling alone, we therefore develop and estimate a structural model of labor market signaling. The model combines three typically distinct modeling approaches: we embed a (1) Spence signaling model---workers invest costly effort to produce noisy signals that positively correlate with their ability in equilibrium---into a (2) discrete choice demand model---employers form indirect expected utilities over application characteristics and their beliefs about ability---which, from the workers' viewpoint, operates as a (3) scoring auction---workers submit applications competing on multiple dimensions to win a contract.

The core mechanism in the model is that higher-ability workers face lower costs of exerting effort to produce signals when applying to jobs than lower-ability workers do. As a result, in equilibrium, higher-ability workers exert more effort and thereby send higher signals on average. Employers thus find these signals informative about worker ability, though they cannot perfectly infer ability since they do not observe effort and signals are noisy. The modeled signaling equilibrium is in the tradition of \citet{spence1973} in that while all workers could increase their chances of being hired by exerting more effort, only higher-ability workers find it worthwhile to do so.

Identifying the model poses two key challenges centered around identifying workers' and employers' equilibrium beliefs. First, we must disentangle workers' costs of undertaking the job from their strategic bidding behavior as a function of their beliefs about their chances of being hired.\footnote{For example, suppose we observe the pattern that workers with high signals also tend to submit high bids. Without identifying a worker's beliefs about her probability of being hired, we cannot disentangle whether this pattern is the result of high-ability workers facing higher costs of undertaking jobs or strategically setting higher prices, believing that employers are willing to pay more for their higher signaled ability.} Second, we need to separately identify employers' beliefs about worker ability as functions of bids and signals from employers' disutility from paying wages and willingness to pay for ability.\footnote{For example, suppose we observe that employers are willing to hire workers with high bids. Without identifying employers' beliefs about worker ability, we cannot disentangle whether this pattern is the result of employers being price insensitive or believing that high-bid workers have high ability.} 

Our identification argument addresses these two challenges by exploiting the information structure of our model. In equilibrium, a worker's belief about her probability of being hired is independent of her private type (cost and ability), conditional on her choice of bid and effort. As a result, worker beliefs equal empirical hiring probabilities as functions of bids and efforts, and are thus directly identified from data on equilibrium hiring decisions. 
Having identified worker beliefs, we can infer a worker’s cost of undertaking the job and her ability from her equilibrium bid and effort.
With the recovered distribution of worker costs and abilities, we can identify employers' beliefs about ability as functions of bids and signals, which allow us to identify employer disutility from paying wages and willingness to pay for worker ability.

We estimate the model on pre-LLM data, following the identification argument closely.\footnote{We do not estimate the model on post-LLM data, since our descriptive results suggest that the signaling equilibrium, on which our identification argument relies, has broken down in that period.} 
We develop a novel simulation-based estimator to recover worker hiring probabilities as functions of bids and efforts, which we use to nonparametrically estimate the joint distribution of worker costs and abilities. Using this estimated distribution of worker types, we then estimate a flexible nonparametric model of employer beliefs about ability and, in turn, recover employer preferences.

Our estimates reveal that employers had a high willingness to pay for workers with high signals in the pre-LLM market because employers highly value worker ability, and the most reliable way to discern worker ability is through signals. We organize our findings into five results. First, employers directly value ability: we estimate that employers are willing to pay \$52.16 on average for a one standard deviation increase in worker ability, which is about 79\% of the standard deviation of bids in our sample. Second, there is significant variation in worker ability: employers value hiring workers at the 80th percentile of the ability distribution \$97 more than they value hiring those at the 20th percentile. Third, workers' observable characteristics---such as their on-platform reputation---poorly predict the ability signaled via workers' proposals: the estimated variation in ability across workers' observable characteristics explains only about 3\% of the total estimated variation in worker ability. Fourth, signals are highly informative about ability in equilibrium: we estimate the correlation between observed signals and estimated ability to be 0.55. Finally, workers with high ability often require high signals to compete with workers who can afford to bid lower than they can: we estimate the correlation between a worker's ability and her cost of undertaking the job to be 0.19.

Having estimated the model, we explore how the elimination of signaling due to LLMs affects equilibrium hiring patterns and welfare. We simulate a counterfactual market equilibrium in which workers do not have access to any signaling technology and only choose bids when applying to jobs.\footnote{Using pre-LLM estimates to simulate this no-signaling counterfactual allows us to hold fixed any other potential effects of LLMs on labor supply and demand.} Accordingly, employers only form beliefs about worker ability based on their observable characteristics.

Compared to the status quo pre-LLM equilibrium with signaling, our no-signaling counterfactual equilibrium is far less meritocratic. Workers in the bottom quintile of the ability distribution are hired 14\% more often, while workers in the top quintile are hired 19\% less often. 

These effects are driven by three mechanisms. First, employers previously relied on signals to make hiring decisions, so losing access to them impinges on their ability to discern worker ability. Second, more indirectly, the significant positive correlation between a worker's ability and cost implies that, when employers lose access to signals and workers are forced to compete more intensely on wages, the prevailing workers with lower bids tend to have lower abilities. Third, since workers' observable characteristics are poor predictors of their ability, employers have little to no information to distinguish between high and low-ability workers. 

These changes to hiring patterns lead to a 5\% reduction in average wages, a 1.5\% reduction in overall hiring rate per posted job, a 4\% reduction in worker surplus, and a small, less than 1\%, increase in employer surplus. Worker welfare losses are driven by both the modest extensive margin decrease in hiring rates and the intensive margin decrease in wages. However, workers' losses are mitigated by the fact that their writing costs are now zero, and by the fact that the counterfactually hired lower-ability workers tend to have lower costs. Employer surplus is virtually unaffected due to the highly competitive nature of the worker-side of the platform: the reduction in wages paid to workers roughly balances out the loss of hiring lower-ability workers. Overall, the market becomes less efficient and significantly less meritocratic: total surplus falls by 1\%, and the new equilibrium favors low-ability workers over high-ability ones.

Our quantitative results highlight that the relationship between worker ability and equilibrium wage offers is a key component in evaluating how LLMs’ disruption of signaling affects equilibrium hiring patterns. Signaling matters in labor markets when employers do not always prefer to hire the cheapest workers. If ability and cost were perfectly negatively correlated, then a market without signaling---where hiring depends solely on wage competition---would result in efficient matches. In our setting, however, high-ability workers tend to face higher costs of undertaking jobs. As a result, when workers can no longer signal their ability, high-ability workers tend to be less able than low-ability workers to compete on wages alone, leading to less meritocratic hiring outcomes.

\paragraph{Related Literature}

We contribute to four distinct literatures. First, there is a recent and growing empirical literature on how generative AI and LLMs affect labor markets (\cite{acemoglu2025simple}, \cite{humlum2025large}, \cite{cui2025effects}, \cite{brynjolfsson2025generative}, \cite{brynjolfsson2025canaries}, \cite{dillon2025shifting}, \cite{hartley2024labor}, \cite{stanton2025benefits}, \cite{stanton2014learning}, \cite{teutloff2025winners}, \cite{eloundou2024gpts}). This literature has been primarily focused on studying how LLMs affect the nature of work and the productivity of workers via surveys and randomized experiments. The economics of AI in these studies is thus largely centered around how AI impacts the supply and demand for labor, though a few papers have begun to explore the implications of AI on signaling (\cite{cowgill2024does}, \cite{wiles2025generative}, \cite{cui2025signaling}\footnote{In a simultaneous paper, \citet{cui2025signaling} also use data from \texttt{Freelancer.com} to study how large language models affect labor market signaling. Building upon our method for measuring signals from application and job‐post text data, they measure the textual similarity between job posts and worker applications as a proxy for signals. Their approach relies on traditional natural language processing methods, such as Term Frequency–Inverse Document Frequency (TF–IDF), rather than LLM‐based measures.}, \cite{gans2024will}, \cite{dhillon2025signal}). We contribute to this literature in multiple ways. Relative to existing papers that study LLMs' effects on signaling via experiments, we move beyond partial equilibrium and provide empirical evidence on how LLMs have disrupted a market-wide signaling equilibrium. We also provide the first quantification of how this disruption of signaling affects equilibrium hiring patterns and welfare. Finally, to the best of our knowledge, our paper presents the first structural empirical model of LLMs' effects on the labor market.

Second, there is a deep literature on the economics of labor market signaling itself. This literature has both been theoretical (\cite{spence1973}, \cite{stiglitz1975theory}, \cite{waldman1984job}) and empirical (\cite{macleod2017big}, \cite{to2018signaling}, \cite{layard1974screening}, \cite{lang1986human}, \cite{bedard2001human}, \cite{tyler2000estimating}, \cite{clark2014signaling}, \cite{fang2006disentangling}).\footnote{There is also a related literature on signaling in educational matching markets such as college admissions (\cite{gandil2022college}) and medical school admissions (\cite{friedrich2024interdependent}). Similar to our results, \citet{gandil2022college} find that written essays play an important role in signaling student quality in college admissions.}
This literature has largely focused, since the seminal work of \citet{spence1973}, on educational choices as signals to the labor market. We contribute to this literature by providing the first empirical structural analysis of labor market signaling, where the signal is the information transmission in the application itself, i.e. the communication between workers and employers. 

Third, there is a growing literature on the economics of digital labor platforms (\cite{wiles2025algorithmic}, \cite{brinatti}, \cite{krasnokutskaya2020role}, \cite{autor2009}, \cite{lehdonvirta}, \cite{horton_2017}, \cite{horton_chapter}, \cite{horton_steering}, \cite{ren_freelancers}, \cite{stanton_thomas}, \cite{pallaisJMP}). This literature has studied a wide variety of topics from platform design to the role of communication in matching. We contribute to this literature by quantifying the potential risks that generative AI poses to the signaling mechanisms that are especially important for matching on congested digital labor platforms.

Finally, there is a small and emerging literature on the structural estimation of signaling models (\cite{kawai2022signaling}, \cite{cai2025information}). This literature has thus far focused on the estimation of signaling models in which the signal takes the form of an explicit market action, such as setting a reserve interest rate in the case of \citet{kawai2022signaling}, or strategically rejecting observable offers in the case of \citet{cai2025information}. We contribute to this literature by providing the first structural estimation of a signaling model in which we directly use data on communication, i.e., text, to measure signals. We also contribute methodologically by developing a model that embeds Spence signaling into a structural model of labor supply and demand.

\paragraph{Paper Outline} The paper proceeds as follows. Section~\ref{data} provides an overview of the \texttt{Freelancer.com} market, a description of the data, and how we construct our sample. Section~\ref{measurement} describes our novel measures of job application's signaling content, workers' signaling effort, and employer's consideration sets. Section~\ref{descriptives} presents new descriptive evidence on how LLMs have disrupted labor market signaling. Section~\ref{model} develops a structural model of labor market signaling and discusses its identification. Section~\ref{estimation} describes how we estimate the model. Section~\ref{est_results} discusses the results of our estimation. Section~\ref{counterfactuals} presents a simulated counterfactual in which LLMs reduce writing costs to zero, prohibiting workers from signaling their ability. Section~\ref{conclusion} concludes.  An \hyperref[appendix_market]{\textcolor{DarkGoldenrod3}{Appendix}} collects additional results and discussions.

\section{Empirical Context}\label{data}

This paper uses data from \texttt{Freelancer.com}, a major digital labor platform (DLP). DLPs bring together two increasingly important pillars of the modern economy. First, they are two-sided platform markets-marketplaces in which a third party facilitates matching between buyers and sellers. Second, they are largely organized around gig work-tasks that firms or individuals outsource to independent workers.

In this section, we provide an overview of the market, describe the data available to us, and explain how we construct the sample used in the analysis. 

\subsection{Market Description}\label{data_market}

\texttt{Freelancer.com} is one of the world's largest DLPs, as measured by both number of users and posted jobs. DLPs such as \texttt{Freelancer.com} and Upwork are global, online marketplaces that match freelance workers with potential employers seeking to outsource digital tasks. These tasks—ranging from software and web development to sales support, marketing, and graphic design—are performed remotely by freelancers hired under short-term contracts. Since its launch in 2009, more than 85 million users from over 247 countries, regions, and territories have registered either to supply or to procure remote work. In fiscal year 2024 alone, employers posted 1.1 million jobs on the platform, generating over \$130 million in payments with an average completed job paying \$334.\footnote{These statistics are based on \texttt{Freelancer.com}'s 2024 Annual Report}

\subsubsection{Market Participants}

Workers on the platform (``freelancers'') are independent contractors seeking short-term digital jobs.\footnote{While some workers rely on freelancing as a primary source of income, most others treat it as supplemental income.} Freelancers are globally distributed, though India is by far the modal country of residence, with the United States being the next most frequent.

Employers are individuals or firms who post jobs on the platform to outsource specific tasks. While we cannot link employers to their off-platform identities, anecdotal evidence suggests they range from individuals outsourcing small tasks for personal projects, to start-ups looking to outsource labor in their initial phases, to larger-scale corporations seeking cheaper labor than they can source in their local labor markets. Similar to the modal countries of residence for workers, the modal countries of origin for employers are India and the United States.

\subsubsection{Types of Jobs Posted}

On \texttt{Freelancer.com}, workers are primarily hired to complete digital tasks in areas such as software development, writing, data entry, and design.\footnote{Each broad occupation can be further disaggregated into detailed categories and associated skills. A list of job categories and examples of their associated skills is provided in Appendix Table~\ref{job_category}.} These jobs can be done remotely and require little physical capital other than a computer and an internet connection.

When looking to outsource a task, employers, for the most part, post either fixed-price or hourly jobs.\footnote{Smaller portions of activity include direct employer searches, contest-style competitions, and more recently a spot market for standardized tasks.} Fixed-price jobs involve a pre-agreed wage for the entire task, while hourly jobs pay based on time worked. We focus our analysis on fixed-price jobs. Between 2021 and 2024, fixed-price jobs were posted 2.7 times as often as hourly jobs. These represent one-off tasks typically completed in a few hours or days. 

\subsubsection{Job Matching Process}

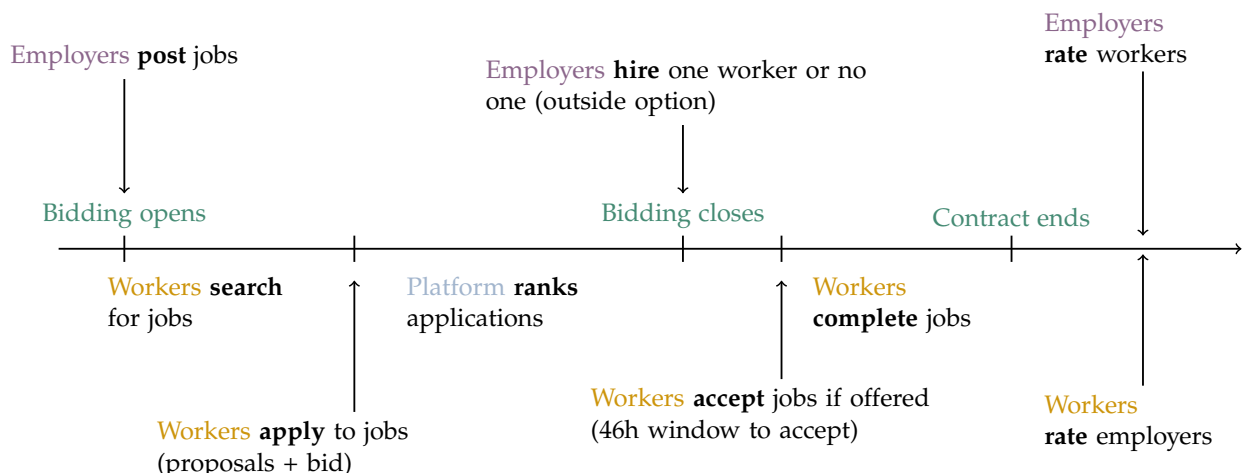
\begin{figure}[h!]
\centering 
\begin{adjustbox}{max totalsize={\linewidth}{\textheight},left}
\begin{tikzpicture}[node distance=1cm, auto] 
\tikzset{ 
    mynode/.style={ellipse, thick, text centered}, 
} 
\draw [thick,->] (0,0) -- (18,0);

\draw[thick] (1,0.2)  -- ++ (0,-0.4) node[above] (bid_opens) {};
\node[above =.1cm of bid_opens] (opens)  {\textcolor{Aquamarine4}{Bidding opens}};
\node[above=2.5cm of bid_opens ] (bidtext) {\textcolor{Plum4}{Employers} \textbf{post} jobs};
\draw [thick,->] (bidtext) -- (opens);

\draw[thick] (4.5,0.2) -- ++ (0,-0.4) node[below,] (bid) {};
\node[below=2cm of bid, text width=6cm] (bid_text)  {\textcolor{DarkGoldenrod3}{Workers} \textbf{apply} to jobs \\ (proposals + bid)};
\draw[->, thick] (bid_text) -- (bid); 
\draw[thick] (3,0.2) node[below] (search) {};
\node[below=0.2cm of search, text width=4.5cm] (infobid)  {\textcolor{DarkGoldenrod3}{Workers} \textbf{search} \\ for jobs};

\draw[thick] (7.8,0.2) node[below] (award) {};
\node[below =0.2cm of award, text width=5cm] (awardtext)  {\textcolor{LightSteelBlue3}{Platform} \textbf{ranks} \\ applications};
\draw[thick] (9.5,0.2)  -- ++ (0,-0.4) node[above, color={Aquamarine4}] (step2) {};
\node[above =.1cm of step2] (step2_text)  {\textcolor{Aquamarine4}{Bidding closes}};
\node[above=1.8cm of step2, text width=6cm] (firm2_text)  {\textcolor{Plum4}{Employers} \textbf{hire} one worker or  no one (outside option)};
\draw[->, thick] (firm2_text) -- (step2_text); 

\draw[thick] (11,0.2) -- ++ (0,-0.4) node[below] (accept) {};
\node[below=1.5cm of accept, text width=5.8cm] (accept_text)  {\textcolor{DarkGoldenrod3}{Workers}  \textbf{accept} jobs if offered \\(46h window to accept)};
\draw[->, thick] (accept_text) -- (accept); 
\draw[thick] (11,0.2) node[below] (complete) {};
\node[below right =0.2cm and 0.2cm of complete, text width=6cm] (completetext)  {\textcolor{DarkGoldenrod3}{Workers} \\ \textbf{complete}  jobs};

\draw[thick] (14.5,0.2)-- ++ (0,-0.4) node[below] (ends) {};
\draw[thick] (16.5,0.2) node[below] (ratings) {};
\node[above = 0.4cm of ends] (ratings_text)  {\textcolor{Aquamarine4}{Contract ends}};
\node[above =2.5cm of ratings, text width=3cm](ratings_firm)  {\textcolor{Plum4}{Employers} \\\textbf{rate} workers};
\node[below =2cm of ratings, text width=3cm](ratings_lancer)  {\textcolor{DarkGoldenrod3}{Workers} \\ \textbf{rate} employers};
\draw[->, thick] (ratings_firm) -- (ratings); 
\draw[->, thick] (ratings_lancer) -- (ratings); 
\end{tikzpicture} 
\end{adjustbox}
\centering
\caption{Matching Process}\label{fig:matching_process}
     \floatfoot{\footnotesize \textit{Notes:} This figure describes the process during which employers and workers match on \texttt{Freelancer.com}. Employers post jobs on the platform; workers search for jobs and submit bids with proposals; the platform ranks applications; employers choose a worker or the outside option (hiring no one); workers  accept offers within 46 hours; if accepted, work starts. Both sides leave ratings post-completion of the job.}
\end{figure}

The job matching process, summarized in Figure \ref{fig:matching_process}, consists of six main steps.

\textbf{1. Employers Post Jobs.} Employers post jobs, tagging required skills and specifying acceptable bid ranges. Posts include a short description of the job, with details on skills, delivery requirements, and other expectations. 

\textbf{2. Workers Search for Jobs.} Workers search for jobs based on skills, language, budget type (fixed or hourly), and budget range. Results are listed by recency. Workers see a preview of job descriptions and must click on the posting for full details. See Appendix Figure \ref{platform_overview} for an example of how job postings are displayed from the point of view of a worker searching for a job.

\textbf{3. Workers Apply to Jobs.} Interested workers submit a written proposal and a bid. For fixed-price jobs, the bid is the amount the employer must pay upon completion.\footnote{The platform enforces payment through its in-built payments system and offers a dispute resolution service for a fee. Payments are not conditional on meeting quality thresholds, but do require ``satisfactory'' work.} 

\textbf{4. Platform Ranks Applications.} Applications are ranked by a proprietary recommendation algorithm based on workers' characteristics (mostly workers' on-platform reputation and prior performance), not bids or proposal content.\footnote{The exact algorithm is proprietary and has changed over time. \texttt{Freelancer.com} provided us with the version in use for the time period our data spans.} Each application receives an \textit{under-the-hood} numerical ``reputation score.''\footnote{The reputation score is not directly visible to employers. However, many of its components, such as work history on the platform, are, and the scores can be loosely inferred from the observed ranking of the applications, which employers do observe.}

\textbf{5. Employers Hire Workers.} Employers review the ``bid list'' (applications ranked by the algorithm), observing each bid, proposal, and worker characteristics (e.g., rating, reviews, completion rate, country). Employers may communicate with applicants via private messages. They can either hire a worker or leave the job unfilled. See Figure~\ref{bid_list} for a screenshot of the platform's appearance from the point of view of employers looking at the bid list.\footnote{Also see Figure~\ref{worker_profile} for an example of a worker's profile page, which employers can click through to from the bid list to learn more about a worker.}

\textbf{6. Workers Accept Offers.} Once an employer makes an offer, the worker has 46 hours to accept or reject. Inaction is treated as rejection. The vast majority of offers are accepted.\footnote{Workers may reject offers if their availability has changed, but they are encouraged to apply only to jobs they can complete.} Upon acceptance, work commences. After completion, both sides are prompoted to leave ratings. Payments may be structured into ``milestones,'' i.e., subcontracts that break the job into separable steps. See Figure~\ref{milestone_payments} for an example of how milestone payments are structured. 

\subsubsection{On-Platform AI-writing Tool}

In April 2023, \texttt{Freelancer.com} introduced an on-platform AI-writing tool that allows workers at a certain tier of paid subscription\footnote{As of October 2025, the AI-writing tool is available to all workers with memberships costing at least \$10 per month, billed monthly.} to the platform to generate applications at the click of a button using a Large Language Model (LLM) that takes as input both the job description and the worker's own profile. Workers with access to the tool can freely choose to use it when applying to any job post. The generated proposals can be edited by workers before submission. Employers cannot directly observe whether a worker used the AI-writing tool to generate their proposal. See Appendix Figure~\ref{ai_bids} for how the AI-writing tool is displayed on the platform and for an example of an AI-written proposal.

\subsection{Data}\label{data_overview}

We have access to the internal database of \texttt{Freelancer.com}, on which the platform stores data on nearly all market operations such as job posting, job applications, employer hiring decisions, job completion, and worker performance. These data include every application submitted to every job post, along with the precise timestamps and content of any on-platform action (stored as click data) taken by any employer or worker on the platform. 

In particular, we observe the complete ``lifecycle'' of every job post: we observe when the job is posted, the exact text of the job title and job description, whether and when the job is awarded to a worker, whether and when the job is completed, the total payment the worker receives for completing the job, and the star rating ($1-5$) the employer gives the worker upon job completion (if any).

We also observe all on-platform worker-employer interactions (save for the content of private messages). We observe whether and when the job poster contacts each worker via the on-platform messaging system (which is always initiated by the job poster), whether and when the job poster clicks through to each worker's profile, and whether and when the job poster clicks on each application in any way (e.g., clicking to expand the proposal text, clicking to ``favorite'' the application, etc.).

Furthermore, we also observe all on-platform actions and information associated with each application. We observe the bid of each application, the text of each application's proposal, the timestamp that each applying worker first views the full job post (i.e., when workers first click on the job post's page), the timestamp that each applying worker submits her application, whether the applying worker used the on-platform AI proposal-writing tool, the country of residence of each applying worker, and the complete work history of each applying worker on the platform, including all prior jobs completed, star ratings received, and reviews received. Finally, we observe each application's reputation score, which, as described above, is the output of the platform's recommendation algorithm. Thus, we observe the dynamically updated ranking of each application on the ``bid list'' page at any point in time.

\subsection{Sample Construction}\label{data_sample}

In this subsection, we describe how we construct our sample. The sampling procedure is shaped both by the financial constraints of large-scale text analysis and by the goals of our empirical study. 

Our analysis relies on and, we argue, requires (see Section~\ref{measurement_signals}) the use of LLMs to process the text of millions of job descriptions and worker proposals. Because this computation is costly, it limits the feasible sample size. This creates a trade-off between capturing heterogeneity on the employer side and on the worker side.

Since our primary goal is to study how employers learn about worker ability, and how this learning is affected by the loss of signaling through written proposals, we prioritize worker-side heterogeneity. To do so, we restrict attention to a relatively homogeneous set of job postings. 

Specifically, we include all jobs posted on the platform between January 1, 2021---before the mass adoption of LLMs---and July 26, 2024---well after mass adoption and, crucially, after the introduction of the on-platform AI proposal-writing tool on April 18, 2023---that satisfy the following criteria:
\begin{itemize}
    \item Fixed-price jobs;
    \item Transactions conducted in USD (bids and payments);
    \item Budget range of \$30–\$250 (the modal range on the platform);
    \item Job description written in English;
    \item Posted by employers with verified identities and payment methods;
    \item Posted by employers who, at any point on the platform (including outside our sample period), have hired and paid at least one worker;
    \item Awarded to at most one worker;
    \item Categorized as involving some form of coding;\footnote{Note that we do not include coding-related skill tags that only appear in the post-LLM sample as to make the types of job posts in both the pre-LLM and post-LLM samples comparable. See Table~\ref{tab:coding_skills} for a list of the top 15 skill tags in our sample.}
    \item Job description containing at least 25 words and at least 182 characters.
\end{itemize}

Applying these restrictions yields a sample of roughly 61,000 job postings, associated with about 2.7 million applications submitted by 212,000 unique workers. 

We divide this dataset into two subsamples: one before the release of ChatGPT on November 30, 2022 (``pre-LLM''), and one after (``post-LLM''). The pre-LLM sample contains approximately 33,000 job postings, 960,000 applications, and 92,000 unique workers. The post-LLM sample contains approximately 28,000 postings, 1.7 million applications, and 135,000 workers. Table~\ref{tab:sample_summary_stats} provides summary statistics for both subsamples.

\input{tables/sample_sum_stats.tex}

\section{Measurement}\label{measurement}

In this section, we provide details on how we measure three key components of our analysis: (1) the signaling effort each worker expends writing her proposal; (2) the ``signaling content'' of each worker's proposal, and (3) which applications are actually viewed and considered by employers.  The first and third use the click and timestamp data available to us from the platform, while the second develops a novel text analysis method that leverages LLMs.

\subsection{Signaling Effort}\label{measurement_effort}

Our analysis requires a measure of how much cognitive effort each worker expends to produce her written application proposal. We proxy for this effort using the time each worker spends reading the job post and writing her proposal. In particular, we measure the time between when each worker first ever clicks on the job post's page and when she submits her application. We refer to this raw measurement as ``bid time,'' and we interchangeably refer to this measure as ``signaling effort'' or simply ``effort.'' 

We successfully measure valid bid times for 70.45\% of all the applications in our sample.\footnote{The pre-LLM sample has a lower fraction of valid bid times (59.76\%) than the post-LLM sample has (76.39\%).} 6.21\% of applications have negative bid times (i.e., the worker's first click occurs after she submits her application), 17.67\% have missing bid times (i.e., we do not observe any clicks by the worker on the job post), 2.54\% have implausibly high bid times (i.e., over 12 minutes), and 3.13\% have implausibly low bid times (i.e., under 4 seconds).

The applications with missing, negative, or implausibly small bid times likely arise from workers accessing the platform's job postings via an application programming interface (API) that allows them to scrape job posts without actually registering a click on the job post. Both from inspecting individual applications with missing or negative bid times and from a thorough statistical analysis of the variables we observe, we find that the proposals associated with these applications do not seem to be statistically distinct from the applications for which we observe valid bid times, conditional on an application's observable characteristics. We treat the incidence of invalid or missing bid times as random, conditional on observables, and we omit these applications from our empirical analyses that directly use bid time, while including them in all other analyses.

\begin{figure}[h!]
    \centering
\includegraphics{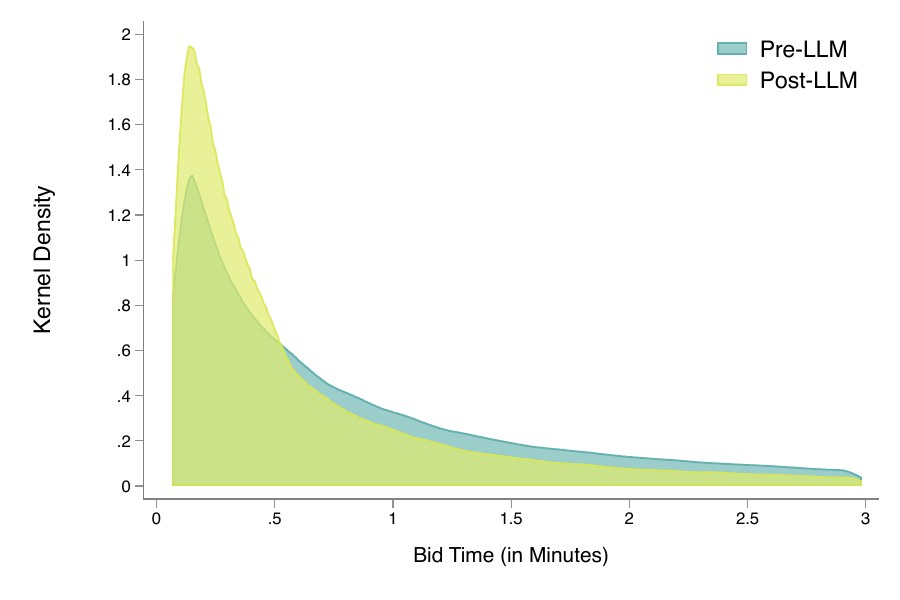}
    \caption{Distribution of Bid Time: Pre-LLM vs. Post-LLM}
    \label{fig:effort_prepost}
     \floatfoot{\footnotesize \textit{Notes:} This figure plots kernel density estimates of the distributions of bid time, defined as the time between when a worker first clicks on a job post and when she submits her application, in both the pre-LLM and post-LLM samples. The post-LLM sample coincides with the subsample of post-LLM job posts beginning in March 26, 2024 for which we compute signal measures. For both samples, we only include applications with bid times between 4 seconds and 12 minutes. The pre-LLM sample has a mean of 1.47 minutes, a median of 0.70 minutes, and a standard deviation of 1.95 minutes, while the post-LLM sample has a mean of 1.08 minutes, a median of 0.43 minutes, and a standard deviation of 1.70 minutes.}
\end{figure}

Our final sample of bid times thus ranges from 4 seconds to 12 minutes, and we display kernel density estimates of both the pre-LLM and post-LLM distributions of bid time in Figure~\ref{fig:effort_prepost}. The average bid time in the pre-LLM sample is 1.47 minutes, which is sensible given the average job post in that period is 93 words long, and the average proposal is 89 words long.

\subsection{Signals}\label{measurement_signals}

Our analysis requires a measure of the ``signal'' contained in workers' written proposals. This paper develops the argument that employers learn about worker ability in equilibrium via worker's expending costly effort to produce job-post-specific proposals. Thus, to measure signals, we use a novel, LLM-based measurement of how customized and relevant each proposal is to the specific job post to which it is being submitted. In this subsection, we first define our measurement, and then we provide further discussion of its design. Details of this measurement procedure can be found in Appendix~\ref{appendix_measurement_signal}, and comparisons to NLP-based alternative measurements can be found in Appendix~\ref{appendix_nlp}.

\subsubsection{LLM-Based Measurement}\label{measurement_signals_definition}

Ideally, to measure how customized and relevant each proposal is to its respective job post, we would have human readers evaluate each proposal relative to the job post according to a standardized rubric. However, doing so would be infeasible for millions of proposals, and thus we turn to LLMs to approximate human judgment at scale. We use \textit{Meta}'s Llama 4 Maverick 17B model to evaluate each proposal relative to its respective job post, one at a time, clearing its context window between each evaluation as to ensure independence between measurements.

To ensure the validity of our measure, we require a clear, reproducible rubric for how to quantitatively evaluate each proposal relative to its respective job post. To satisfy this requirement, we pose a series of nine questions to the LLM assessing each proposal on nine different criteria. In an effort to reduce the ``black box'' effect of using LLMs, we design our criteria to mimic a classification exercise.\footnote{In other words, we wanted to define a measure whose ``ground truth'' would be easy to assess, so we could compare the outputs of the LLM to a human reader's answers, which we did on a small subset of our data to validate our measure. We compared the degree of variability in scores between human and LLM assessments to those of scores between two different human readers. We found that the LLM produces scores that are no more different from a given human reader's score than the two humans' scores are from one another. Details can be found in Appendix~\ref{appendix_measurement_signal}} We restrict the answer to each question to ternary outcomes, 0, 1, and 2 with explicit definitions for each of the three classifications. While generally, one can think of the 0, 1, and 2 of each criterion corresponding to ``poor,'' ``adequate,'' and ``good,'' the details of how we pose these nine questions, the exact rubric we give the LLM on which to assess them, and the resulting score distributions for each criterion can be found in Appendix~\ref{appendix_measurement_signal}.

We separate our nine criteria into two categories. The first set directly aims to capture the extent to which proposals are customized to and engage with the details of the specific job post to which they are being submitted. These criteria assess whether the proposal contains clear, demonstrable evidence that the worker:
\begin{enumerate}
    \item Has read the details of the job post;
    \item Has tailored the proposal to the job post, avoiding boilerplate language;
    \item Understands the goal of the task described in the job post;
    \item Understands the complexity of the task described in the job post;
    \item Has demonstrated initiative to complete the task described in the job post.
\end{enumerate}

The second set of criteria were designed to capture elements of the proposal that may signal quality without necessarily being customized to the specific job post. These criteria assess whether the proposal:
\begin{enumerate}
    \item Directly mentions the requisite skills to complete the task described in the job post;
    \item Directly provides evidence of the worker possessing relevant experience;
    \item Is written in clear and mostly grammatically correct English;
    \item Is written using a professional, non ``spammy'' tone.
\end{enumerate}

We label the first set of five criteria ``custom'' criteria, and the second set of four criteria ``generic'' criteria.

\subsubsection{Copy-Pasting Correction}\label{measurement_signals_copypaste}

In order to mitigate the risk of false positives on the custom criteria, we adjust the scores based on whether we observe that a given proposal is a direct copy and paste of another proposal submitted by the same worker.

To address this concern, we compute the normalized minimum Levenshtein distance (i.e., ``edit-distance'') between the proposal being scored and all other proposals submitted by the same worker that we can observe, even beyond our estimation sample described above. These distances range from 0 to 1 and can be interpreted as the minimum percentage of words that would need to change in the job post to perfectly match the closest proposal submitted by that same worker. Our preferred measure of signal sets all customized criteria scores to 0 if the minimum distance computed above is below 4\%.

\subsubsection{Formal Definition}\label{measurement_signals_formaldef}

Indexing our five custom criteria by $(\text{custom}_{1}, \ldots, \text{custom}_{5})$, our four generic criteria by $(\text{generic}_{1}, \ldots, \text{generic}_{4})$, and denoting the normalized minimum Levenshtein distance by $d^{\text{edit}}$, we define our preferred signal measure $s$ for each proposal as:
$$
s = \frac{18}{28} \cdot \left(2\cdot \sum_{i=1}^{5} \text{custom}_i \cdot \mathbbm{1}\{d^{\text{edit}} \geq 0.04\} + \sum_{k=1}^{4} \text{generic}_k\right),
$$
where $d^{\text{edit}} = 1$ for any proposal for which we cannot compute an edit-distance due to not observing another application submitted by the same worker. Note that we place twice the weighting on the custom criteria to align with our a priori beliefs that customization carries greater signal of effort than relevance or writing quality. Also note that we scale our score by $\frac{18}{28}$ to ensure that it is on the same scale as a raw sum of the nine criteria.

\begin{figure}[h!]
    \centering
\includegraphics{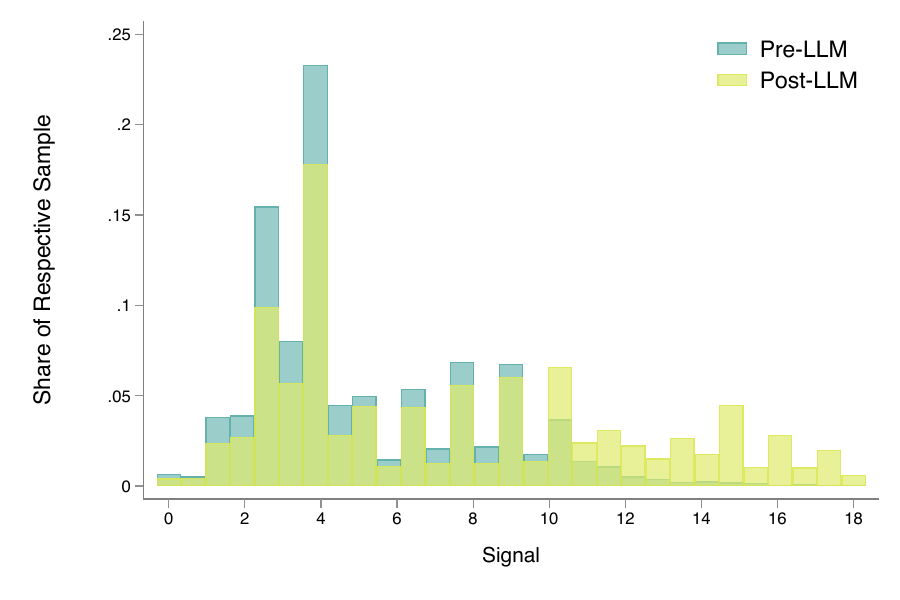}
    \caption{Distribution of Signals: Pre-LLM vs. Post-LLM}
    \label{fig:signals_prepost}
     \floatfoot{\footnotesize \textit{Notes:} This figure plots histogram estimates of the distributions of signals in both the pre-LLM and post-LLM samples. The post-LLM sample corresponds to the subsample of post-LLM job posts that were posted after March 26, 2024. This figure uses our preferred measure of signal, which places twice the weight on custom criteria than on generic criteria, and sets custom scores to zero if the application is more than 96\% copy-pasted from another proposal from the same worker. The pre-LLM sample has a mean of 5.16, a median of 3.86, and a standard deviation of 2.96, while the post-LLM sample has a mean of 7.50, median of 6.43, and a standard deviation of 4.63.}
\end{figure}

Our final sample of signals thus ranges from 0 to 18. Due to cost constraints, we could not use the LLM to grade every proposal in our data. Since we estimate our model on the pre-LLM sample, we grade all proposals in that period, which is around 960,000 applications, while only including around 39\% of applications in the post-LLM sample. To select the subsample of post-LLM applications to grade, we grade all applications submitted after March 26, 2024 (around 435,000 applications), and we randomly sample job posts from the period between November 30, 2022 and March 26, 2024, grading all applications submitted to those job posts (around 236,000 applications).\footnote{We choose these subsamples for two reasons. First, we wanted to have a large enough sample of graded applications close to the end of our sample period, so that we could argue the equilibrium has had time to settle after the introduction of the platform's AI proposal-writing tool in April 2023, when estimating our reduced-form logit model in Section~\ref{descriptives}. Second, we wanted to have a random sample of job posts from the entire post-LLM period to be able to compute the event-study analysis in Section~\ref{descriptives}.} Figure~\ref{fig:signals_prepost} displays histograms of the distribution of signals in both the pre-LLM and the post March 26, 2024 post-LLM samples. The average signal in the pre-LLM sample is 5.16 with a standard deviation of 2.96, while the average signal in the post March 26, 2024 post-LLM sample is 7.50 with a standard deviation of 4.63. See Appendix~\ref{appendix_measurement_signal_examples} for examples of low and high scoring proposals.

\subsubsection{Discussion of Signal Measurement}\label{measurement_signals_discuss}

Our main goal in developing our measure of signal is to quantitatively capture the extent to which each workers' written proposals may be signaling their abilities to employers in equilibrium. The challenge in doing so is that we need to construct a measure of signaling that is invariant to the equilibrium informational environment and thus is an equally valid measure in both the pre-LLM and post-LLM periods. In particular, our research question relies on the idea that signals, especially those that are on average produced in under 2 minutes of writing, do not themselves contain intrinsic information about worker ability. Instead, the information contained in these signals comes from the equilibrium in which workers with higher abilities on average send better signals. Thus, we are not simply trying to extract information from the text of the proposals but instead are aiming to measure something about these proposals that could only be increased with \textit{application-specific} costly effort before the introduction of LLMs. 

This challenge rules out two main methods of text analysis typically used by other researchers that may have otherwise seemed sensible solutions to measuring signals. First, we cannot train a machine learning model on successful application outcomes in the pre-LLM period and use this model to build a scoring metric since whether certain elements of the proposal text are predictive of successful applications would be endogenous to that period's market equilibrium.

Second, we cannot use more traditional measurements of writing quality, which typically involve Natural Language Processing (NLP) techniques that, for example, enumerate grammatical errors or score syntactic complexity. While of course it may require effort for workers to produce text free of grammatical errors or with complex syntax, especially given a large fraction of our sample of workers reside in non-majority English-speaking countries, these efforts need not be application-specific. In other words, if workers spend considerable effort to produce one perfectly written proposal that they then copy and paste when applying to multiple job posts, then the writing quality of those proposals may not reflect application-specific effort.

However, we do include the generic criteria in our measure of signal to ensure that we do not build a measurement that automatically satisfies our hypothesized signaling mechanism. Specifically, others in the literature have found evidence that directly contradicts the Spence-signaling framework when studying similar contexts (\cite{wiles2025algorithmic}). Including the generic criteria allows us to falsify whether it is the signal produced by application-specific effort that matters to employers, or whether it is simply the overall quality of the writing and relevance of the proposal that matters.

In support of our Spence-signaling hypothesis, Tables~\ref{tab:logit_signal_custom} and \ref{tab:logit_signal_generic} show that employers have a high willingness to pay (WTP) for the sum of the custom criteria scores in the pre-LLM sample, and their WTP for the sum of generic criteria scores is not statistically significantly different from zero. However, we also find that effort is a statistically significant positive predictor of both sums in the pre-LLM sample. These findings suggest that both the custom and generic criteria may take effort to produce, yet only the custom criteria serve as credible signals of employers.

Despite what these results suggest, in an effort not to define our measure based on its relationship with successful application outcomes, we include all nine criteria in our final measure of signal via a weighted sum that places twice the weight on the custom criteria than on the generic criteria, reflecting our a priori beliefs about the relative importance of the custom criteria to the signaling content of the proposals.

\subsection{Consideration Sets}\label{measurement_consideration}

In this subsection, we describe how we use click, bid ranking, and timestamp data to build our best proxy of which applications job posters actually see and consider.

Many applications submitted to job posts on DLPs are never read by employers, as verified by click data and anecdotal reports. Without measuring which applications are considered by employers, we may introduce significant bias into our demand estimation. For example, if unconsidered applications are more likely than considered ones to have lower bids, then including these unconsidered applications in our demand estimation would introduce attenuation bias into our estimate of employers' price sensitivity.

Our approach to measuring consideration sets revolves around three key assumptions. First, we assume employers are likely to consider applications that they engage with directly, such as clicking to expand the proposal if its text does not fit within the default bid list display window, clicking to ``pin'' an application to the top of their bid list, or sending a direct message to the worker who submitted the application. Second, we assume that if an employer engages with an application ranked $n$th at time $t$, then they are likely to consider all applications ranked above it, i.e., with ranks $< n$, at time $t$. Third, we assume that employers are more likely to consider applications the earlier they arrive and the higher they are ranked.

We define consideration sets via an algorithm that combines the three assumptions above. We assign each application a ``consideration score'' that is a function of (1) whether the employer directly engaged with the application, (2) the reputation score of the application, (3) the time of the application's submission, and (4) whether the application was scrolled past by the employer before engaging with a lower ranked application. We first define any application as considered if it has a nonzero consideration score, and then adjust these definitions to have stricter consideration score thresholds if a job post has implausibly many or few considered applications. Details of this algorithm can be found in Appendix~\ref{appendix_measurement_consideration}.

As per our research agreement with \texttt{Freelancer.com}, we do not report the exact fraction of applications that are considered in our sample, but we can report that our measured consideration sets are reasonably sized with about 12 applications per job post on average in the pre-LLM period and about 17 applications per job post on average in the post-LLM period. Reflecting the increase in the number of applications per job post in the post-LLM period, the average fraction of all applications that are considered decreases in the post-LLM period by about 14 percentage points.

See Table~\ref{tab:consideration_sum_stats} for summary statistics on the size and composition of consideration sets in the pre-LLM and post-LLM samples.

\section{Descriptive Evidence}\label{descriptives}

In this section, we present descriptive evidence suggesting that prior to the widespread adoption of LLMs, written proposals functioned as Spence-like signals of worker ability; however, following the widespread adoption of LLMs, this signaling mechanism collapsed. We organize this evidence into three main findings. First, signals are strong predictors of employer demand in the pre-LLM period. Second, signals are strong predictors of effort in the pre-LLM period, and, conditional on being hired, effort is a strong predictor of positive job completion outcomes. Third, in the post-LLM period, signals are much weaker predictors of employer demand and, for the proposals written with the platform's AI-writing tool, signals actually negatively predict effort.

We argue that these three findings taken together provide suggestive evidence that prior to the mass adoption of LLMs, written proposals functioned as a noisy, yet informative, signal of effort, which, in equilibrium, was a signal for a worker's propensity to successfully complete a job, but that these signals can no longer be relied on to predict job outcomes once LLMs dramatically lower the cost of producing such written proposals.

\subsection{Employers Demand Signals}\label{descriptives_1}

\begin{figure}[h!]
    \centering
\includegraphics{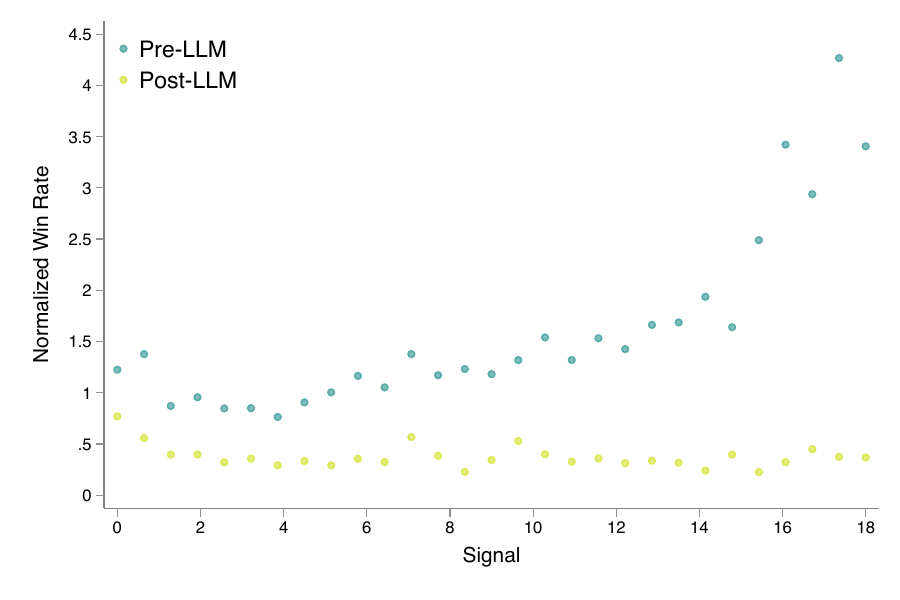}
    \caption{Binscatter Win Rate (i.e., Divided by Pre-LLM Mean) on Signal: Pre-LLM vs. Post-LLM}
    \label{fig:win_vs_signal_prepost}
     \floatfoot{\footnotesize \textit{Notes:} This figure plots a binscatter of whether the applying worker is hired on signal in both the pre-LLM and post-LLM samples. As per our research agreement with \texttt{Freelancer.com}, we do not disclose the level of hiring rates, so we normalize (i.e., divide) the $y$-axis by the unconditional pre-LLM overall win rate. The post-LLM sample corresponds to the subsample of post-LLM job posts that were posted after March 26, 2024. This figure uses our preferred measure of signal, which places twice the weight on custom criteria than on generic criteria, and sets custom scores to zero if the application is more than 96\% copy-pasted from another proposal from the same worker.}
\end{figure}

Figure \ref{fig:win_vs_signal_prepost} displays a binscatter of whether an application was ultimately awarded the job on our signal measure. As per our research agreement with the platform, we do not disclose the level of hiring rates, so we display win rates on the $y$-axis normalized (i.e., divided) by the pre-LLM overall win rate. Before taking a more formal econometric approach that acknowledges the inherent correlation between the outcomes of applications to the same job post, we first present this visual evidence of the demand response to signals. Note that worker competition is intense, with a very low hiring rate per application. 

Despite these low hiring rates, we see that pre-LLM applications in the right tail of the signal distribution have nearly four times higher hiring rates than the average application in the pre-LLM sample. These patterns, which are unconditional on consideration sets, hold despite an application's signal and inclusion in the consideration set having a negative correlation coefficient of -0.078 in the pre-LLM sample.

While suggestive of our signaling hypothesis, this pattern could be driven by other factors than a direct demand response to signals, such as the competition each application faces or other confounding variables correlated with signal.

Table~\ref{tab:logit_signal} presents the results of a series of estimated multinomial logit models that provide a reduced-form approach to capturing employer demand. These estimated models use a variety of explanatory variables in addition to signal including bid, reputation score, log arrival time (i.e., the time between the job posting and when a worker first clicks on the job post), and an applicant's country of residence. For each specification, we estimate a version of the model treating all applications as being in the consideration set and a version including only the applications we measured to be in the consideration set.

In both pre-LLM specifications used in Table~\ref{tab:logit_signal}, we find that signals are strong statistically significant predictors of employer demand. Given that the scale of our signal measure is arbitrarily defined (without loss), we interpret the coefficients on signal in Table~\ref{tab:logit_signal} by computing the decrease in an application's bid, all else equal, needed to achieve the same predicted increase in demand as a standard deviation increase in signal.

Using the estimates of Column (2) in Table~\ref{tab:logit_signal}, which only include considered applications, we find that in the pre-LLM sample, a one standard deviation increase in signal has the same predicted increase in demand, all else equal, as a \$25.67 decrease in bid, which is about 39\% of a standard deviation of considered bids in that sample.

\input{tables/logit_table_signal_edited.tex}

Since of course employers do not directly value workers' proposals themselves, these results suggest that prior to the mass adoption of LLMs, employers were either using workers' proposals as signals of attributes they do value, or signals were correlated with an attribute employers value that we do not observe. Since we observe virtually everything the employer sees about an application and the worker sending the application, and since over 99.9\% of applications in our sample are sent by workers who have never worked with the job poster before, we argue that the most likely explanation for these results is that employers were using workers' proposals as a signal of something they directly value.

\subsection{Signals Predict Effort, Which Predicts Job Completion}\label{descriptives_2}

\begin{figure}[h!]
    \centering
\includegraphics{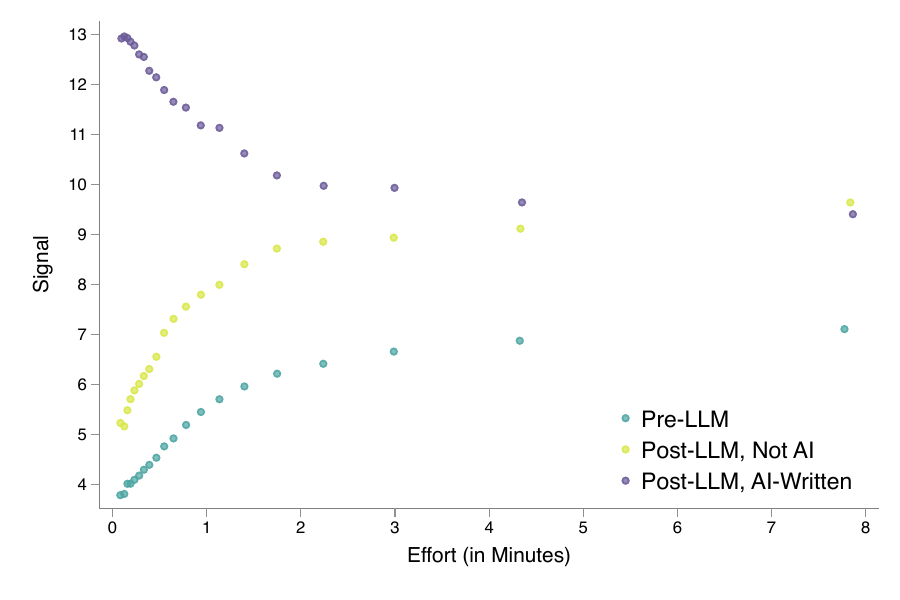}
\captionsetup{justification=centering}
\caption{Binscatter Signal on Effort\\ \small Pre-LLM vs. Post-LLM, Not AI vs. Post-LLM, AI-Written}
\label{fig:signal_vs_effort_prepostai}
 \floatfoot{\footnotesize\raggedright \textit{Notes:} This figure plots a binscatter of signal on effort in both the pre-LLM and post-LLM samples, splitting the latter by on-platform AI usage. The post-LLM sample corresponds to the subsample of post-LLM job posts that were posted after March 26, 2024. This figure uses our preferred measure of signal, which places twice the weight on custom criteria than on generic criteria, and sets custom scores to zero if the application is more than 96\% copy-pasted from another proposal from the same worker.}
\end{figure}

We argue that proposals primarily function as signals of effort expended by workers in reading the job post and crafting a written response. Figure~\ref{fig:signal_vs_effort_prepostai} presents binscatters of signal on effort. In the pre-LLM sample, we see a clear increasing and concave relationship between effort and signal. This relationship suggests that signals are produced with effort, i.e., the longer a worker spends reading the job post and/or writing her proposal, the more customized and relevant the proposal is to the job post.

While the visual evidence displayed in Figure~\ref{fig:signal_vs_effort_prepostai} suggests a possible causal relationship between effort and signal in the Pre-LLM period, one might question whether the pattern is produced simply by the fact that workers who spend more time reading and writing are also the types of workers to produce better written content. Thus, key to the argument that the relationship is causal is whether the pattern holds true within workers: when a given worker spends more time writing a proposal, does she produce a higher signal?

Table~\ref{tab:reg_s_on_log_e} presents Ordinary Least Squares (OLS) estimates that confirm this within-worker relationship between effort and signal, by regressing signal on log effort with worker fixed effects. Of the roughly 92,000 distinct workers we observe sending applications in the pre-LLM sample, around 32,000 of them send multiple applications. The average worker among those 32,000 workers, sends approximately 28 applications in the pre-LLM period. Furthermore, of the workers we see sending multiple applications, the average standard deviation of effort within worker is 1.27 minutes, which is about 65\% of the standard deviation of effort in the entire pre-LLM sample. Thus, we argue we have sufficient within-worker variation in effort to credibly test whether the relationship between signal and effort holds within worker.

\input{tables/reg_s_on_log_e.tex}

Column (2) of Table~\ref{tab:reg_s_on_log_e} shows that, even within worker, effort is a statistically significant positive predictor of signal. These results suggest that workers' written proposals function as noisy signals of effort.

\input{tables/reg_outcomes_on_signal.tex}

Next, we ask why employers might demand signals of effort. Table~\ref{tab:reg_5stars_on_signal} presents OLS estimates regressing whether a worker completes a job with a 5-star rating conditional on being hired on signal, effort, and various controls.\footnote{Visual evidence of this descriptive relationship between job completion outcomes and signals and effort is presented in Figure~\ref{fig:five_star_completion_binscatters}.} Columns (1) and (2) suggest that signals predict job completion outcomes, even after controlling for other variables that employers may use to select workers. Column (3) shows that once we control for effort, the coefficient on signal falls by about 41\% and that log effort itself is a strong predictor of job completion outcomes.\footnote{Table~\ref{tab:reg_5stars_on_signal_effcorr} re-estimates column (3) from Table~\ref{tab:reg_5stars_on_signal} with a corrected measure of effort, accounting for worker fixed effects in signal production, that we implement when estimating the model presented in Section~\ref{model}. Column (3) of Table~\ref{tab:reg_5stars_on_signal_effcorr} shows that once we control for corrected effort, signals are insignificant predictors of job completion outcomes. This result further motivates our claim that proposals are only used as signals in that they are predictors of effort, and that the information contained within signals, conditional on effort, do not predict worker ability.} This result implies that while signals do predict job completion outcomes, they do so in large part because they are correlated with effort, which itself is a strong predictor of job completion outcomes.

The regressions presented in Table~\ref{tab:reg_5stars_on_signal} are susceptible to selection bias since we only observe job completion outcomes for workers who are hired. For example, a worker selected with a low signal may be selected on some other unobservable, to the econometrician, component that positively predicts job completion outcomes. While we maintain that we observe virtually everything the employer sees about an application and the worker sending the application, this selection bias would bias our estimates of the relationship between signal and job completion outcomes towards zero. Thus, we argue that the coefficients on signal and effort in columns (1)-(3) of Table~\ref{tab:reg_5stars_on_signal} are most likely lower bound estimates of the true relationships.

Taking these results together, we argue that they provide suggestive evidence that prior to the mass adoption of LLMs, workers' written proposals functioned as Spence-like signals of effort, which employers cannot observe, and that effort itself predicts a worker's propensity to successfully complete a job. Thus, in other words, we have presented suggestive evidence that written proposals functioned as Spence-like signals of worker ability in the pre-LLM period.

\subsection{Signaling Breaks Down With Mass Adoption of LLMs}\label{descriptives_3}

The preceding two subsections built the argument that prior to the mass adoption of LLMs, workers' written proposals functioned as Spence-like signals of effort, and in equilibrium, functioned as signals of worker ability. In this subsection, we present descriptive evidence that suggests that this signaling story breaks down in the post-LLM period. To do so, we now examine the same empirical patterns presented in the preceding two subsections for the post-LLM period and, in particular, study the subset of applications written with the on-platform AI proposal-writing tool.

\subsubsection{Signals Shift Rightwards in the Post-LLM Period}

\begin{figure}[h!]
    \centering
\includegraphics[width=0.7\linewidth]{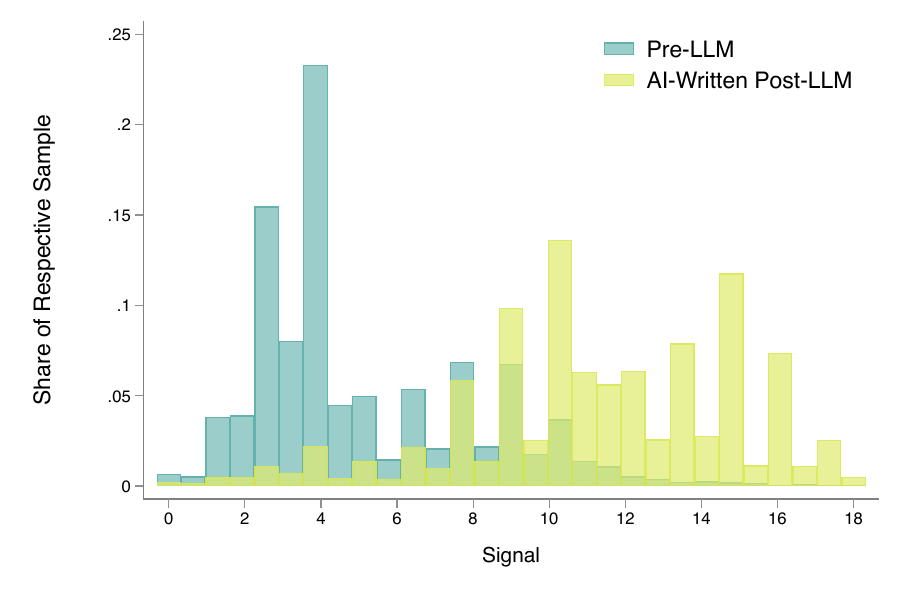}
    \caption{Distribution of Signals: Pre-LLM vs. Post-LLM, AI-Written}
    \label{fig:signals_preai}
     \floatfoot{\footnotesize \textit{Notes:} This figure plots histogram estimates of the distributions of signals in both the pre-LLM and post-LLM samples, where the latter is subsetted to only include the applications using the on-platform AI-writing tool. The post-LLM sample corresponds to the subsample of post-LLM job posts that were posted after March 26, 2024. This figure uses our preferred measure of signal, which places twice the weight on custom criteria than on generic criteria, and sets custom scores to zero if the application is more than 96\% copy-pasted from another proposal from the same worker. The pre-LLM sample has a mean of 5.16, a median of 3.86, and a standard deviation of 2.96, while the post-LLM, AI-written sample has a mean of 11.88, median of 12.21, and a standard deviation of 3.62.}
\end{figure}
\begin{figure}[h!]
    \centering
\includegraphics[width=0.7\linewidth]{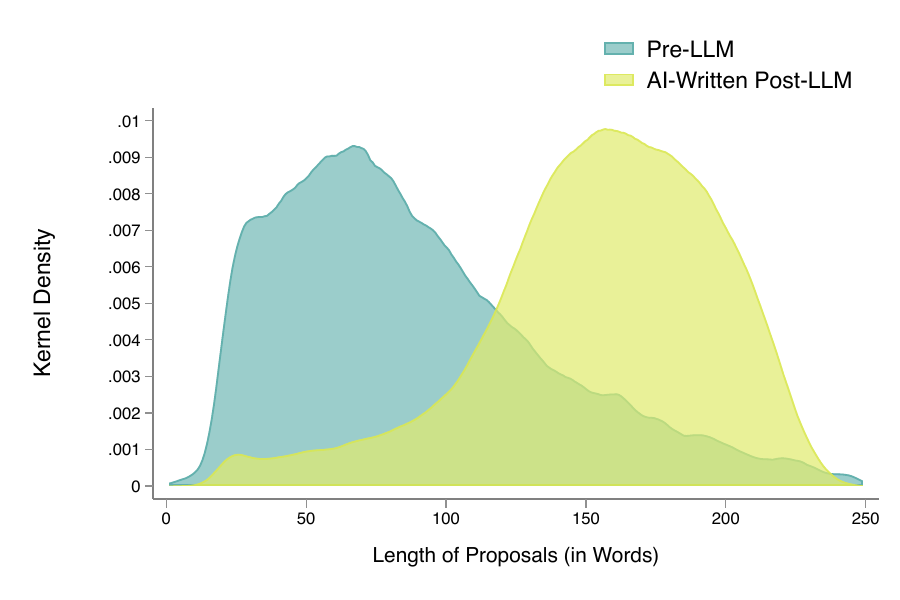}
    \caption{Distribution of Proposal Word Counts: Pre-LLM vs. Post-LLM, AI-Written}
    \label{fig:proposal_length_preai}
     \floatfoot{\footnotesize \textit{Notes:} This figure plots kernel density estimates of the distributions of proposal word counts in both the pre-LLM and post-LLM samples, where the latter is subsetted to only include the applications using the on-platform AI-writing tool. The post-LLM sample corresponds to the subsample of post-LLM job posts that were posted after March 26, 2024. The pre-LLM sample has a mean of 89 words, a median of 79 words, and a standard deviation of 50 words, while the post-LLM, AI-written sample has a mean of 155 words, a median of 159 words, and a standard deviation of 43 words.}
\end{figure}

Figure~\ref{fig:signals_prepost} shows that the distribution of signals has shifted rightwards in the post-LLM period, but that figure alone does not tell us whether that shift is driven by the use of LLMs to write proposals. Figure~\ref{fig:signals_preai}, however, shows that when workers use the platform's AI-writing tool, the distribution of signals is dramatically shifted to the right relative to both the pre-LLM sample: the average signal measurement in the pre-LLM sample is 5.16, and the average signal measurement of applications using the on-platform AI in the post-LLM sample is 11.88. Figure~\ref{fig:proposal_length_preai} shows a similar rightward shift in the distribution of proposal lengths of applications using the on-platform AI-writing tool.

Of particular note in Figure~\ref{fig:signals_preai} is the high density of proposals in the AI-written post-LLM sample that receive scores between 12 and 18, which were the scores that exhibited the sharp increase in hiring rates in the pre-LLM sample in Figure~\ref{fig:win_vs_signal_prepost}.

However, one might question whether this rightward shift in the distribution of signals displayed in Figure~\ref{fig:signals_prepost} is driven by some other factor than the use of LLMs. For example, it might be the case that the types of workers who use the platform's AI-writing tool are simply more likely to write better proposals.

\begin{figure}[h!]
    \centering
\includegraphics{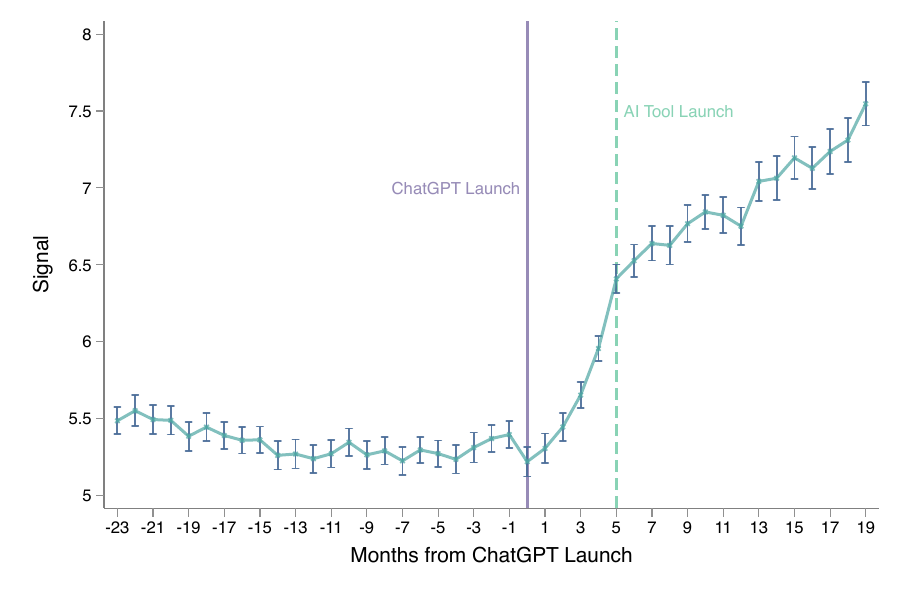}
    \caption{Event Study of Signals: Pre-LLM vs Post-LLM}
    \label{fig:signals_event_study}
     \floatfoot{\footnotesize \textit{Notes:} This figure plots event study estimates of the average signal over time residualizing out worker fixed effects and clustering standard errors at the worker-month level. The solid vertical line indicates when ChatGPT was released in November 2022, and the dashed vertical line indicates when the platform's AI-writing tool was introduced in April 2023.}
\end{figure}

To address this concern, Figure~\ref{fig:signals_event_study} presents an event study of the average signal over time residualizing out worker fixed effects, which differences out all time-invariant heterogeneity in attributes like writing ability, language, and prior experience. This figure shows that the average signal in the pre-LLM period is stable, and that there is a sharp increase in the average signal precisely when the platform's AI-writing tool was introduced in April 2023. Furthermore, we see that the average signal steadily increases over time in the post-LLM period, which is consistent with either the platform's AI-writing tool improving over time or more workers in general adopting the use of AI tools to help write their proposals. We argue that this event study provides strong evidence that the rightward shift in the distribution of signals displayed in Figures~\ref{fig:signals_prepost} and \ref{fig:signals_preai} is driven by the use of LLMs to write proposals.

\subsubsection{Demand for Signals Falls in the Post-LLM Period}

\begin{figure}[h!]
    \centering
\includegraphics{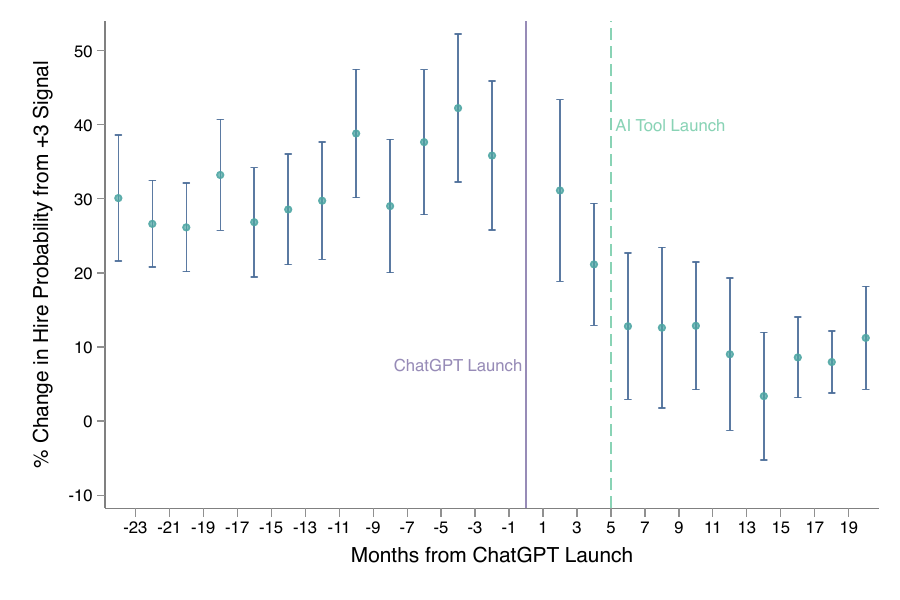}
    \caption{Event Study of Signals' Predicted Effect on Hiring Probability: Pre- vs Post-LLM}
    \label{fig:signals_beta_event_study}
     \floatfoot{\footnotesize \textit{Notes:} This figure plots the predicted percentage increase in hiring probability from a 3 unit increase in signals (approximately one standard deviation in the pre-LLM sample) from regressions of hiring on signal, bid, reputation score, log arrival time, and an applicant's country of residence estimated in two-month rolling windows over time, only including applications we measure to be in the consideration set. The percentage change measured on the $y$-axis is computed relative to that two-month window's baseline hiring probability for considered applications. Also plotted are 95\% confidence intervals around the point estimates. The solid vertical line indicates when ChatGPT was released in November 2022, and the dashed vertical line indicates when the platform's AI-writing tool was introduced in April 2023.}
\end{figure}

Despite the fact that the distribution of signals has a wider variance in the post-LLM sample than in the pre-LLM sample, Figure~\ref{fig:win_vs_signal_prepost}, Figure~\ref{fig:signals_beta_event_study}, and Table~\ref{tab:logit_signal} all show that signals are much weaker predictors of employer demand in the post-LLM period. Figure~\ref{fig:win_vs_signal_prepost} shows that the hiring rate is flat across the entire signal distribution in the post-LLM sample. To be clear, this figure does \textit{not} restrict the sample to applications using the on-platform AI tool and thus presents visual evidence of the lack of demand response to signals for any application in the post-LLM sample.

To investigate how the market adjusted to the new equilibrium, Figure~\ref{fig:signals_beta_event_study} presents an event study of the predicted percentage increase in hiring probability from a 3 unit increase in signals (approximately one standard deviation in the pre-LLM sample) from regressions of hiring on signal, bid, and controls estimated in two-month rolling windows over time, only including applications we measure to be in the consideration set. This figure shows that, prior to the introduction of LLMs, a 3 unit increase in signal predicted between a 30\% and 40\% increase in hiring probability, conditional on being considered. However, after the release of ChatGPT in November 2022, this predicted percentage change in hiring probability begins to fall, and after the platform's AI-writing tool was introduced in April 2023, this predicted percentage change in hiring probability falls to about 10\%. Note that the overall hiring rate is lower in the post-LLM period, so a 10\% increase in the post-LLM period corresponds to an even smaller absolute percentage point increase than a 10\% increase in the pre-LLM period. This figure provides evidence that employers begin to adjust their beliefs about signals shortly after the release of ChatGPT, and that these beliefs adjust downwards even further after the platform's AI-writing tool is introduced.

Column (4) of Table~\ref{tab:logit_signal}, which uses our preferred specification estimated only on applications we measure to be considered by the employer, confirms the visual evidence presented in Figures~\ref{fig:win_vs_signal_prepost} and \ref{fig:signals_beta_event_study} that the demand response to signals is dramatically weaker in the post-LLM sample. Comparing the coefficient on signal in Column (4) to that in Column (2), and noting that the coefficients on bid are essentially the same across both specifications, we find that the coefficient falls by almost two thirds from 0.0967 to 0.0340.

Although this finding alone shows that the demand response for the absolute levels of our signal measure falls, it does not necessarily indicate that proposals are relied on less as signals in the post-LLM period. For example, if the standard deviation of signals increased so much that the same change in the level of signal did not produce the same demand response as the pre-LLM period, it may still be that a one standard deviation increase in post-LLM signals produces the same demand response as a one standard deviation increase in pre-LLM signals. However, as we can see from the results in column (4) of Table~\ref{tab:logit_signal}, a one standard deviation increase in post-LLM signal has the same predicted increase in demand, all else equal, as a \$14.85 decrease in bid, as compared to a \$25.67 decrease in bid in the pre-LLM sample. Thus, we find that even when accounting for the increased variance in signals, signals are still much weaker predictors of employer demand in the post-LLM period.

One might have assumed that the introduction of LLMs made all proposals ``look the same.'' As Figures~\ref{fig:signals_prepost} and \ref{fig:signals_preai} show, this assumption does not hold true and that there is actually more variance in the signal distribution in the post-LLM sample than in the pre-LLM sample. Thus, we conclude that the demand response to signals has fallen in the post-LLM period not because one can no longer differentiate between proposals, but instead because employers no longer value the information contained in the variation in proposals. 

In the following subsection, we provide evidence for why the demand response to signals collapses: effort no longer predicts signal post‑LLM, and the correlation between signal and effort even turns negative for AI‑written proposals.

\subsubsection{Effort No Longer Predicts Signals in the Post-LLM Period}

To see why employers may no longer value the information contained in signals, we point to the findings in Figure~\ref{fig:signal_vs_effort_prepostai} and Table~\ref{tab:reg_s_on_log_e}. Figure~\ref{fig:signal_vs_effort_prepostai} shows that in the post-LLM sample, there is a clear decreasing and concave relationship between effort and signal for applications written with the platform's AI-writing tool. In other words, when writing without the help of an LLM in the pre-LLM period, higher effort is associated with higher signals; but, when writing with the help of the on-platform AI tool, higher effort is associated with lower signals, while still having a higher average signal at all levels of effort than in the pre-LLM sample. Appendix Figure \ref{NLP_effort_vs_signal} shows that these trends are robust across a wide range of alternative NLP measures of signals (TF-IDF, Average Words Embeddings, Sentence-BERT). These visual results are confirmed by the estimates presented in Table~\ref{tab:reg_s_on_log_e}. Moreover, column (4) of Table~\ref{tab:reg_s_on_log_e} shows that, when controlling for worker fixed effects, even for those applications not using the platform's AI-writing tool, effort is a weaker predictor of signal in the post-LLM period than in the pre-LLM period.

In the pre-LLM sample, just over 77.51\% of the applications with valid effort measurements are being produced by workers reading the job post, crafting a written proposal, and submitting their applications in under 2 minutes, with an average effort of 1.47 minutes.\footnote{Figure~\ref{fig:bidtimes_preai} displays how the distribution of efforts differs between the pre-LLM sample and the applications written with the on-platform AI tool in the post-LLM sample.} One need only use an LLM to costlessly generate a written proposal that is as informative as one a worker could have produced with just two minutes of reading and writing, to have the potential to disrupt the existing signaling equilibrium on the platform. Thus, we argue it should be clear why LLMs have the potential to significantly disrupt the signaling value of written communication on the platform. See Appendix~\ref{appendix_measurement_signal} for examples of proposals written with relatively low and high effort, the first pair from the pre-LLM period and the second pair from the post-LLM period. These examples illustrate how a difference in signals between a proposal written in 30 seconds and another written in 2 minutes is much more pronounced in the pre-LLM period than in the post-LLM period.

\subsubsection{Signals No Longer Predict Job Completion Outcomes in the Post-LLM Period}

Columns (4)-(6) of Table~\ref{tab:reg_5stars_on_signal} shed light on why employers may no longer value signals in the post-LLM period. All three columns show that signals are no longer statistically significant predictors of job completion outcomes in the post-LLM period, even before controlling for effort. Column (4)'s coefficient on signal is an order of magnitude smaller than that in Column (1), and column (5)'s coefficient on signal falls by about 80\% relative to that in Column (2). Finally, column (6) shows that effort itself is still a statistically significant predictor of job completion outcomes in the post-LLM period, with a slightly lower coefficient than in the pre-LLM period. However, given that signals can no longer reliably predict effort in the post-LLM period, employers can no longer rely on signals to predict job completion outcomes.

\subsection{Need for a Model}

Taken together, this section presents suggestive evidence that prior to the mass adoption of LLMs, workers' written proposals functioned as Spence-like signals of effort, which employers could rely on to predict a worker's propensity to successfully complete a job. In the post-LLM period, however, this signaling mechanism appears to have collapsed: signals no longer predict effort or job completion outcomes, and thus employers no longer value signals.

Although this descriptive evidence is suggestive, it does not allow us to quantify how LLMs’ lowering of the cost of writing proposals has affected the distribution of ability among hired workers, hiring patterns, or employer and worker welfare. Moreover, descriptive evidence alone cannot isolate LLMs’ effects on signaling from other changes LLMs may have induced by altering the nature of work.

To both quantify and isolate the effect of LLMs' disruption of labor market signaling, we need to (1) identify the joint distribution of worker abilities and costs of undertaking jobs, (2) identify how employers' hiring decisions change when LLMs erode the informativeness of signals, and (3) know how workers of differing abilities would change their bidding behavior in an equilibrium without signaling.

To see why these three inferences are necessary, suppose we observe the pattern that workers sending better signals tend to bid higher. First, consider the case in which this pattern results from high-ability workers knowing they can signal their higher-than-average value to employers and thus bidding high to capture more surplus, rather than high-ability workers having higher costs than low-ability workers do. In this case, if LLMs eliminate signaling, high-ability workers would lower their bids to compete with lower-ability workers, and employers may end up hiring a similar distribution of workers as they would with signaling. Now, consider the case in which this pattern results from high-ability workers having higher costs of undertaking the job than low-ability workers do. In this case, if LLMs eliminate signaling, high-ability workers may not be able to lower their bids enough to compete with low-ability workers, and employers may end up hiring far fewer high-ability workers.

We therefore build a model of labor market signaling in the pre-LLM market, which will allow us to identify these three necessary components. Once we have estimated the model, we can simulate a counterfactual market equilibrium without signaling, thereby allowing us to quantify the isolated effect of LLMs disrupting signaling on hiring patterns and welfare.

\section{A Model of the Pre-LLM Market}\label{model}

We develop a static model that combines three typically distinct modeling approaches: we embed a (1) Spence signaling model---workers invest costly effort to produce noisy signals that positively correlate with their ability in equilibrium---into a (2) discrete choice demand model---employers form indirect expected utilities over application characteristics and their beliefs about ability---which, from the workers' point of view, operates as a (3) scoring auction---workers submit applications competing on multiple dimensions to win a contract.

The overall structure and timing of the model is as follows. Workers arrive at job postings, learn their own ability and cost of undertaking the job, and apply to these jobs choosing both a price bid and an effort level to expend on signaling. Signals are realized as a noisy function of worker effort, and consideration sets are formed based on the observable characteristics of the workers. Each job poster then hires the worker in her consideration set that maximizes her expected utility, provided that that worker provides a greater expected utility than the job poster's outside option. By hiring a worker, the job poster commits to pay that worker equal to her bid, and the worker commits to do the work for which she was hired to the best of her ability. 
The rest of this section presents the details of our model.

\subsection{Job Posting and Worker Arrival}\label{model_arrival}
We model jobs being posted and workers arriving at job posts as being a random exogenous process, and we assume that all workers who arrive at a job post submit an application.\footnote{See \cite{krasnokutskaya2020role} for a paper studying a similar DLP context in which worker entry to a job post is endogenous. This assumption significantly changes and complicates identification and estimation and is at the core of the empirical challenges overcome by their approach.}

There is a large number, $I$, of ex-ante identical employers, each posting one job on a monopolistic platform, hoping to outsource a task. We use $i$ to interchangeably index both the job post and the employer posting the job for notational convenience. Each job post, $i$, has set of $N_i$ workers applying to it, indexed by $j \in \{1, \dots, N_i\}$.

Each worker $j$ bidding on job post $i$ has a vector of discrete observables $x_{i j} \in X$ that workers do not control.\footnote{Note that we include the $i$ subscript on $x$ for two distinct reasons. First, since workers are indexed from 1 to $N_i$ for each job post $i$, we need a way to distinguish, for example, worker $j = 3$ applying to job post $i = 1$ from worker $j = 3$ applying to job post $i = 2$. Second, although the model presented in this section has each worker only appearing once, in our empirical implementation we see workers applying to multiple job posts, and since some dimensions of $x_{i j}$, namely a worker's arrival time and a worker's platform reputation at the time of applying to job post $i$, can change over time, it is useful to distinguish between a given worker's characteristics when applying to one job post and that same worker's characteristics when applying to another.} In our empirical implementation of this model, this vector contains where the worker is from, when the worker arrives at the job post relative to when the job is posted, and the worker's current reputation on the platform. We discuss the precise construction of this vector in Section~\ref{estimation}.

Since these observable characteristics are discrete and $X$ is finite, each worker belongs to one of finitely many observable types, $|X|$. Thus, we can provide an ex-ante characterization of each job post as an $|X|$-dimensional vector, $J_i$, of natural numbers that define the number of each $x$-type worker that arrives at job $i$.\footnote{Recall that we constructed our sample so that job posts are homogenous on most observable dimensions. Thus, we do not include employer or job post observables in our characterization of each job post.} We define $\Gamma$ to be a discrete distribution over the set of possible ex-ante worker arrivals, $J_i$, that describes how workers jointly arrive at job posts. Worker arrival is completely determined by their observable type $x$, and thus worker arrival is neither endogenous nor selected on unobservables.

\subsection{Worker Types}\label{model_worker_types}
Upon arrival, workers draw a two-dimensional type $(c_{i j}, a_{i j}) | x_{i j} \sim $ i.i.d. from a distribution $F_{c \times a | x}$, which we allow to have any differentiable joint distribution over a compact connected support in $\mathbb{R}^2_+$. We define $c_{i j}$ to be worker $j$'s opportunity cost of taking the job $i$ and thus represents the amount she would need to be paid to be exactly indifferent between taking the job and not taking the job. We define $a_{ij}$ to be worker $j$'s ability to do the task involved in job post $i$.

\subsection{Worker Actions and Ex-Post Payoffs}\label{model_worker_actions}
After observing their unobservable type $(c_{i j}, a_{i j})$, workers choose two actions: a bid $b_{i j}$ and a signaling effort $e_{i j}$. The latter entails a cost $C(e_{i j}; a_{i j})$, where the cost function satisfies the following properties:
\begin{enumerate}[label=(\roman*)]
\item $C(e; a)$ is twice continuously differentiable in both arguments for all $e > 0$ and all $a$;
\item $\lim_{e \to 0} C(e; a) = 0$ for all $a$;
\item $\frac{\partial C(e; a)}{\partial e} > 0$ for all $e > 0$ and all $a$;
\item $\frac{\partial^2 C(e; a)}{\partial e^2} > 0$ for all $e > 0$ and all $a$;
\item $\frac{\partial C(e; a)}{\partial a} < 0$ for all $e > 0$ and all $a$.
\item $\frac{\partial^2 C(e; a)}{\partial e \partial a} < 0$ for all $e > 0$ and all $a$.
\end{enumerate}

This cost function implies that higher-ability workers can produce any given level of signaling effort at lower cost than lower-ability workers can. Moreover, higher-ability workers face lower marginal costs of increasing effort at any given level of effort.

Workers pay this effort cost upfront, regardless of whether they are hired, but they only receive the payment $b_{i j}$ and incur the opportunity cost of working $c_{i j}$ if they are awarded the job. Thus, a worker's ex-post utility from bidding $b_{i j}$ and expending effort $e_{i j}$ is given by:
\begin{align}
    U^W(b_{i j},e_{i j};c_{i j},a_{i j},w_{i j})\equiv
    \underbrace{w_{i j}}_{\text{indicator for winning}}\, \cdot \, \underbrace{(b_{i j}-c_{i j})}_{\text{net benefit of working}} - \underbrace{C(e_{i j};a_{i j})}_{\text{upfront cost of effort}} 
\end{align}
where $w_{i j}$ is an indicator for whether worker $j$ is awarded job $i$.

\subsection{Signal Production}\label{model_signals}
Workers' signaling effort $e_{ij}$ is combined with signaling noise $\varepsilon^s_{ij}$ to produce a signal $s_{ij}$, conditional on their observable type $x_{ij}$, according to signal production function
\begin{align}
    s_{ij} = s(e_{i j}; \varepsilon^s_{i j}, x_{i j}) \equiv f^s(e_{i j}; x_{i j}) + \varepsilon^s_{i j},
\end{align}
where:
\begin{enumerate}[label=(\roman*)]
    \item $\varepsilon^s_{i j}$ is independent of $(c_{ij},a_{ij},e_{ij})$ conditional on $x_{i j}$;
    \item $\varepsilon^s_{i j}$ is distributed i.i.d. according to $F_{\varepsilon^s | x}$ with $\mathbb{E}[\varepsilon^s_{i j} | x_{i j}] = 0$ for all $x_{i j} \in X$;
    \item for all $e > 0$ and all $x \in X$, $f^s(e; x)$ is continuously differentiable in $e$;
    \item for all $e > 0$ and all $x \in X$, $\frac{\partial f^s(e; x)}{\partial e} > 0$.
\end{enumerate}

By modeling signal production in this way, we are making two key assumptions. First, we are assuming that workers know how much effort they put into a proposal, and they know the distribution of signals that this effort might produce, but they do not know what that realization will be. Second, we are assuming that a worker's ability $a_{ij}$ only affects the cost of producing signals, not the production of signals themselves.

\subsection{Consideration Sets}\label{model_consideration}
A key feature of our empirical setting is that employers do not consider all workers who apply to their job posts. Similar to our model of worker arrival, we model consideration sets as being a random exogenous process.
In particular, our model incorporates a very specific notion of consideration: consideration is a binary status such that an employer observes an application if and only if it is considered.

We model consideration set formation as being jointly dependent on all the observables of the applications that job post receives, $\{x_{ij}\}_{j \in N_i}$.\footnote{For example, a worker's reputation score determines her positioning on the bid list only insofar as it compares to her competitors' reputation scores.}

Formally, we model the probability that a worker $j$ is considered by job poster $i$ with the function
\begin{align}
    Q(x_{ij}, x_{i, -j}) \equiv \mathbbm{P}(q_{i j} = 1 | x_{ i j}, x_{i, -j}): X^{N_i} \rightarrow (0, 1),
\end{align}
where $x_{i, -j} \equiv \Big(x_{i k} \Big)_{k \in \{1, ..., N_i\} \text{ and } k \neq j}$ is the vector of observable characteristics of all the other workers applying to job post $i$, and $q_{i j}$ is the indicator for whether employer $i$ considers worker $j$. Furthermore, let $Q_i \equiv \Big\{j \in \{0, 1, ..., N_i\} \mid q_{i j} = 1\Big\}$ be employer $i$'s consideration set, where $j = 0$ represents the outside option discussed below.

\subsection{Employer Choices and Ex-Post Payoffs}\label{model_employer_expost}
We model employer labor demand following the tradition of differentiated goods discrete choice demand models. In particular, we model each employer $i$'s indirect utility of hiring worker $j$ as\footnote{Employers can only hire those applications in their consideration sets, but their utility from hypothetically hiring an unconsidered worker who applies to their job post is still well-defined.}:
\begin{align}\label{eq:employer_utility}
    U^E(b_{i j},  a_{i j}, \nu_{i j}; x_{i j}) = T(x_{i j}) + \beta \cdot    a_{i j} - \alpha \cdot  b_{i j} + \nu_{i j},
\end{align}
where $T(x)$ is the utility for hiring a worker with observables $x$, holding fixed the worker's ability and bid, and $\nu_{i j}$ is a stochastic utility shock distributed i.i.d.\ according to $F_{\nu}$.  

Moreover, employer $i$ receives utility $U_{i 0} \equiv U_0 + \nu_{i 0}$ from selecting the outside option---e.g., doing the work themselves, or simply not hiring anyone---where $\nu_{i 0}$ is also drawn i.i.d.\ according to $F_{\nu}$, and we normalize $U_{i0}$ to 0.

We interpret $T(x)$ as potentially including both worker ability, common to all workers from group $x$, that does not enter signaling costs and a taste-based utility from hiring a member of that group.\footnote{We are not be able to recover non-signaled worker ability, since we cannot separately identify ability from taste-based utility in $T(x)$, but this poses no challenge to estimating our counterfactuals since we are only changing how workers are able to specifically signal their signaling-relevant ability $a_{ij}$.} For ease of exposition, we refer to the signaling-relevant ability $a_{i j}$ simply as ``ability'' in the remainder of the paper.

In addition, we interpret $\nu_{ij}$ as an employer-worker-specific idiosyncratic taste shock or match effect. Our key assumption is that workers do not observe $\nu_{ij}$ before making their bid and effort choices.\footnote{For example, the taste shock could be derived from looking at a worker's profile picture or the aesthetics of their writing style. Most likely, given we observe essentially everything that an employer observes about an application, other than a worker's profile picture and name, this term either captures pure randomness in employer decision-making (e.g., salience bias for certain applications) or it captures elements about the text signal that we do not capture in our measure of signal, i.e., it contains measurement error.} 

Finally, we assume that employers exogenously abandon their job post, completely independent of the workers who apply to it, with probability $(1-\pi)$, where $\pi \in (0, 1)$. This assumption helps rationalize why employers sometimes choose their outside options even when they have very good workers in their consideration sets.

\subsection{Worker Information and Strategies}\label{model_worker_strategies} 

Workers all simultaneously choose their actions after observing their own abilities $a_{i j}$, costs $c_{i j}$, and observable types $x_{i j}$ without observing anything about their competitors, including how many of them there are. Thus, the strategy of each worker $j$ applying to job post $i$ is a function
\begin{align}
    \sigma_{i j}: \left(\mathbb{R}^2 \times X\right) \rightarrow \mathbb{R}^2_+
\end{align}
that maps their type $(c_{i j}, a_{i j}; x_{i j})$ to their actions $(b_{i j}, e_{i j})$. Where useful, we define the  functions $\sigma_{i j}^b(c, a; x)$ and $\sigma_{i j}^e(c, a; x)$ to denote the bid and effort components of worker strategies, respectively, such that $\sigma_{i j}(c, a; x) = \left(\sigma_{i j}^b(c, a; x), \sigma_{i j}^e(c, a; x)\right)$.

\subsection{Employer Information and Strategies}\label{model_employer_strategies} 

Each employer chooses which worker to hire, or the outside option, after observing only the bids, signals, observable characteristics of the workers in their consideration sets, as well as the utility shocks of those workers and the outside option. Thus, the strategy of each employer $i$ with consideration set $Q$ is a function
\begin{align}
\tau_i: \left(\mathbbm{R}^{3 \cdot\left|Q\right| -2} \times X^{|Q|-1}\right) \rightarrow Q
\end{align}
that maps the vector of considered bids $\mathbf{b}$, signals $\mathbf{s}$, observable characteristics $\mathbf{x}$, and utility shocks $\boldsymbol{\nu}$ to a discrete choice of which worker in the consideration set to hire, or the outside option.\footnote{The term $3\cdot |Q| - 2$ in the domain of $\tau_i$ comes from the fact that the outside option has no bid or signal but does have a utility shock, i.e., we have $|Q| - 1$ bids, $|Q| - 1$ signals, and $|Q|$ utility shocks.} 

\subsection{Equilibrium Definition}\label{model_eqm_def}

We restrict attention to \textit{type-symmetric pure-strategy Perfect Bayesian Equilibria} (psPBE), which we define in our setting to be a worker strategy profile $\sigma^*$, an employer strategy profile $\tau^*$, and a system of beliefs\footnote{We omit discussion of off-path beliefs, and we further restrict our equilibrium concept to one in which all bids in the allowed range for the job posts in our empirical sample, i.e., $[\$30, \$250]$ are on-path. Moreover, we note that the signaling noise in each worker's signal production yields a full support of on-path signals for any on-path effort. This feature of our model in combination with the assumption that employers do not observe worker effort implies beliefs need not be defined for off-path effort.} for employers $\mu^*$ such that, for all possible realizations of worker types, competitor sets, consideration sets, signaling noise, and utility shocks:
\begin{enumerate}
    \item Employers form correct beliefs\footnote{Note that because employer utility (Equation~\ref{eq:employer_utility}) is linear in ability, the equilibrium only depends on the mean of employers' posterior beliefs, rather than the whole posterior belief itself.} according to Bayes' rule:
    \begin{align*}
        \mu^*\left(b, s; x, \sigma^*\right) = \mathbbm{E}_{c, a, \varepsilon^s} \left[ a \: \Big| \: \sigma^{*b}(c, a; x) = b, \:\: s\left(\sigma^{*e}(c, a; x); \varepsilon^s, x\right) = s; \; x\right]
    \end{align*}
    \item Employers decide whom to hire to maximize their expected utility:
    {\notsosmallbutbigger
        \begin{align*}
            \tau^*\left(\mathbf{b}, \mathbf{s}, \boldsymbol{\nu}; \mathbf{x}, Q\right) =
            \begin{cases}
            \displaystyle \argmax_{k\in Q \setminus \{0\}} \left\{U^E\left(b_k, \mu^*(b_k, s_k; x_k, \sigma^*),\nu_k; x_k\right) \right\}
            & \hspace{-1.5mm} \text{if} \displaystyle \max_{k\in Q \setminus \{0\}} \left\{U^E\left(b_k, \mu^*(b_k, s_k; x_k, \sigma^*),\nu_k; x_k\right)\right\} > \nu_0\\
            0 & \hspace{-1.5mm} \text{if otherwise}
            \end{cases}
        \end{align*}
    }
    \item Workers choose bids and efforts to maximize their expected utility:
        \begin{align*}
            \sigma^*\left(c, a; x\right) = \argmax_{(b, e) \in \mathbb{R}^2_+} \left\{p\left(b, e; x, \sigma^*, \tau^*\right)\cdot (b - c ) - C(e; a ) \right\},
        \end{align*}
        where $p\left(b, e; x, \sigma, \tau\right)$ is the ex-ante probability that worker $j$ is hired:
        {\notsosmall
        \[
        \mathbb{P}_{c_{-j}, a_{-j}, x_{-j}, \boldsymbol{\varepsilon^s}, \boldsymbol{\nu}, Q}
        \left(
        \begin{aligned}
        & j \in Q \text{ and } \\
        & j =
        \tau\left(b,\; \sigma^b_{-j}(c_{-j}, a_{-j}; x_{-j}), \; s(e; \varepsilon^s_j, x),\; s_{-j}\left(c_{-j}, a_{-j}; \varepsilon^s_{-j}, x_{-j}, \sigma^e\right), \;
         \boldsymbol{\nu}; \; x, \, x_{-j}, \; Q\right)
        \end{aligned}
        \,\middle|\, b,e;\, x,\sigma,\tau
        \right)
        \]
        }
        with
    \begin{flalign*}
    &\sigma^b_{-j}(c_{-j}, a_{-j}; x_{-j}) \equiv \left(\sigma^b(c_k, a_k; x_k)\right)_{k\in Q \setminus \{j,\, 0\}}, &&\\
    &s_{-j}\left(c_{-j}, a_{-j}; \varepsilon^s_{-j}, x_{-j}, \sigma^e\right) \equiv \left(s\left(\sigma^e(c_k, a_k; x_k);\; \varepsilon^s_k, x_k\right)\right)_{k\in Q \setminus \{j,\, 0\}}, &&\\
    &c_{-j} \equiv (c_k)_{k\in Q \setminus \{j,\, 0\}}, \;
    a_{-j} \equiv (a_k)_{k\in Q \setminus \{j,\, 0\}}, \;
    x_{-j} \equiv (x_k)_{k\in Q \setminus \{j,\, 0\}}, \text{ and }
    \varepsilon^s_{-j} \equiv (\varepsilon^s_k)_{k\in Q \setminus \{j,\, 0\}}.&&
    \end{flalign*}

\end{enumerate}

\subsection{Identification}\label{id}

Identifying the model poses two key challenges. First, to identify workers' costs and abilities from their observed equilibrium bids and efforts, we must first identify their ex-ante equilibrium beliefs about their chances of being hired as functions of their bids, efforts, and observables.\footnote{For example, suppose we observe a worker submitting a high bid. Without identifying that worker's beliefs about her probability of being hired, we cannot disentangle whether she is bidding high because she has a high cost of undertaking the job or because she believes she has a high chance of winning even with a high bid.} Second, to identify employers' preferences from their observed equilibrium choice probabilities, we must first identify their equilibrium beliefs about worker ability as functions of bids, signals, and observables.\footnote{For example, suppose we observe that employers are willing to hire workers with high bids. Without identifying employers' beliefs about worker ability, we cannot disentangle whether this pattern is the result of employers being price insensitive or believing that high-bid workers have high ability. Additionally, suppose we see that employer demand is highly responsive to signals. Without identifying how employers form beliefs about ability as a function of those signals, we cannot disentangle whether this responsiveness is due to employers having a high WTP for ability or due to employers' beliefs about ability being highly responsive to signals.}

We overcome these two challenges sequentially by exploiting the information structure of our model. First, we identify worker beliefs from observed conditional hiring probabilities. Because we observe everything that is commonly observable to both workers and firms, a worker's ex-ante belief about her probability of being hired is independent of her private type (cost and ability), conditional on her bid, effort, and observables. As a result, worker beliefs equal empirical hiring probabilities as functions of bids, efforts, and observables, and thus they are directly identified from data on employer choices. Once we have identified worker beliefs, we can invert workers' first order conditions (FOCs) to infer their costs and abilities from their observed bids and efforts.

Second, having identified worker costs and abilities, and thus having identified the joint distribution of worker private types, actions, and observables, we can identify employer beliefs about ability as functions of bids, signals, and observables. These identified employer beliefs enable us to identify employer demand through the standard discrete choice arguments.

The remainder of this subsection presents the preceding argument in more detail.

\subsubsection{Identifying Supply}\label{id_supply} 

The key to identifying labor supply is that a worker's private types neither affect her chance of being hired, conditional on her bid and effort, nor the realization of her signal, conditional on her effort. Thus, observing hiring probabilities as functions of bids, efforts, and observables directly identifies worker beliefs about hiring probabilities. We denote this equilibrium hiring probability function $P^*(b, e; x)$, which is the reduced form of $p\left(b, e; x_{ij}, \sigma^*, \tau^*, \mu^*\right)$.

We can write workers' first order conditions for optimal bids and efforts as a function of $P^*(b, e; x)$ and its derivatives\footnote{We conjecture that due to the noise in signal production and the taste shock $\nu$, the reduced-form win-probability $P^*(b, e; x)$ is differentiable in both bids and efforts over $\mathbb{R}_+^2$.}:
\begin{align}
    P^*(b, e; x) + \left(\frac{\partial}{\partial b} P^*(b, e; x)\right)\cdot\left(b - c\right) = 0;\label{eq:foc_bid} \\
    \left(\frac{\partial}{\partial e} P^*(b, e; x)\right) \cdot \left(b - c\right) - \frac{\partial C(e; a)}{\partial e} = 0.\label{eq:foc_effort}
\end{align}

To achieve identification of the joint distribution of worker types $F_{c\times a | x}$, we assume that each worker's observed bid and effort are interior solutions to her optimization problem\footnote{We thus assume that each worker choosing a bid of \$30 or \$250 is globally optimizing over $\mathbb{R_+}$.} and that Equations \ref{eq:foc_bid} and \ref{eq:foc_effort} hold with equality for all applications $(i, j)$. Thus, we can invert the FOCs, following the tradition of \citet{GPV}, and write them as:
\begin{align}
    c_{i j} &=
    b_{i j} + P^*(b_{i j}, e_{i j}; x_{ij}) \cdot \left(\frac{\partial}{\partial b} P^*(b, e; x_{i j})\Big|_{(b, e) = (b_{i j}, e_{i j})}\right)^{-1};\label{eq:foc_bid_invert} \\
    \frac{\partial C(e; a_{ij})}{\partial e}\Big|_{e = e_{i j}} &=
    -P^*(b_{i j}, e_{i j} ; x_{i j})\cdot\left(\frac{\partial}{\partial e} P^*(b, e; x_{i j})\Big|_{(b, e) = (b_{i j}, e_{i j})}\right)\cdot\left(\frac{\partial}{\partial b} P^*(b, e; x_{i j})\Big|_{(b, e) = (b_{i j}, e_{i j})}\right)^{-1}.\label{eq:foc_effort_invert}
\end{align}

The right-hand sides of Equations \ref{eq:foc_bid_invert} and \ref{eq:foc_effort_invert} are both identified from having identified $P^*(b, e; x)$. Thus, we have identified the joint distribution of worker costs and equilibrium marginal costs of effort.

To translate these marginal costs to abilities, we must specify a functional form for the effort cost function $C(e; a)$ with the additional property that for any $k > 0$ and $e' > 0$, there exists a unique $a^*$ in the support of $a$ such that $\frac{\partial C(e; a^*)}{\partial e}\Big|_{e = e'} = k$. Note that $a$ is serving as the free parameter in the cost function, and since it varies at the $(i, j)$ level, we cannot identify any additional parameters in the cost function.

Finally, note that, because ability enters linearly in employer utility, any affine rescaling of ability in the effort cost function can be absorbed by the corresponding utility coefficient.\footnote{For example, if we defined ability to enter the cost function as $2a$ rather than $a$, then we would simply estimate a demand parameter on ability that is half as large, and the economic interpretation of our estimates would be unchanged.} Hence, we normalize this transformation to the identity without loss of generality.

\subsubsection{Identifying Demand}\label{id_demand} 

In equilibrium, workers choose bids and signaling efforts, as functions of their costs and abilities. Therefore, employers' choice probabilities, as functions of bids and signals, do not directly reveal their disutility for wages or their willingness to pay for ability, since bids and signals encode equilibrium information about workers' abilities.

However, having identified the equilibrium joint distribution of bids, signals, and abilities, we identify the equilibrium belief expectation function $\mathbb{E}[a | b, s; x]$, which is the reduced form of $\mu^*\left(b, s; x, \sigma^*\right)$.

Thus, substituting identified employer beliefs into the employer indirect utility function from Equation \ref{eq:employer_utility}, we can write employers' indirect expected utility as a function of bids, signals, observables, and taste shocks:
\begin{align}\label{eq:employer_expected_utility}
    EU^E(b_{i j},  s_{i j}, \nu_{i j}; x_{i j}) \equiv T(x_{i j}) + \beta \cdot    \mathbb{E}[a_{ij} | b_{ij}, s_{ij}; x_{ij}] - \alpha \cdot  b_{i j} + \nu_{i j}.
\end{align}
Recall the key assumption that workers do not observe their taste shock $\nu_{i j}$ before applying. 

Therefore, observing how conditional hiring probabilities vary with bids, observables, and the now-identified equilibrium beliefs about ability identifies the demand parameters $\alpha$, $\beta$, and $T(x)_{x\in X}$.

\section{Estimation}\label{estimation}

We estimate the model using a simulation-based estimator whose structure mirrors the identification argument. The estimator finds the supply and demand parameters $\theta$ that maximize the likelihood of observed employer choices---functions of bids, observables, consideration sets, and beliefs---subject to the constraints that (i) workers’ simulated first-order conditions hold exactly as functions of $\theta$, and (ii) employers’ posterior mean beliefs about worker ability are consistent with the joint distribution of types implied by $\theta$.

Estimation proceeds in three stages. First, we invert simulated worker FOCs---using flexible estimates of the signal production functions and employer choice probabilities as functions of bids, signals, and observables---to recover worker costs and abilities. Second, we nonparametrically estimate employers’ equilibrium mean beliefs about worker ability from the recovered joint distribution of worker types. Finally, we estimate labor demand by maximizing the likelihood of observed employer choices as functions of bids, beliefs, and worker observables.

The remainder of this section summarizes the key details of our estimation procedure, the full details of which we present in Appendix~\ref{appendix_estimation}.

\subsection{Estimating Supply}\label{estimation_supply} 

This subsection describes how we estimate labor supply, which includes how we stratify our data by worker observables, how we estimate signal production functions, and how we simulate worker FOCs to recover worker costs and abilities.

\subsubsection{Data Stratification by Worker Observables}\label{estimation_xgroups} 

To maintain nonparametric flexibility when estimating labor supply, we stratify our data by observable characteristics. We construct groups using Cartesian products of workers' country, arrival times, and reputations. By arrival time, we specifically mean the time elapsed between the job being posted and the worker first clicking on the job post.\footnote{We impute arrival time as the time elapsed between the job being posted and the worker's application being submitted for each worker whose first click is either missing or measured to be after their application was submitted. Note, however, that these workers are dropped from any analysis that requires use of a worker's effort.} By reputation, we mean the one-dimensional ``reputation score'' that the platform assigns to each worker in order to rank them in the bid list.\footnote{Employers observe many of the inputs to the reputation scores, they see the ranking of the workers in the bid list, but they do not see the actual scores. Using the platform's reputation scores as a proxy for one-dimensional ``reputation'' is a convenient way to collapse what is otherwise high-dimensional data about each worker. Conditional on choosing to collapse these worker-level data into one dimension, we argue it makes sense to trust the platform's algorithms over one we could devise on our own, especially given we do not have access to all the detailed data that the platform uses for its algorithm.} This process results in 56 groups. The exact definitions of and statistics on these groups can be found in Appendix~\ref{appendix_stratification}.

\subsubsection{Estimating Signal Production Functions}\label{estimation_signal_production} 

Although the deterministic portion of our signal production function $f^s(e; x)$ is nonparametrically identified, we impose a parametric functional form to estimate it. In particular, we assume that:
\begin{align}\label{eq:signal_production}
    s_{i j} = K^s(x_{i j}) + \gamma^s(x_{i j}) \cdot \log(e_{i j}) + \varepsilon^s_{i j},
\end{align}
where $K^s(\cdot)$ and $\gamma^s(\cdot) > 0$ are functions from $X$ to $\mathbb{R}$ that map the observable type $x_{i j}$ to the signal production function's intercept and slope, respectively. Furthermore, we assume that the distribution of signaling noise $F_{\varepsilon^s|x} = N(0, V_{\varepsilon^s}(x))$, where $V_{\varepsilon^s}(x)$ is a function from $X$ to $\mathbb{R}$ that describes the variance of the signaling noise conditional on the observable type $x_{ij}$.\footnote{In principle, we could allow $V_{\varepsilon^s}$ to vary with effort $e$, i.e., we could allow for heteroskedasticity.} 

\vspace{-2mm}
\paragraph{Effort Correction}
Before using our effort measurements in our estimation procedure, we implement a correction that uses an empirical Bayes estimator of worker random effects in signal production that addresses a key concern with our measurement. Some workers may be more time-efficient at expending effort than others, and thus two workers spending the same amount of time reading and writing may not be expending the same amount of effort. In words, this effort correction can be interpreted as follows: if worker $j$, who is observed spending $t$ minutes of raw effort, produces on average the same signals as the average worker spending $t'$ minutes, then worker $j$’s true effort when spending $t$ minutes is defined to be $t'$. The precise details of this correction can be found in Appendix~\ref{appendix_effort_correction}.\footnote{Table~\ref{tab:reg_5stars_on_signal_effcorr} re-estimates Tables~\ref{tab:reg_5stars_on_signal} and yields the model-predicted result that when regressing job completion rates on signals, controlling for corrected effort, the coefficient on signals becomes insignificant, while the coefficient on corrected effort remains strongly significant. In other words, signals are only predictive of worker ability in that they proxy for effort.}

Once we have corrected effort, which we simply refer to as ``effort'' for the remainder of the paper, we estimate Equation~\ref{eq:signal_production} separately for each $x$-type using OLS.

\subsubsection{Estimating Worker Beliefs}\label{estimation_worker_beliefs} 

\paragraph{Overview}
To estimate labor supply, we must estimate worker beliefs, i.e., the equilibrium win-probability function $P^*(b, e; x)$. To do so, we simulate many job posts and signal realizations, compute the probability of a fixed worker of type $x$ winning each simulated job post as a function of $(b, e)$, and then numerically integrate over all simulated job posts, including the signal realizations, to obtain an estimate of $P^*(b, e; x)$.

\paragraph{Simulating Job Posts}
For this numerical integration to accurately estimate worker beliefs, we need to ensure that we sufficiently cover the support of both signaling noise and competing workers' bids and signals. Therefore, we first bootstrap consideration sets---vectors of counts of each $x$-type considered at each job post---to expand our simulated sample well beyond the size of our data. Then, we leverage the fact that workers choose bids and efforts based only on their own types and the fact that signaling noise is drawn i.i.d. conditional on each worker's observable type to bootstrap many draws of bids and signals at the individual worker level conditional on $x$ to fill in the support of bids and signals for each $x$-type. Finally, we simulate many draws of signaling noise from our estimated $\widehat{F}_{\varepsilon^s|x}$ for the worker whose win-probability we are simulating to fill in the support of her own signal realizations conditional on her observable type.

\paragraph{Ex-Interim Equilibrium Win-Probability Approximation}
Before performing our numerical integration, we need to estimate a worker's ex-interim equilibrium win-probability at each job post as a function of her bid and signal, given the bids and signals of all other considered competitors at that job post. We do so by first assuming that $F_\nu$ is Type I Extreme Value, which implies that employers choose workers according to the standard logit formula. 

Not yet knowing employer beliefs but knowing they must be some function of $b$, $s$, and $x$, we approximate the employer expected utility in Equation~\ref{eq:employer_expected_utility} by a linear function of bids, signals, and observables, which we denote $\tilde{\delta}(b_{ij}, s_{ij}; x_{ij}, \tilde{\theta})$, where $\tilde{\theta}$ are the parameters of this approximation. We estimate $\tilde{\theta}$ by maximizing the standard logit likelihood of observed employer choices as functions of bids, signals, and observables.

Once we have estimated $\tilde{\theta}$, we can compute an $x$-type worker's ex-interim equilibrium win-probability at a simulated job post $m$ as a function of her bid $b$, effort $e$, and simulated signaling noise $\varepsilon^s$, and simulated consideration status $q_{mj}$, conditional on the employer not exogenously abandoning the job post, as:
\begin{align}
    \widehat{P}^*_{m j}(b, e) \equiv q_{m j} \cdot \left(\frac{\tilde{\delta}\left(b, \widehat{f}^s(e; x) + \varepsilon^s_{m j}; x, \widehat{\tilde{\theta}}\right)}{1 + \tilde{\delta}\left(b, \widehat{f}^s(e; x) + \varepsilon^s_{m j}; x, \widehat{\tilde{\theta}}\right) + \sum_{k \in Q_m \setminus \{0, j\}} \tilde{\delta}\left(b_{m k}, s_{m k}; x_{m k}, \widehat{\tilde{\theta}}\right)}\right)
\end{align}

Finally, we compute the ex-ante equilibrium win-probability function as:
\begin{align}
    \widehat{P}^*(b, e; x) = \pi \cdot \left(\sum_{m = 1}^M N_{m x} \right)^{-1} \sum_{m = 1}^M \sum_{j = 1}^{N_m} \widehat{P}^*_{m j}(b, e) \cdot \mathbbm{1}\{x_{m j} = x\},
\end{align}
where $M$ is the number of simulated job posts, $N_m$ is the total number of workers simulated at job post $m$, $N_{m x}$ is the number of workers of observable type $x$ simulated at job post $m$, and recall that $(1-\pi)$ is the exogenous probability that an employer abandons a job post. We similarly compute the derivatives of $\widehat{P}^*(b, e; x)$ with respect to $b$ and $e$ using the analytic derivatives of $\widehat{P}^*_{m j}(b, e)$.

\paragraph{Inverting Worker FOCs}
To simulate worker FOCs, we need to specify a functional form for the effort cost function $C(e; a)$. We assume that\footnote{This exponential functional form assumption helps us fit our data better than a linear function would and helps us rationalize why some workers find it profitable to expend very little effort (e.g., 10 seconds) while others find it profitable to spend an order of magnitude or more effort (e.g., 12 minutes) even at convex cost.}:
\begin{align}
    C(e_{i j}; a_{i j}) = \frac{e_{i j}^2}{2\cdot \exp\left(a_{i j}\right)}
\end{align}
For each worker in our data with a valid effort measurement, we invert the simulated FOCs by plugging $\widehat{P}^*(b, e; x)$ and its derivatives, as well as the parametric form of $C(e; a)$, into Equations \ref{eq:foc_bid_invert} and \ref{eq:foc_effort_invert}, which yields estimates of each worker's cost $\widehat{c}_{i j}$ and ability $\widehat{a}_{i j}$. Thus, we have a nonparametric estimate of the joint distribution of worker costs and abilities $\widehat{F}_{c \times a | x}$.

\subsection{Estimating Employer Beliefs}\label{estimation_beliefs} 

To make our estimation of employer beliefs more feasible, we make the assumption that employers form beliefs about workers' abilities based only on workers' signals and observables, and not their bids.\footnote{This step is where our assumption that bids do not enter employer beliefs simplifies things the most. It is far easier to fit a nonparametric model of a conditional expectation, separately by observable group, in one dimension, rather than two. In principle, our procedure could be modified to allow for bid information to be included in employer beliefs, but this would either require placing more structure on these beliefs or coarsening worker observable groups even further.} 
This assumption is not costless, but we believe it is a reasonable one in our empirical setting.\footnote{We argue the reasonability of this assumption in three points. First, the cognitive load employers would face in order to form beliefs based on two-dimensional worker strategies separately for each observable group is quite high, and thus it is plausible that employers would simplify their inference problem by focusing on the dimension of applications more saliently and directly related to ability. Second, given the wide distribution of worker's countries of residence, employers may not have a good sense of how each worker's bid relates to her opportunity cost, making it difficult for them to infer ability from bids. Third, anecdotally when reading how workers discuss application strategies on online forums, they rarely mention considering how their bid may be viewed as a signal of ability.} Thus, we assume that employers' equilibrium belief function takes the form $\mu^*(s; x) \equiv \mathbb{E}[a | s; x]$.

To estimate employer beliefs, we use the estimated joint distribution of worker types $\widehat{F}_{c \times a | x}$ to fit a flexible nonparametric model of the conditional expectation $\mathbb{E}[a | s; x]$ with the shape restriction that it is weakly increasing in $s$ for each $x$---a condition which our model's equilibrium implies.

We operationalize this nonparametric estimation by first estimating an isotonic regression of ability on signal, separately by observable group $x$, and then fitting a Piecewise Cubic Hermite Interpolating Polynomial (PCHIP) to the isotonic regression. This approach allows us to capture data-driven non-linearities, while also ensuring that the estimated employer belief function is continuous, differentiable, and monotonically increasing.

\subsection{Estimating Demand}\label{estimation_demand} 

Once we have estimated the employer belief function $\widehat{\mu}(s; x)$, we apply it to each application in our data to obtain pseudo-data on employer beliefs.\footnote{Note that while we could only estimate labor supply using the subset of workers with valid effort measurements, we can estimate labor demand using all applications in our data since we have observed bids and signals for all applications.} We can then estimate labor demand parameters $\alpha$, $\beta$, and $T(x)_{x\in X}$ by maximizing the standard logit likelihood of observed employer choices as functions of bids, beliefs, and observables.

\section{Estimation Results}\label{est_results} 

In this section, we present the key results of our estimation procedure. We first present selected results from the estimated joint distribution of worker costs and abilities, followed by selected results from employer demand estimates. The rest of our estimation results can be found in Appendix~\ref{appendix_extra_results}. 

\subsection{Supply Estimates}\label{est_results_supply} 

Table~\ref{tab:supply_estimates} presents summary statistics of the estimated marginal distributions of worker costs, abilities, and writing costs. To make our estimates of ability more interpretable, we present them in terms of dollars of employers' willingness to pay (WTP) computed as $\frac{\widehat{\beta}}{|\widehat{\alpha}|}\cdot\widehat{a}$, where $\widehat{\beta}$ is the estimated coefficient on ability and $\widehat{\alpha}$ is the estimated coefficient on bid, both from estimated employer demand.

\input{tables/supply_estimates_sum_stats.tex}

We find there to be substantial heterogeneity in both worker costs and abilities, with a range between P20 and P80 of \$121.64 and \$97.45 respectively. We estimate that over 25\% of workers have negative costs, meaning they would be willing to pay to be hired. This finding almost surely arises from the fact that our model ignores dynamics, and thus we are not capturing a worker's continuation value of being hired for a job, which may be particularly important for workers with low reputations who are trying to build up their on-platform reputation.

\begin{figure}[h!]
    \centering
\includegraphics{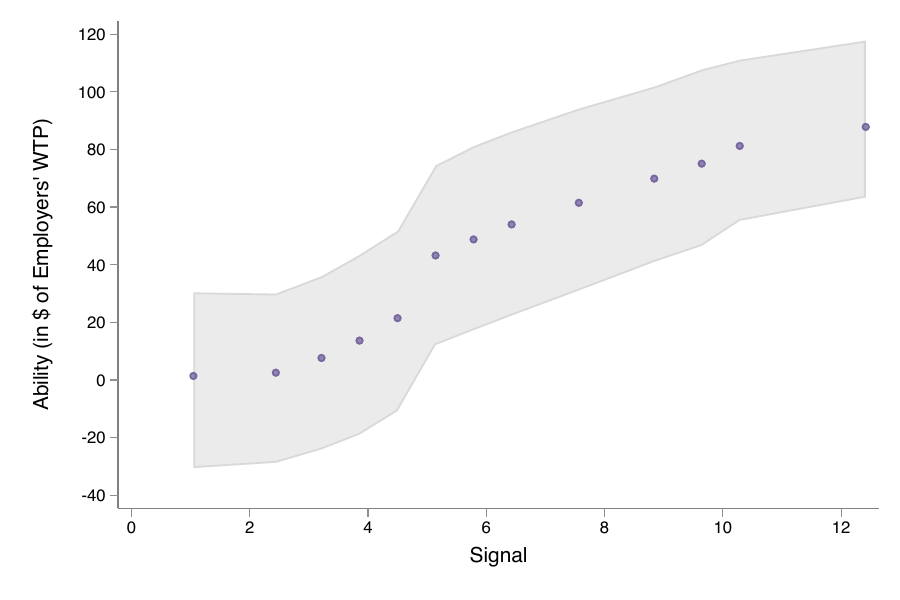}
    \caption{Binscatter Ability on Signal}
    \label{fig:ability_vs_signal}
     \floatfoot{\footnotesize \textit{Notes:} This figure plots a binscatter of estimated worker ability on signal, where the former is measured in dollars of employers' willingness to pay (WTP) computed as $\frac{\widehat{\beta}}{|\widehat{\alpha}|}\cdot\widehat{a}$, where $\widehat{\beta}$ is the estimated coefficient on ability and $\widehat{\alpha}$ is the estimated coefficient on bid, both from estimated employer demand. Also plotted are 25th and 75th percentile bands of ability within each signal bin. The correlation coefficient between signal and ability is 0.547.}
\end{figure}

The wide range of estimated abilities highlights the importance of signaling in this market. Though Table~\ref{tab:supply_estimates} shows unconditional distributions of costs and abilities, we find that within-group variances of costs and abilities dwarf between-group variances with between group variances only explaining around 2\% and 3\% of the total variances of costs and abilities respectively. Thus, even conditional on observables, an employer hiring a worker without further information on ability could end up with a wide range of abilities.

\begin{figure}[h!]
    \centering
\includegraphics{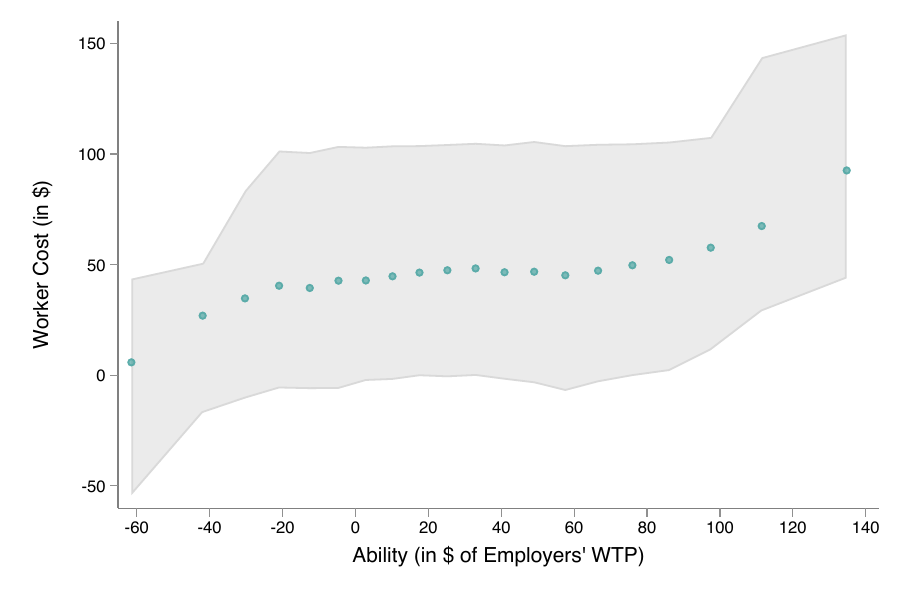}
    \caption{Binscatter of Worker Cost on Ability}
    \label{fig:worker_types}
     \floatfoot{\footnotesize \textit{Notes:} This figure plots a binscatter of estimated worker cost on estimated worker ability, where the latter is measured in dollars of employers' willingness to pay (WTP) computed as $\frac{\widehat{\beta}}{|\widehat{\alpha}|}\cdot\widehat{a}$, where $\widehat{\beta}$ is the estimated coefficient on ability and $\widehat{\alpha}$ is the estimated coefficient on bid, both from estimated employer demand. Also plotted are 25th and 75th percentile bands of ability within each cost bin. The correlation coefficient between cost and ability is 0.193.}
\end{figure}

Further highlighting the importance of signaling, Figure~\ref{fig:ability_vs_signal} presents a binscatter of estimated worker ability on signal, with 25th and 75th percentile bands of ability within each signal bin. This figure shows a strong positive correlation (0.547) between signals and abilities, and both the slope of ability on signal and the relatively tight interquartile range show that signals are good predictors of ability in the equilibrium being played in the data.

Finally, Figure~\ref{fig:worker_types} presents a binscatter of estimated worker cost on estimated worker ability, with 25th and 75th percentile bands of ability within each cost bin. This figure highlights a key feature of our estimates that impact our counterfactuals: worker costs and abilities are positively correlated. Even though the correlation is not particularly strong (0.193), this correlation generally implies that higher-ability workers also have higher costs. Thus, in a market without signaling, when price competition increases, higher-ability workers may be priced out of the market, when they may otherwise have been hired in a market in which they could signal their ability. 

More detailed results on the estimated distributions of worker costs and abilities, equilibrium strategies, and signal production functions can be found in Appendix~\ref{appendix_extra_results}.

\subsection{Demand Estimates}\label{est_results_demand} 

\input{tables/demand_estimates_sum_stats.tex}

Table~\ref{tab:demand_estimates} presents our main demand estimates. Our price coefficient generally aligns with our estimated price coefficient from our reduced-form demand estimation, the results of which are presented in Table~\ref{tab:logit_signal}. Our estimated coefficient on ability implies that a one standard deviation increase in ability, all else equal is valued at about \$52.16 by employers, as our estimates showed in Table~\ref{tab:supply_estimates}. Importantly, this estimate of the WTP for ability should be interpretted as specifically being the WTP for a standard deviation increase in the ability \textit{that can be signaled}. As we made clear in Section~\ref{model}, we cannot separate group-level non-signaled ability and taste-based group preferences in our estimates of $T(x)$.

Finally, note that one should only interpret the dollar value estimates of $T(x)$ relative to each other, or relative to the mean outside option value, which is noramlized to zero plus the dollar-valued mean of the T1EV distribution. Our estimates of $T(x)$ broadly fit our expectations from anecdotal evidence from the market. All else equal, employers prefer workers with high ratings over those with middle and low ratings, and they slightly prefer workers with poor ratings that have some experience over those that have no experience on the platform. Employers prefer workers who take their time to arrive at job posts, rather than those that arrive quickly upon the job being posted. Finally, employers prefer workers from non-majority-English-speaking European countries the most, then workers from the large ``Other'' grouping, then workers from English Speaking countries, and lastly they have a sharp dislike for workers from South Asia. Note that these $T(x)$ estimates are separate from the group level prior beliefs, i.e., before signaling, about ability, i.e., $\mathbb{E}[a | x]$.

\section{Counterfactuals}\label{counterfactuals}

In this section, we describe our simulated counterfactuals. Our main counterfactual imagines a counterfactual state of the world in which LLMs have changed nothing in the market except for bringing writing costs to zero. In such a world, workers could each write arbitrarily customized proposals for free, thus rendering signaling uninformative in equilibrium. As such, our main counterfactual is a ``no-signaling'' (NS) equilibrium, in which workers only choose bids, and employers only use worker observables to form beliefs about worker abilities. We compare the equilibrium market outcomes of this counterfactual to that of the ``status-quo'' (SQ) equilibrium we have estimated in the data, in which workers signal with costly effort.

As a benchmark for our welfare estimates, we also simulate a ``full-information'' (FI) equilibrium, in which employers have perfect information about worker abilities, and thus workers again only choose bids. This counterfactual helps us put the welfare differences between the no-signaling equilibrium and the equilibrium we estimate in the data into context.

In the NS counterfactual, workers choose bids $b$ only according to their costs $c$, and employers only use a worker's observables $x$ to predict their ability $a$ according to group means $\mathbb{E}[a | x]$.\footnote{Note that without a second dimension of worker's actions to discipline the relationship between bids and abilities, and without a perfect negative correlation between costs and abilities, there is no pure-strategy equilibrium in which employers use bids to form beliefs about ability, other than one in which all the workers pool on the same bid. Additionally, maintaining the assumption used in our estimation that prices are not used as signals helps us compare `apples to apples' when comparing the no-signaling equilibrium to the equilibrium we estimate in the data.} In the FI counterfactual, workers submit bids $b$ according to their costs $c$ \textit{and} abilities $a$, and employers observe each worker's true ability $a$. In both counterfactuals, there is no signaling, and thus no cost of effort. 

We solve for the equilibrium of each of the NS and FI counterfactuals using bootstrapped job posts, worker observables, consideration sets, and worker types from our data and from our supply estimates.\footnote{We do not prove the uniqueness of these counterfactual equilibria, but we do note that the scope for multiplicity is far more limited than in our full model, given that workers only choose one-dimensional bids and there is no signaling, and thus no equilibrium belief formation.} For the simulated SQ equilibrium, we similarly use bootstrapped job posts, worker observables, consideration sets, and worker types, and we additionally bootstrap bids and efforts from our estimated equilibrium strategies that map worker types to bids and efforts. We then simulate signals according to our estimated signal production functions and our bootstrapped efforts. In all three equilibria, we simulate employer choices according to our estimated demand parameters, using our estimated employer belief functions for the SQ equilibrium. 

\subsection{Hiring Patterns}

\begin{figure}[h!]
    \centering
\includegraphics{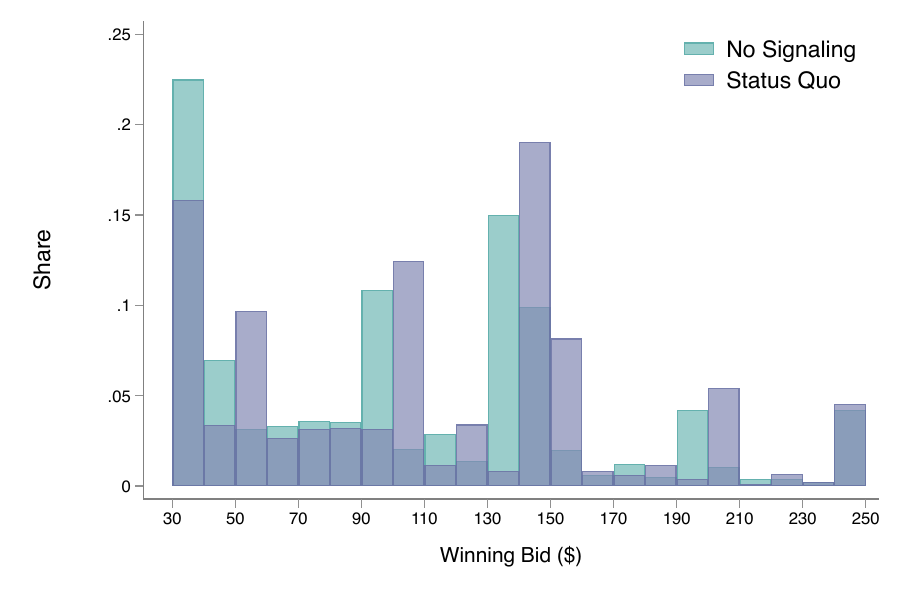}
    \caption{Winning Bids by Counterfactual Equilibrium}
    \label{fig:winning_bids_no_vs_sq}
     \floatfoot{\footnotesize \textit{Notes:} This figure plots histogram estimates of the distribution of winning bids in both the status-quo (SQ) and no-signaling (NS) equilibria}
\end{figure}

In this subsection, we compare the hiring patterns of the NS and SQ equilibria to shed light on how losing access to signaling changes both hiring rates and the composition of hired workers. Broadly, we find that without signaling, as compared to the status quo, employers hire fewer high-ability workers and more low-ability workers, and they hire slightly fewer high-cost workers and slightly more low-cost workers. Accordingly, the market becomes more price-competitive. The average winning bid in the NS equilibrium is \$102.68 as compared to \$108.02 in the SQ equilibrium, a decrease of 4.94\%. Figure~\ref{fig:winning_bids_no_vs_sq} presents the distribution of winning bids in both the SQ and NS equilibria, showing that the distribution of winning bids in the NS equilibrium is shifted to the left, with a higher mass of winning bids at lower levels.

\subsubsection{Extensive Margin Changes}

Firstly, we find a very modest extensive margin response. The hiring rate---the probability that a posted job hires a worker---decreases from 40.86\% in the SQ equilibrium to 40.23\% in the NS equilibrium. This small decrease of 0.63 percentage points is somewhat driven by the fact that we estimate that 42.51\% (i.e., $1 - \widehat{\pi}$) of job posts are exogenously abandoned. Conditional on an employer not abandoning the job post, we find that the hiring rate decreases from 71.08\% in the SQ equilibrium to 69.98\% in the NS equilibrium, a decrease of 1.10 percentage points, or a 1.5\% drop. Given how small this change in even the conditional hiring rate is, we focus the rest of our analysis on the intensive margin.

\subsubsection{Changes in Hiring by Ability and Cost Quantiles}

\begin{figure}[h!]
    \centering
\includegraphics{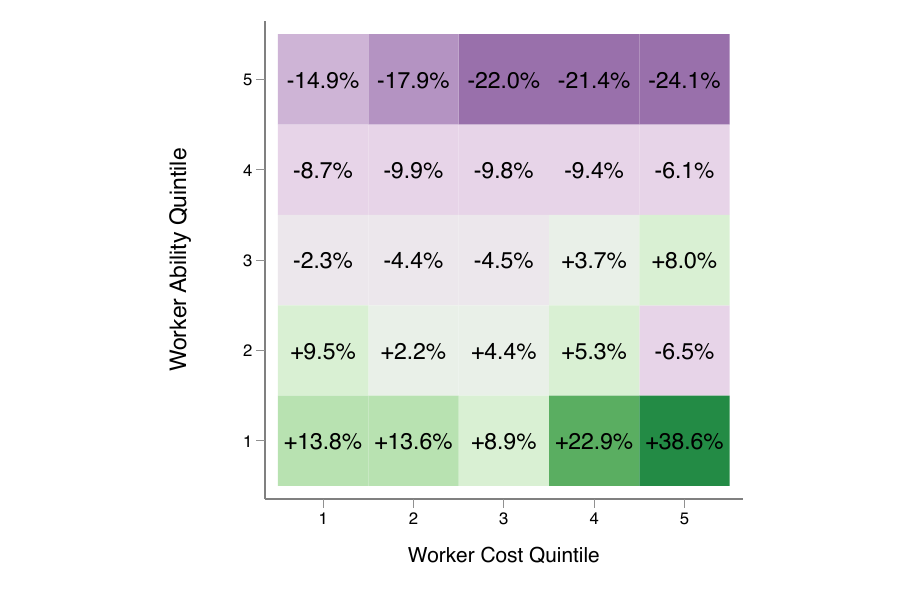}
\captionsetup{justification=centering}
    \caption{Percentage Changes in Hiring Rates by Ability and Cost Quintiles: \\ No-Signaling vs. Status-Quo}\label{fig:ability_cost_quintile_changes_sq_vs_no}
     \floatfoot{\footnotesize \raggedright  \textit{Notes:} This figure plots a heatmap of the percentage change in hiring rates for each combination of worker ability and cost quintiles when moving from the status-quo (SQ) equilibrium to the no-signaling (NS) equilibrium. The color of each cell represents the percentage change in hiring rates for workers in that ability-cost quintile cell when moving from the SQ to the NS equilibrium. Purple shades represent decreases in hiring rates, while green shades represent increases. The numbers in each cell present the exact percentage change in hiring rates for that cell. The top left cell represents the highest ability, lowest cost quintile, while the bottom right cell represents the lowest ability, highest cost quintile.}
\end{figure}

The impact of workers losing the signaling mechanism is clearest when examining the changes to the hiring rates by worker type of the form $\mathbb{P}(\text{Hired} | \text{worker type})$. Figure~\ref{fig:ability_cost_quintile_changes_sq_vs_no} presents a heatmap of the percentage change in hiring rates for each combination of worker ability and cost quintiles when moving from the SQ equilibrium to the NS equilibrium. One clear pattern from this figure is that as it moves from high-ability workers to low-ability workers-moving down the rows of the figure-the percentage changes in hiring rates start very negative and eventually become positive for second quintile ability workers and very positive for the lowest ability quintile.

This pattern shows that higher-ability workers relied on signals to distinguish themselves from lower-ability workers, and thus when signaling is removed, their hiring rates drop. This shift in hiring towards lower-ability workers is also consistent with the fact that there is far more variation in ability within observable groups than across them. In other words, high-ability workers cannot simply rely on their observable characteristics to distinguish themselves from lower-ability workers, and thus signaling is key for these workers to be hired.

Another instructive finding that emerges from Figure~\ref{fig:ability_cost_quintile_changes_sq_vs_no} is the change in hiring rates for workers in the corners of the figure.

\vspace{-4mm}
\paragraph{High-Ability, Low-Cost Workers.}
First, consider the top left corner, which contains workers in the top 20\% of the ability distribution and the bottom 20\% of the cost distribution, i.e., the ``best'' type of workers for welfare. The hiring rate for these workers drops by 14.9\%. Although these workers can compete on price, they no longer receive the large boost to their hiring rates that they received from being able to signal their ability in the SQ equilibrium.

\vspace{-4mm}
\paragraph{High-Ability, High-Cost Workers.}
Second, consider the top right corner, which contains workers in the top 20\% of the ability distribution and the top 20\% of the cost distribution, i.e., high-ability and high-cost workers. The hiring rate for these workers falls by 24.1\%---a larger decrease than any of the 24 other quantile-by-quantile groups. These workers are among the least competitive on price, but being able to signal their ability allows them to overcome this deficit in the SQ equilibrium. However, in the NS equilibrium, without the ability to signal, these workers have similar hiring rates to the rest of the highest-cost workers, which in combination with their ability to signal in the SQ equilibrium, leads to the largest percentage drop in hiring rates.

\vspace{-4mm}
\paragraph{Low-Ability, Low-Cost Workers.}
Third, consider the bottom left corner, which contains workers in the bottom 20\% of the ability distribution and the bottom 20\% of the cost distribution, i.e., low-ability and low-cost workers. The hiring rate for these workers increases by 13.8\%. These workers can compete well on price in both equilibria, but in the SQ equilibrium, they are heavily penalized for being low ability and thus struggle to get hired relative to workers with similarly low costs and higher abilities. However, in the NS equilibrium, these workers can blend in with high-ability workers and thus shed their penalty for being low ability, leading to a large percentage increase in their hiring rates.

\vspace{-4mm}
\paragraph{Low-Ability, High-Cost Workers.}
Finally, consider the bottom right corner, which contains workers in the bottom 20\% of the ability distribution and the top 20\% of the cost distribution, i.e., the ``worst'' type of workers for welfare. The hiring rate for these workers increases by 38.6\%---a larger increase than any of the 24 other quantile-by-quantile groups. These workers are among the least competitive on price, and they also get heavily penalized for being low ability in the SQ equilibrium. However, in the NS equilibrium, these workers can also blend in with high-ability workers and thus shed their penalty for being low ability, leading to the largest percentage increase in their hiring rates, even if their hiring rates remain low in absolute terms. See Figure~\ref{fig:ability_cost_quintile_sq_vs_no} for the levels of hiring rates by ability and cost quintiles in both equilibria.

\begin{figure}[h!]
    \centering
\includegraphics[width=1\linewidth]{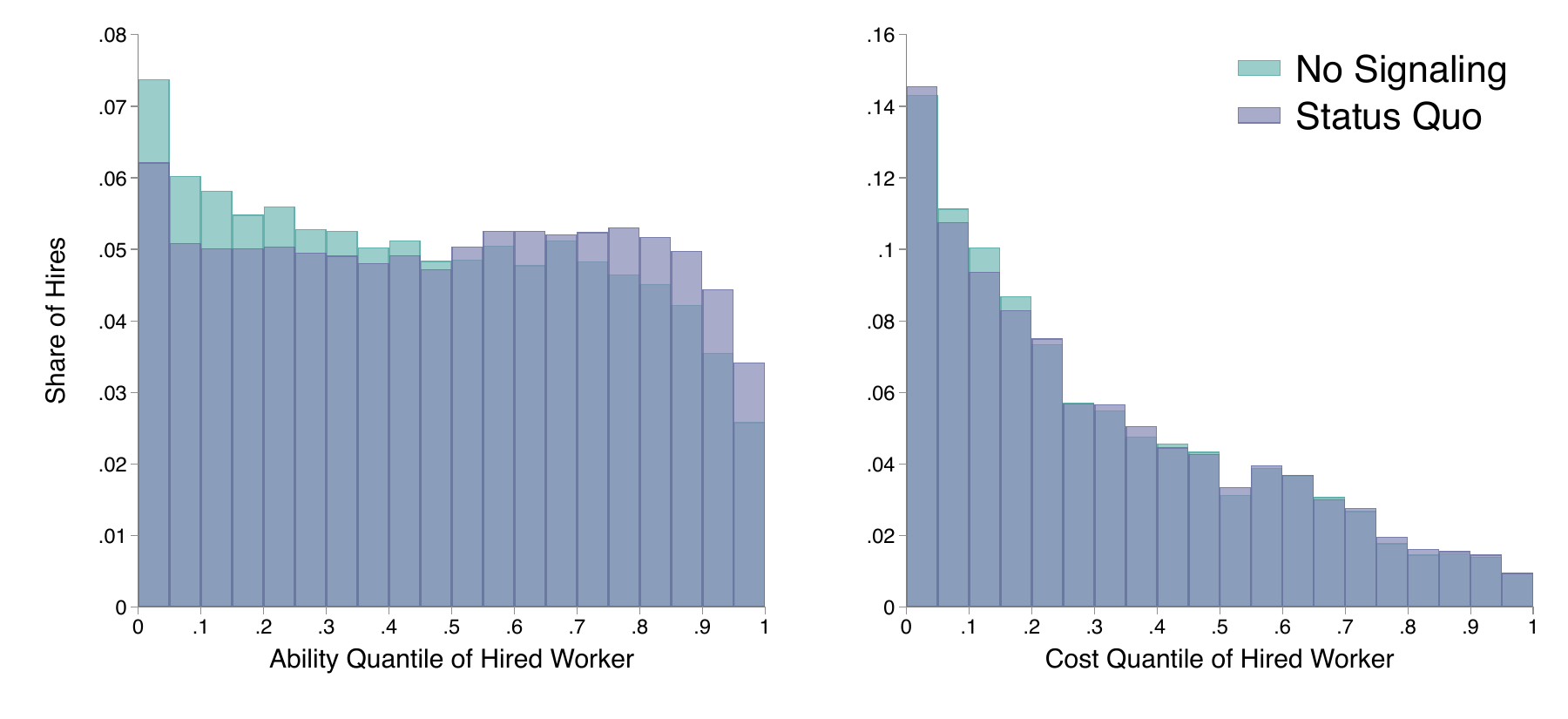}
\captionsetup{justification=centering}
    \caption{Histograms of Hired Workers by Ability and Cost Quintiles: \\ No-Signaling vs. Status-Quo}
    \label{fig:cost_ability_quantile_changes_sq_vs_no}
     \floatfoot{\footnotesize \raggedright \textit{Notes:} This figure plots histogram estimates of the distribution of quantiles of the abilities of hired workers in both the status-quo (SQ) and no-signaling (NS) equilibria}
\end{figure}

While these changes in hiring rates by ability and cost quintiles are stark, the changes in hiring rates by observable groups are far more muted. See Table~\ref{tab:percent_changes_hiring} for results on the changes in hiring rates by observable groups.

We also investigate the changes in the composition of hired workers by ability and cost quantiles, i.e., $\mathbb{P}(\text{worker type} | \text{Hired})$. Figure~\ref{fig:cost_ability_quantile_changes_sq_vs_no} presents histograms of the distribution of hired workers by ability and cost quantiles in both the SQ and NS equilibria, which both helps us visualize the changes in hired worker composition and how the joint distribution of changes in hiring rates averages out in the marginal distributions.

The most striking change is that the share of hired workers in the bottom half of the ability distribution increases sharply, while the share of hired workers in the top half of the ability distribution correspondingly decreases. In particular, the share of hired workers coming from the bottom 20\% of the ability distribution, whose ability is valued on average by employers at \$-38.67, increases by 14.06\% in the NS equilibrium compared to the SQ equilibrium. In contrast, the share of hired workers coming from the top 20\% of the ability distribution, whose ability is valued on average by employers at \$107.75, decreases by 18.74\% in the NS equilibrium compared to the SQ equilibrium.

Note that the changes in the composition of hired workers by cost quantiles are far more muted when looking at the marginal distribution of hired worker costs, masking the heterogeneity in changes in hiring we saw by ability and cost quintiles in Figure~\ref{fig:ability_cost_quintile_changes_sq_vs_no}.

Figure~\ref{fig:cost_ability_changes_sq_vs_no} presents kernel density and histogram estimates of the distributions of ability and cost levels, respectively, of hired workers in both the SQ and NS equilibria. 

\paragraph{Discussion.} Overall, these patterns show that when workers cannot signal their ability, employers hire far fewer high-ability workers in favor of low-ability workers. While the direction of this change may have been a predictable consequence of our model, the magnitude is not. Because abilities and costs are positively correlated, once the market relies more on bids to determine hiring, the composition of hired workers naturally shifts  towards lower-ability workers. If, for example, we had found from our supply estimation that costs and abilities were negatively correlated, then the losses due to losing signaling could have been largely mitigated. In other words, in any equilibrium in which workers cannot signal their ability, workers with lower bids, and thus lower costs, are hired more often. The implications of this change on the distribution of hired ability then depends heavily on who those workers with the lowest bids are. In our case, they tend to be workers with lower abilities, exacerbating the welfare losses from LLMs disrupting labor market signaling.

\subsection{Welfare Changes}

\begin{figure}[h!]
    \centering
    \includegraphics[width=0.87\linewidth,trim=0 0 0 80,clip]{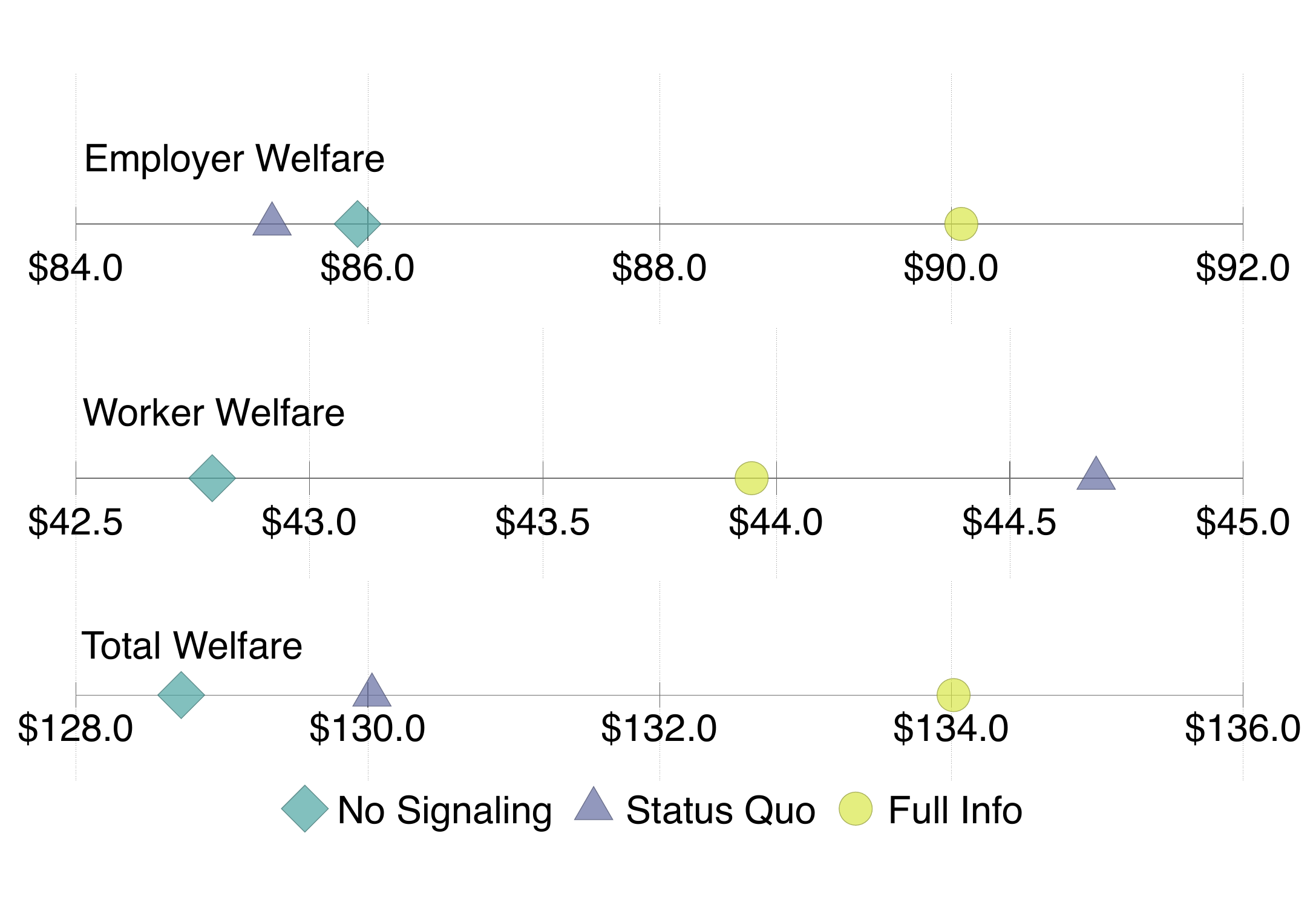}
    \caption{Changes in Welfare by Counterfactual}\label{fig:welfare_changes}
\end{figure}

These changes to hiring patterns mostly harm workers and leave employers largely unaffected. As we can see in Figure~\ref{fig:welfare_changes}, these changes lead to a 4\% reduction in worker surplus, and a small, less than 1\%, increase in employer surplus. Worker welfare losses are driven by both the modest extensive margin decrease in hiring rates, and the intensive margin decrease in wages. Their losses are mitigated both by the elimination of writing costs, and by the fact that the lower-ability workers tend to have lower costs. Employer surplus is virtually unaffected because the reduction in wages paid to workers roughly balances out the loss of hiring lower-ability workers. Overall, the market becomes less efficient, as total surplus falls by 1\%, and significantly less meritocratic, as the new equilibrium favors lower-ability workers over higher ability ones.

\section{Conclusion}\label{conclusion}

This paper studies how LLMs disrupt labor market signaling, and the implications of this disruption for hiring patterns and welfare. We develop a novel measure of how customized a given application is to a given job post. Using this measure and data from a large digital labor platform \texttt{Freelancer.com}, we document new evidence on how LLMs disrupt signaling. We show that prior to the advent of generative AI, employers on the platform had significant demand for workers who customized their applications. We then provide suggestive evidence that they do so because that customization is a noisy signal of a worker's application-specific effort, which itself is a predictor of whether workers successfully complete jobs. Having argued that application customization functions as a Spence-like signal in the pre-LLM period, we then show suggestive evidence that this signaling equilibrium erodes in the post-LLM market. Employers no longer demand signals; signals no longer predict job completion; and for applications we know were written using generative AI tools, signals no longer predict application-specific effort.

With these findings, we built and estimated an equilibrium model of labor market signaling. In doing so, we find that, in this market, workers' abilities and costs are positively correlated. Using the estimated model, we simulate a counterfactual in which workers lose their ability to signal due to LLMs lowering writing costs to zero, while holding fixed everything else in the pre-LLM market. This counterfactual exercise allows us to isolate the equilibrium effects of losing labor market signaling via written communication on both hiring patterns and welfare.

Our counterfactual estimates suggest that, in the absence of signals of ability in the form of written applications, employers divert hiring away from higher-ability workers towards lower-ability ones. This finding is driven by three key mechanisms. First, and most obviously, employers can no longer identify high-ability workers, conditional on observables and bids. Second, because workers' abilities and costs are positively correlated, when employers rely more on bids to determine hiring, the composition of hired workers naturally shifts towards lower-ability workers. Third, because there is far more variation in ability within observable groups than across them, high-ability workers cannot simply rely on their observable characteristics to distinguish themselves from lower-ability workers. These patterns, taken together, imply losses in worker welfare and overall welfare, but virtually no effect on employer welfare, as employers are compensated by the lack of hired ability with having to pay lower wages.

These results imply that many markets that rely on costly written communication may face significant welfare and meritocratic threats from generative AI's ability to cheaply produce expertly-written text. Either firms within or designers of these markets may seek to mitigate these threats by investing in more effective screening technology that cannot be gamed by generative AI, or by redesigning labor contracts to incentivize more exploratory short-term hiring that allows employers to learn about workers' abilities on the job, rather than relying on pre-hire signals.

More broadly, these adverse effects of generative AI on the effectiveness communication and signaling are not exogenous to market conditions. One could imagine that when LLM-based tools are deployed on digital labor platforms on the employer side to assist in writing job descriptions, they may be successful in helping employers write more thorough job descriptions that improve matching. The key difference between this use of generative AI and that which we study in this paper is the underlying incentives behind the communication task that the LLMs are augmenting. When a sender sends signals in order to persuade, LLMs can disrupt natural occurring signaling mechanisms in which the costs of communication provide separation of types, thereby sorting the market efficiently. When a sender sends information in order to inform, LLMs can help improve the quality of that information by lowering the costs of producing that information.

Therefore, in a market less congested than the typical digital labor platform, in which workers can more easily find jobs, LLMs may actually improve the efficiency of labor market signaling by enhancing sorting on horizontal preferences. Workers can use an LLM to signal what \textit{kinds} of jobs they want, rather than just how qualified they are for \textit{any} job. However, when labor market competition is fierce, this paper finds that workers are incentivized to use this technology to send the best signals possible, thus rendering previously costly communication as nothing more than cheap talk.

We believe fruitful future work would be to understand how we can design markets in order to incentivize the use of generative AI to enhance communication, horizontal matching, and ultimately welfare.

\newpage 
\vspace*{-5ex}
\nocite{*} 
\vspace*{-5ex}
\printbibliography

\appendix
\newpage
\section*{\huge \centering Appendix}\label{appendix}
\renewcommand{\thesubsection}{\Alph{subsection}}
\renewcommand{\thesection}{\Alph{section}}

\renewcommand{\thefigure}{A\arabic{figure}} 
\setcounter{figure}{0}
\renewcommand{\thetable}{A\arabic{table}} 
\setcounter{table}{0}
\renewcommand{\thesubsection}{A.\arabic{subsection}}
\setcounter{subsection}{0}
\section{Additional Descriptive Tables and Figures}\label{appendix_descriptives}


\input{tables/coding_skill_tags.tex}

\newpage
\begin{figure}[H]
    \centering
        \begin{minipage}{0.85\linewidth}
        \centering
        \includegraphics[width=1\linewidth]{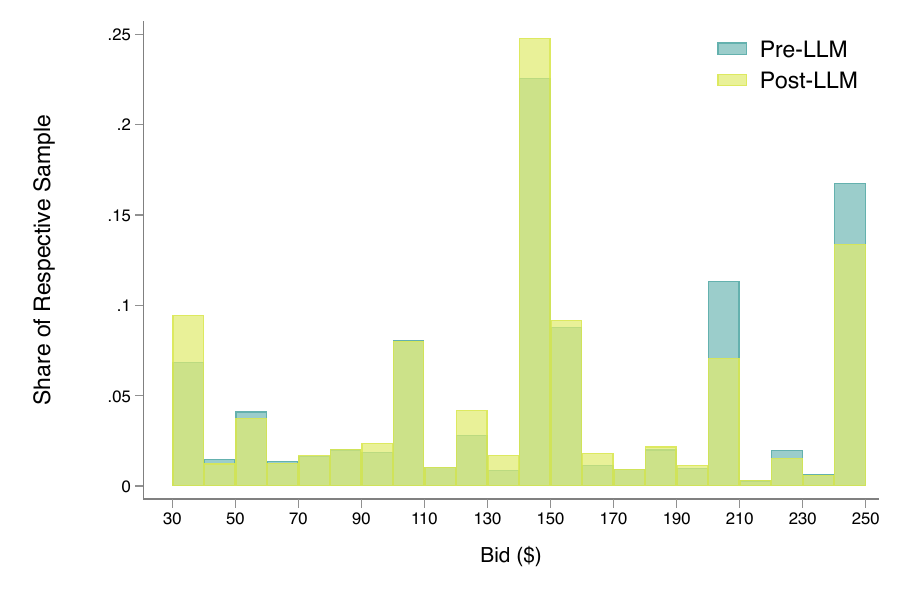}
    \end{minipage}

    \begin{minipage}{0.85\linewidth}
        \centering
        \includegraphics[width=1\linewidth]{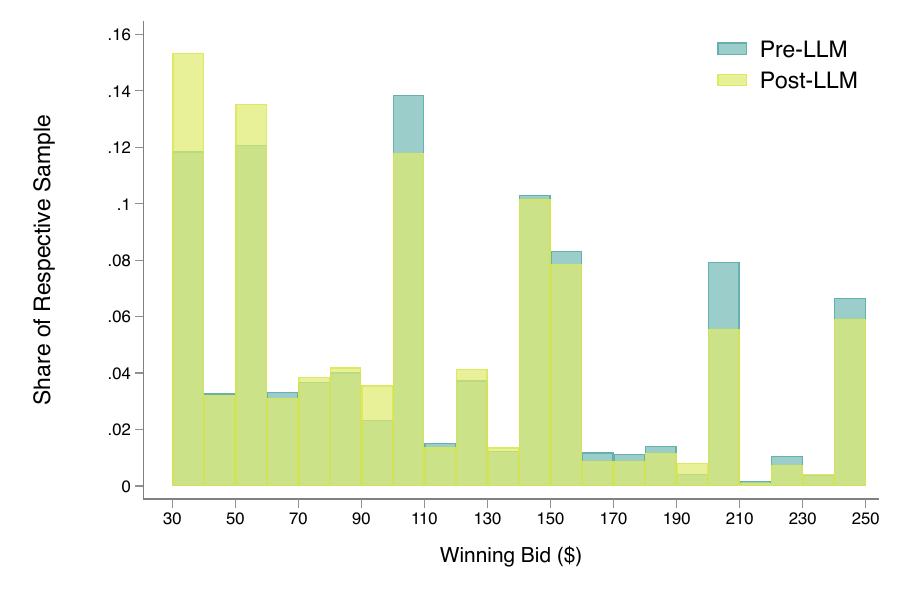}
    \end{minipage}
    \vspace{2mm}
    \caption{Distributions of Bids and Winning Bids: Pre-LLM vs. Post-LLM}
    \label{fig:bids_and_winning_bids_prepost}
    \floatfoot{\footnotesize \textit{Notes:} This figure plots histogram estimates of the distribution of bids and winning bids in both the pre-LLM and post-LLM samples. The post-LLM sample corresponds to job posts after March 26, 2024.}
\end{figure}
\newpage

\input{tables/considered_sum_stats.tex}

\input{tables/logit_table_custom_edited.tex}

\input{tables/logit_table_generic_edited.tex}

\newpage
\begin{figure}[H]
    \centering
    \begin{minipage}{0.85\linewidth}
        \centering
        \includegraphics[width=1\linewidth]{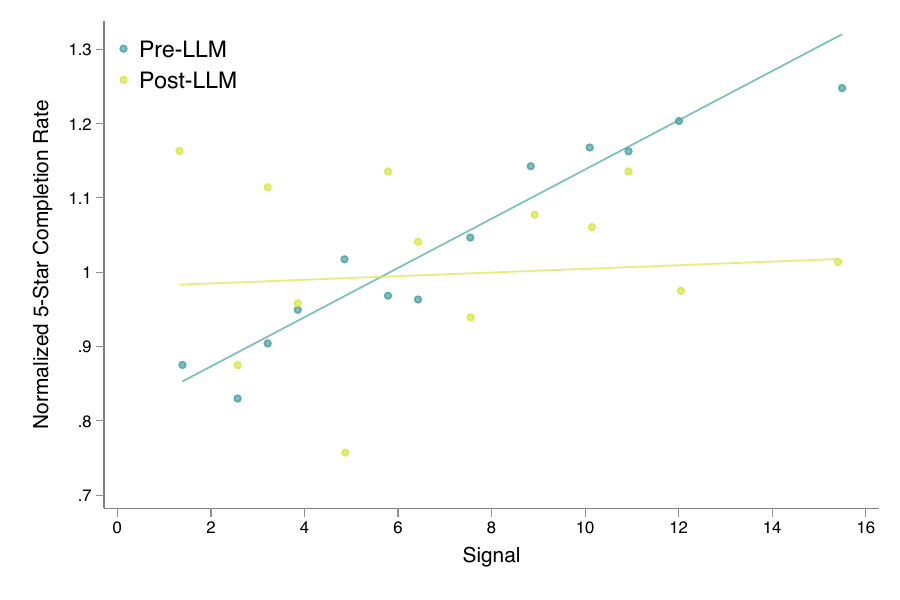}
    \end{minipage}

    \begin{minipage}{0.85\linewidth}
        \centering
        \includegraphics[width=1\linewidth]{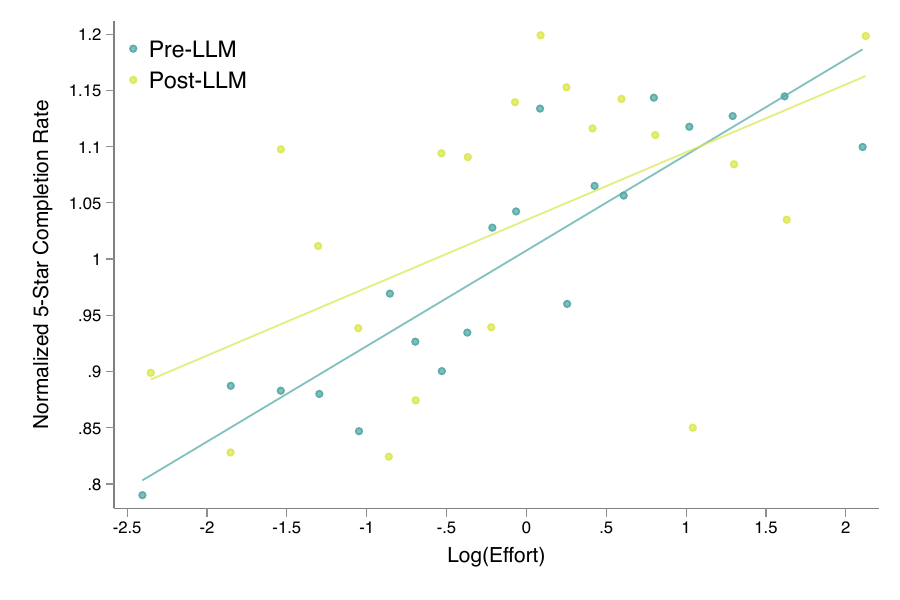}
        \vspace{1mm}
    \end{minipage}
    \vspace{2mm}
    \captionsetup{justification=centering}
    \caption{Binscatters of 5‑Star Completion Rate (i.e., Divided by Pre-LLM Means) \\ on Signal and Log Effort: Pre‑LLM vs. Post‑LLM}
    \label{fig:five_star_completion_binscatters}
    \floatfoot{\footnotesize \raggedright \textit{Notes:} This figure plots binscatters of whether the hired worker completes the job with 5 stars on signal and log effort in the pre‑LLM and post‑LLM samples. As per our research agreement with \texttt{Freelancer.com}, we do not disclose the level of completion rates, so we normalize (i.e., divide) the $y$-axis by the unconditional pre-LLM overall 5-star completion rate. The post‑LLM sample corresponds to the subsample of job posts after March 26, 2024.}
\end{figure}

\newpage
\begin{figure}[H]
    \centering
\includegraphics{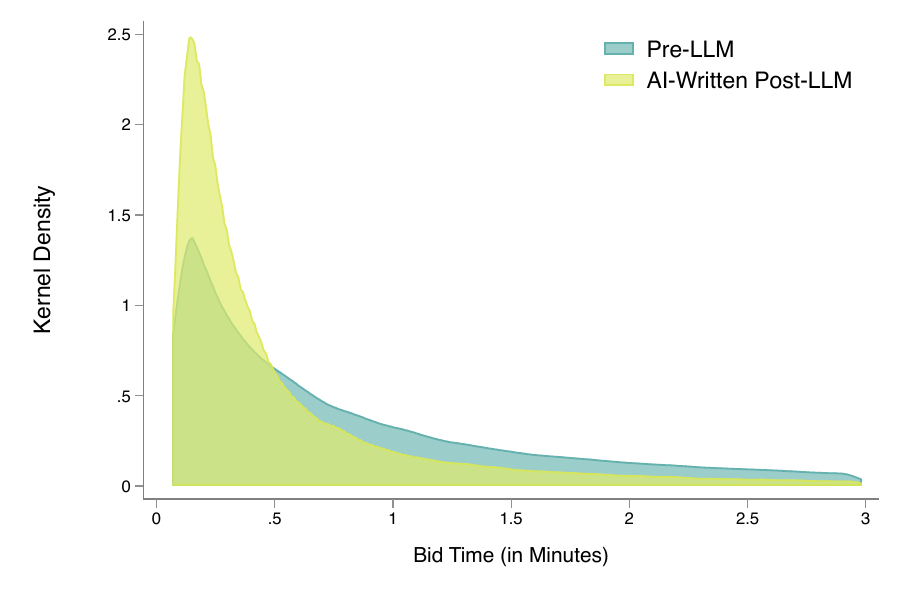}
    \caption{Distribution of Bid Times: Pre-LLM vs. Post-LLM, AI-Written}
    \label{fig:bidtimes_preai}
     \floatfoot{\footnotesize \textit{Notes:} This figure plots kernel density estimates of the distributions of bid time, defined as the time between when a worker first clicks on a job post and when she submits her application, in both the pre-LLM and post-LLM samples, where the latter is subsetted to only include the applications using the on-platform AI-writing tool. The post-LLM sample corresponds to the subsample of post-LLM job posts that were posted after March 26, 2024. For both samples, we only include applications with bid times between 4 seconds and 12 minutes. The pre-LLM sample has a mean of 1.47 minutes, a median of 0.70 minutes, and a standard deviation of 1.95 minutes, while the post-LLM, AI-written sample has a mean of 1.00 minutes, a median of 0.37 minutes, and a standard deviation of 1.70 minutes.}
\end{figure}

\input{tables/reg_outcomes_on_signal_effcorr.tex}

\newpage
\FloatBarrier
\renewcommand{\thefigure}{B\arabic{figure}} 
\setcounter{figure}{0}
\renewcommand{\thetable}{B\arabic{table}} 
\setcounter{table}{0}
\renewcommand{\thesubsection}{B.\arabic{subsection}}
\setcounter{subsection}{0}
\section{Additional Results}\label{appendix_extra_results}

\subsection{Additional Model Estimates}\label{appendix_extra_results_signal_production}

\begin{figure}[H]
    \centering
\includegraphics[width=1\linewidth]{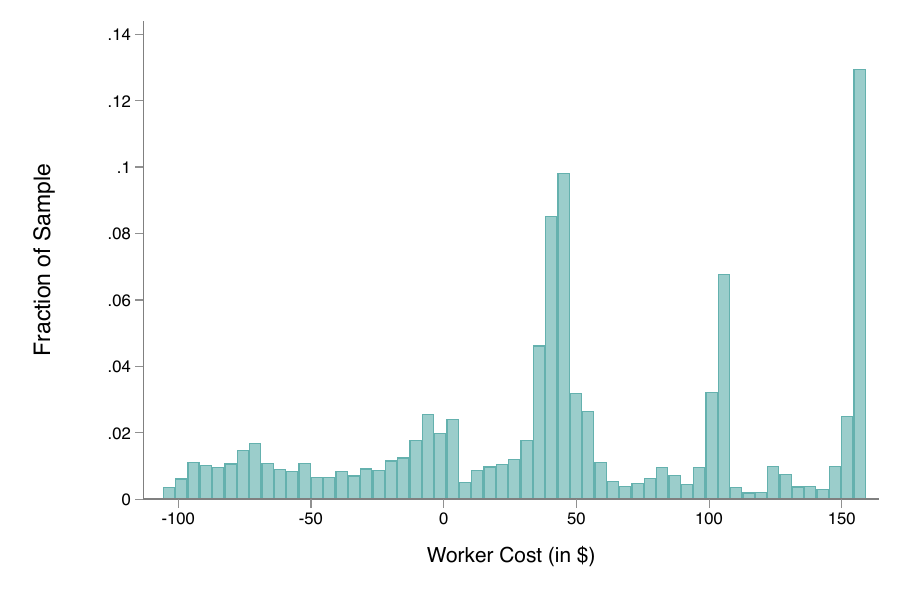}
    \caption{Histogram of Cost Estimates}\label{fig:hist_cost}
     \floatfoot{\footnotesize \textit{Notes:} This figure plots histogram estimates of the distribution of worker costs estimated from the model.}
\end{figure}

\newpage

\begin{figure}[H]
    \centering
\includegraphics[width=1\linewidth]{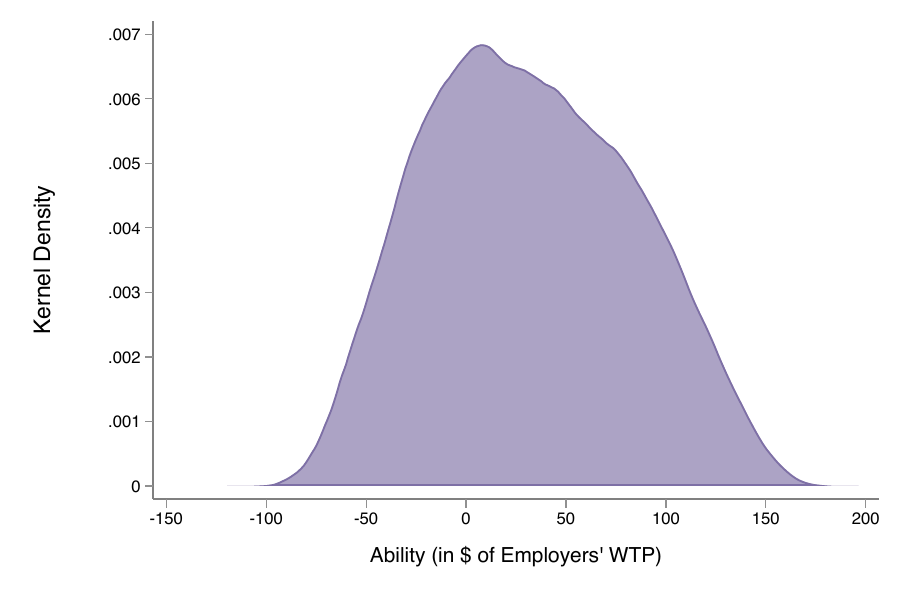}
    \caption{Kernel Density Estimate of Worker Abilities}\label{fig:kdens_ability}
     \floatfoot{\footnotesize \textit{Notes:} This figure plots kernel density estimates of the distribution of worker abilities estimated from the model. Ability estimates are reported in dollars of employer's WTP, i.e., $\frac{\widehat{\beta}}{|\widehat{\alpha}|}\cdot\widehat{a}$.}
\end{figure}

\newpage

\begin{figure}[H]
    \centering
\includegraphics[width=1\linewidth]{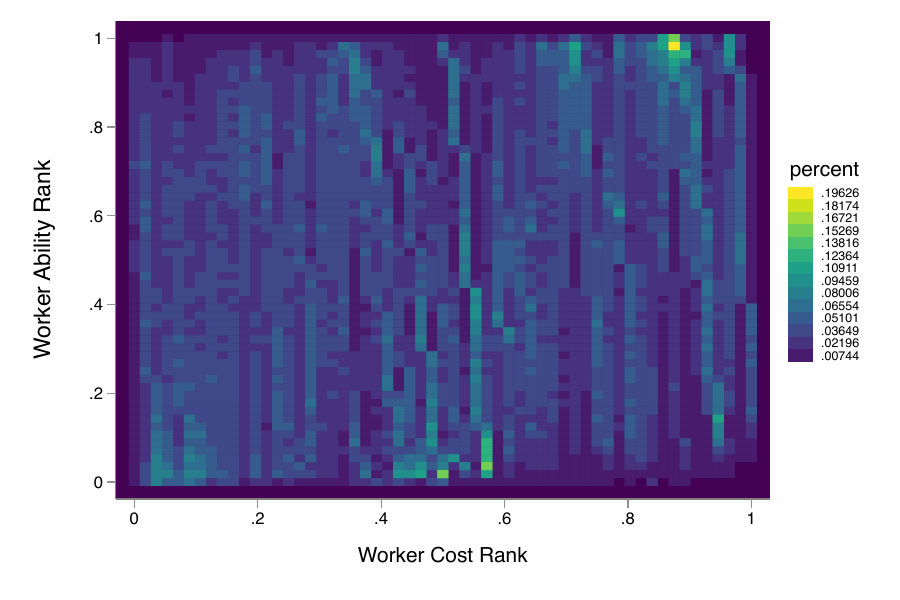}
    \caption{Joint Distribution of Worker Abilities and Costs in Ranks}\label{fig:heatmap_cost_ability}
     \floatfoot{\footnotesize \textit{Notes:} This figure plots a heatmap estimate of the joint distribution of estimated worker abilities (in ranks) and costs (in ranks) estimated from the model.}
\end{figure}

\newpage

\input{tables/cost_ability_by_x.tex}

\newpage

\begin{figure}[!ht]
    \centering
\includegraphics[width=1\linewidth]{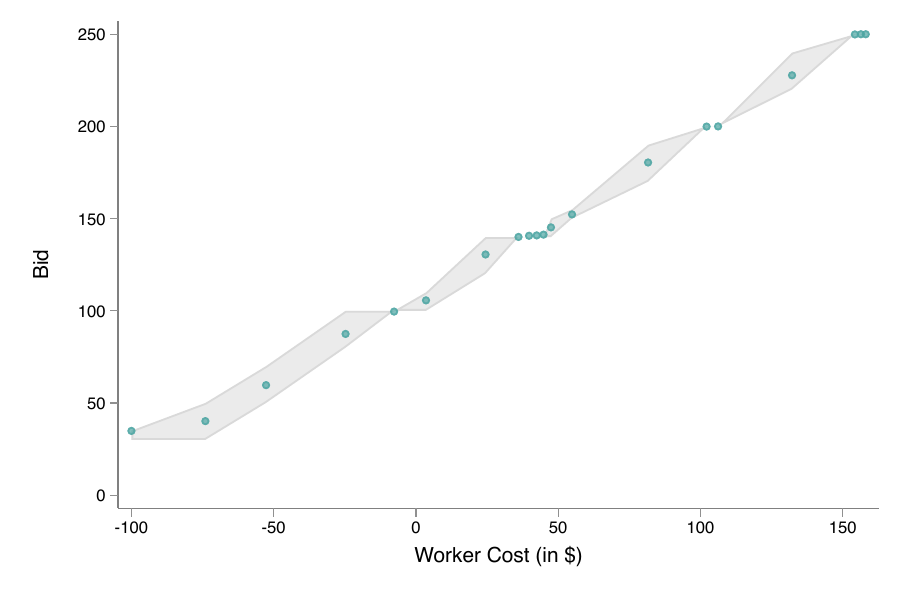}
    \caption{Estimated Equilibrium Strategy: Bid vs. Cost}\label{fig:bid_vs_cost}
     \floatfoot{\footnotesize \textit{Notes:} This figure plots a binscatter of equilibrium (i.e., observed) bids on estimated costs. Also plotted are 25th and 75th percentile bands of equilibrium bid within each cost bin.}
\end{figure}

\newpage

\begin{figure}[!ht]
    \centering
\includegraphics[width=1\linewidth]{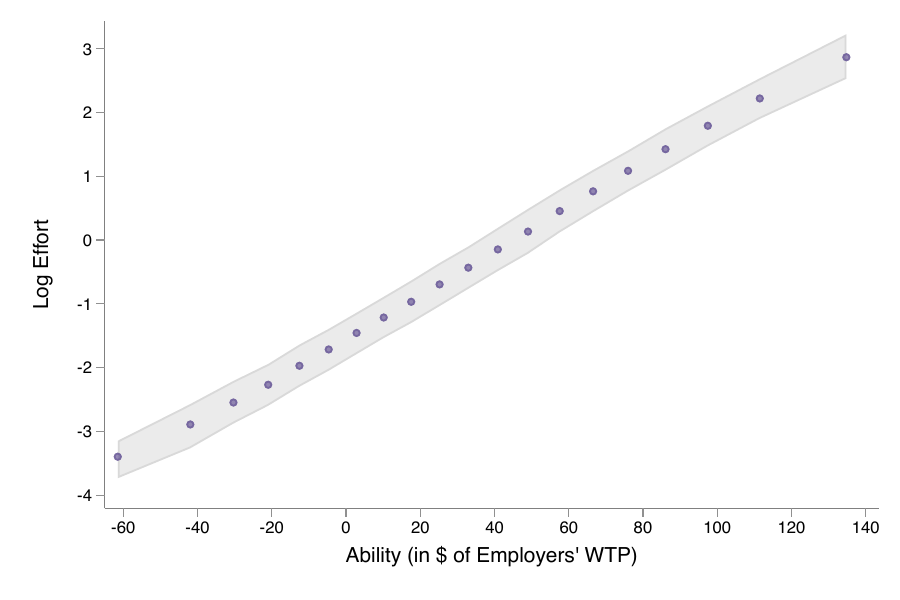}
    \caption{Estimated Equilibrium Strategy: Log Effort vs. Ability}\label{fig:log_effort_vs_ability}
     \floatfoot{\footnotesize \textit{Notes:} This figure plots a binscatter of equilibrium (i.e., observed) log efforts on estimated abilities. Also plotted are 25th and 75th percentile bands of log effort within each ability bin. Ability is reported in dollars of employers' WTP, i.e. $\frac{\widehat{\beta}}{|\widehat{\alpha}|}\cdot\widehat{a}$.}
\end{figure}

\newpage

\begin{figure}[!ht]
    \centering
\includegraphics[width=1\linewidth]{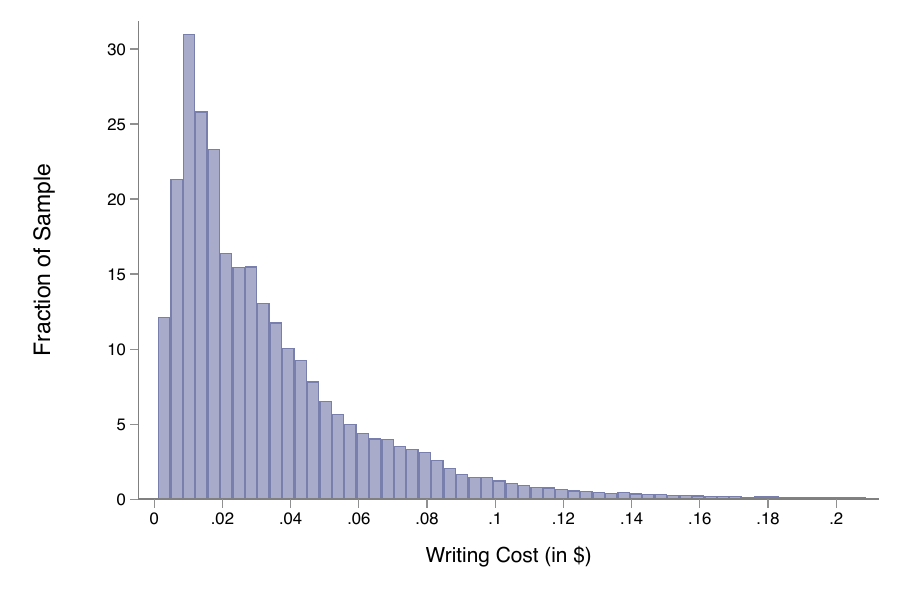}
    \caption{Histogram of Equilibrium Writing Cost Estimates}\label{fig:hist_writing_cost}
     \floatfoot{\footnotesize \textit{Notes:} This figure plots histogram estimates of the distribution of worker writing costs of the observed effort choices, with abilities estimated from the model, i.e., $\frac{1}{2}\cdot\exp(-\widehat{a}_{i j})\cdot e_{ij}^2$.}
\end{figure}

\newpage

\input{tables/signal_production_estimates.tex}

\newpage

\input{tables/tx_by_x.tex}

\newpage

\FloatBarrier
\subsection{Additional Counterfactual Results}\label{appendix_extra_results_cfs}

\begin{figure}[H]
    \centering
\includegraphics{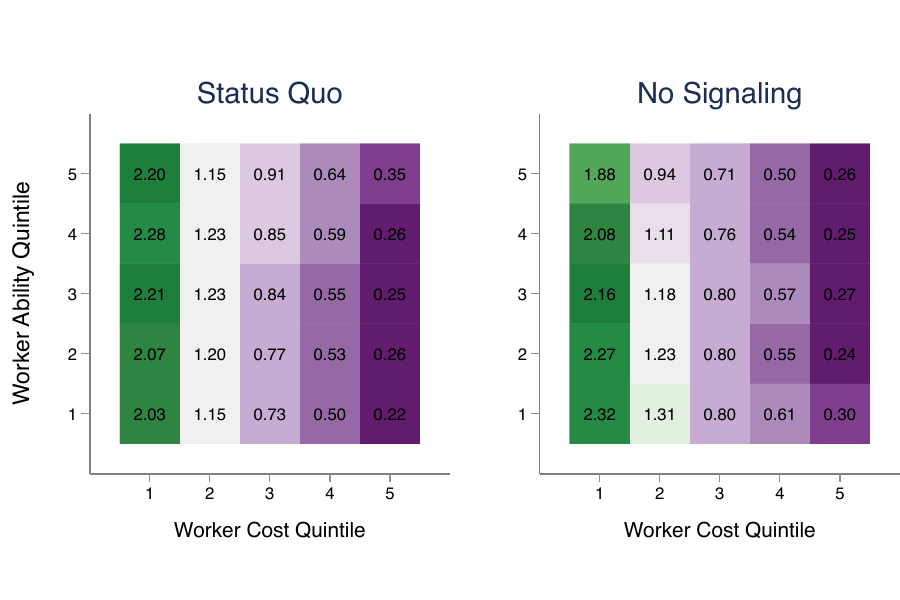}
\captionsetup{justification=centering}
    \caption{Normalized Hiring Rates by Ability and Cost Quintiles: \\ No-Signaling vs. Status-Quo}\label{fig:ability_cost_quintile_sq_vs_no}
     \floatfoot{\footnotesize \raggedright \textit{Notes:} This figure plots heatmaps of the normalized win rates for each combination of worker ability and cost quintiles in both the status-quo (SQ) and no-signaling (NS) equilibria. The color of each cell represents the normalized win rate for workers in that ability-cost quintile cell. The numbers in each cell present the exact normalized win rate for that cell. Win rates are normalized (i.e., divided) by the overall win rate in the SQ equilibrium, so for example, the number reported in the top left corner of the heatmap on the right should be interpreted as workers in the bottom quintile of the cost distribution and the top quintile of the ability distribution are hired 1.88 times more in the NS equilibrium than the average worker is hired in the SQ equilibrium.}
\end{figure}

\newpage
\begin{figure}[!htbp]
    \centering
\includegraphics[width=1\linewidth]{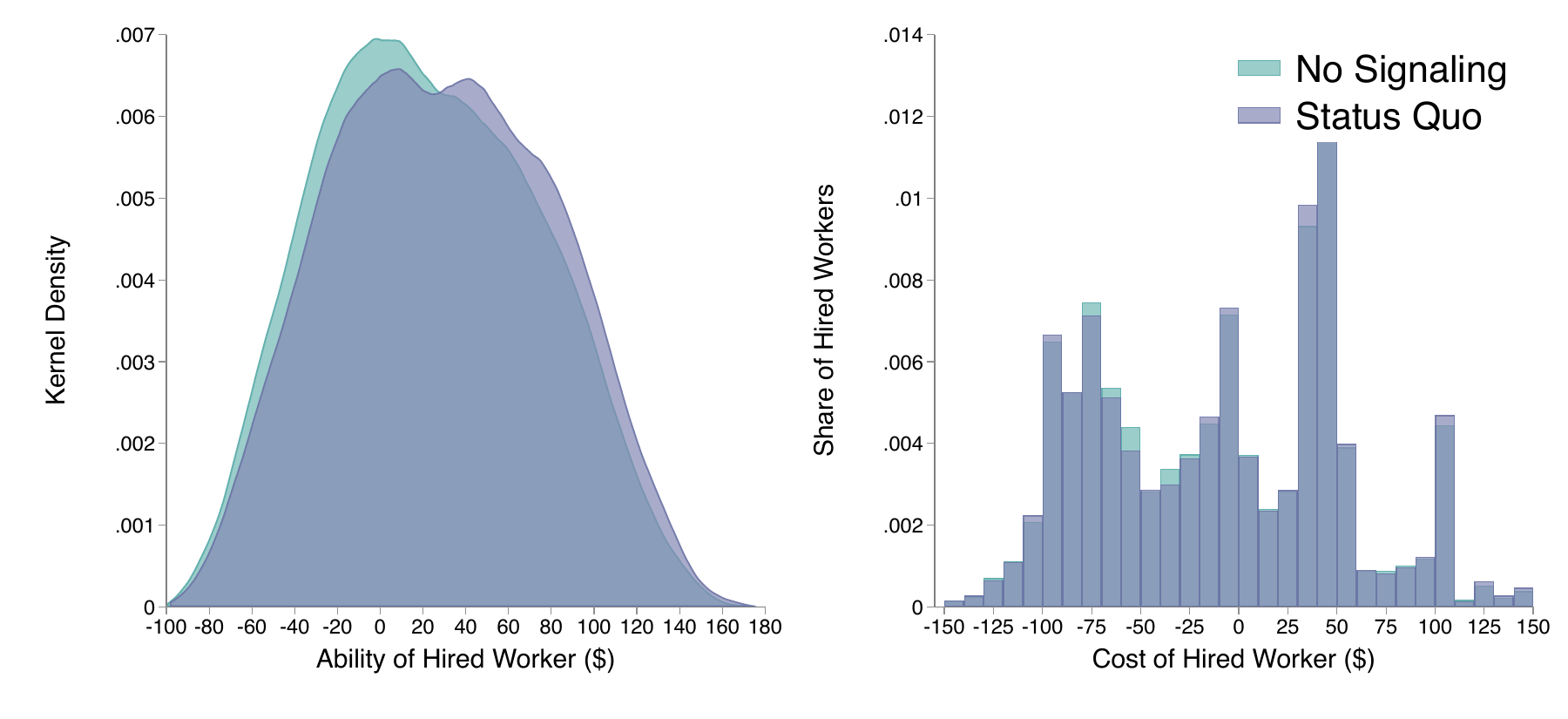}
\captionsetup{justification=centering}
    \caption{Histograms of Hired Workers' Ability and Cost Quantiles: \\ No-Signaling vs. Status-Quo}\label{fig:cost_ability_changes_sq_vs_no}
     \floatfoot{\footnotesize \raggedright \textit{Notes:} This figure plots histogram estimates of the distribution of quantiles of the abilities and costs of hired workers in both the status-quo (SQ) and no-signaling (NS) equilibria.}
\end{figure}

\newpage
\input{tables/pct_diff_sq_vs_no_by_obs.tex}

\newpage
\vspace{1mm}
\newpage
\FloatBarrier
\renewcommand{\thefigure}{C\arabic{figure}} 
\setcounter{figure}{0}
\renewcommand{\thetable}{C\arabic{table}} 
\setcounter{table}{0}
\renewcommand{\thesubsection}{C.\arabic{subsection}}
\setcounter{subsection}{0}
\section{Measurement Details}\label{appendix_measurement}

\subsection{Signal Measurement Details}\label{appendix_measurement_signal}

In this subsection, we describe our procedure for measuring signals. We begin with an overview of the measurement procedure, then provide excerpts of the exact prompt we used, and finally we provide four examples of applications and job posts along with their measured efforts and signals.

\subsubsection{Signal Measurement Overview}\label{appendix_measurement_signal_overview}
Our general procedure was to process each application job post (including job name) pair one at a time, clearing the LLM's context between each pair. We used \textit{Meta}'s Llama 4 Maverick 17B model to assess each pair on nine different criteria, five of which we characterize as directly concerning customization, and four of which we characterize as concerning relevance and generic writing quality. For each criterion, we prompt the LLM to output a 0, 1, or 2, each defined according to a specific rubric detailed in Section~\ref{appendix_measurement_signal_prompt}.

We prompt the LLM to provide reasoning for its answers by citing direct evidence or lack thereof for why it chose the response it did. This ``reasoning-step'' is useful to enhance the model's internal logic and also allows us to audit the model's responses for a small sample of applications to ensure that the model is following the intended rubric. We do not use the text of the reasoning in constructing our measure of signal. We also provide the model with a few examples for each call of how to apply the rubric before asking it to evaluate the actual application job post pair.

After scoring each application on the nine criteria above, we compute our copy-pasting correction. In particular, we compute a normalized Levenshtein (i.e., ``edit'') distance between an application and all other applications sent by the same freelancer in our sample. This normalized edit distance can be interpreted as the percent an application would need to change to make it equal to the other. We then take the minimum normalized edit distance across all other applications by the same freelancer. If that minimum distance is below 4 percent, we argue that the worker cannot have customized the application to the associated job post. Thus, we reasonably conclude that if the LLM assigned any of the custom criteria to be above 0, then those scores are incorrect, and we adjust them to zero.

After applying the copy-pasting correction, we sum all nine criteria scores, placing twice the weight on the five custom criteria, and we normalize the resulting sum to be between 0 and 18. This normalized sum is our final measure of signal.

\begin{figure}[H]
    \centering
\includegraphics[width=.74\linewidth]{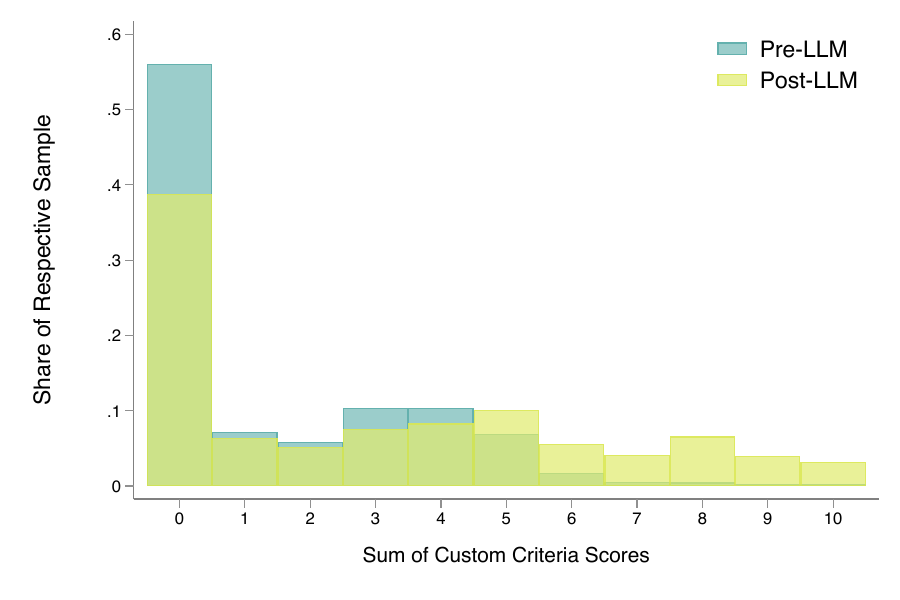}
    \caption{Distribution of Sum of Custom Criteria Scores: Pre-LLM vs. Post-LLM}
    \label{fig:custom_signals_prepost}
     \floatfoot{\footnotesize \textit{Notes:} This figure plots histogram estimates of the distributions of the sum of copy-paste adjusted custom criteria scores in both the pre-LLM and post-LLM samples. The post-LLM sample corresponds to the subsample of post-LLM job posts that were posted after March 26, 2024.}
\end{figure}

\begin{figure}[H]
    \centering
\includegraphics[width=.74\linewidth]{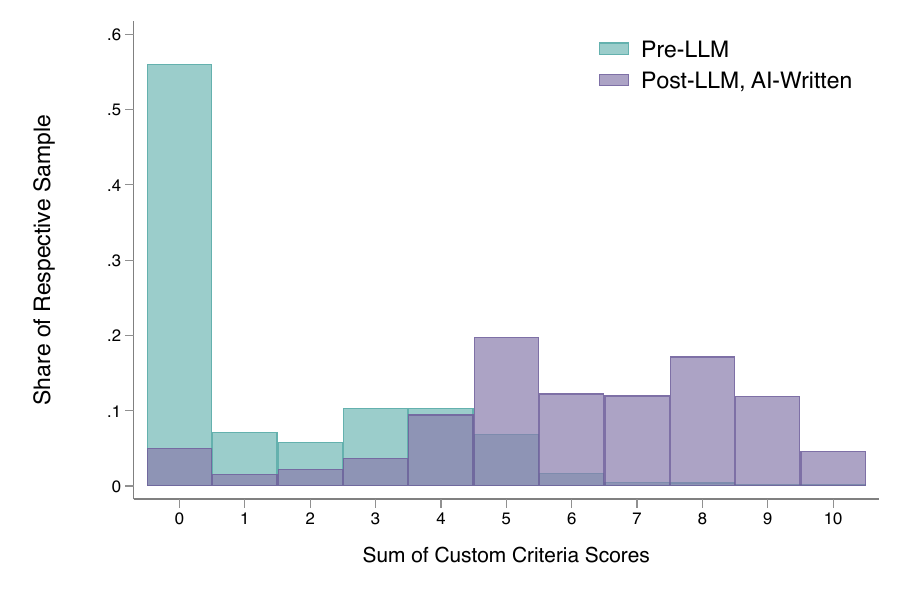}
    \caption{Distribution of Sum of Custom Criteria Scores: Pre-LLM vs. AI-Written Post-LLM}
    \label{fig:custom_signals_pre_vs_ai}
     \floatfoot{\footnotesize \textit{Notes:} This figure plots histogram estimates of the distributions of the sum of copy-paste adjusted custom criteria scores in both the pre-LLM and post-LLM samples, where the latter is subsetted to only include the applications using the on-platform AI-writing tool. The post-LLM sample corresponds to the subsample of post-LLM job posts that were posted after March 26, 2024.}
\end{figure}

\begin{figure}[H]
    \centering
\includegraphics[width=.74\linewidth]{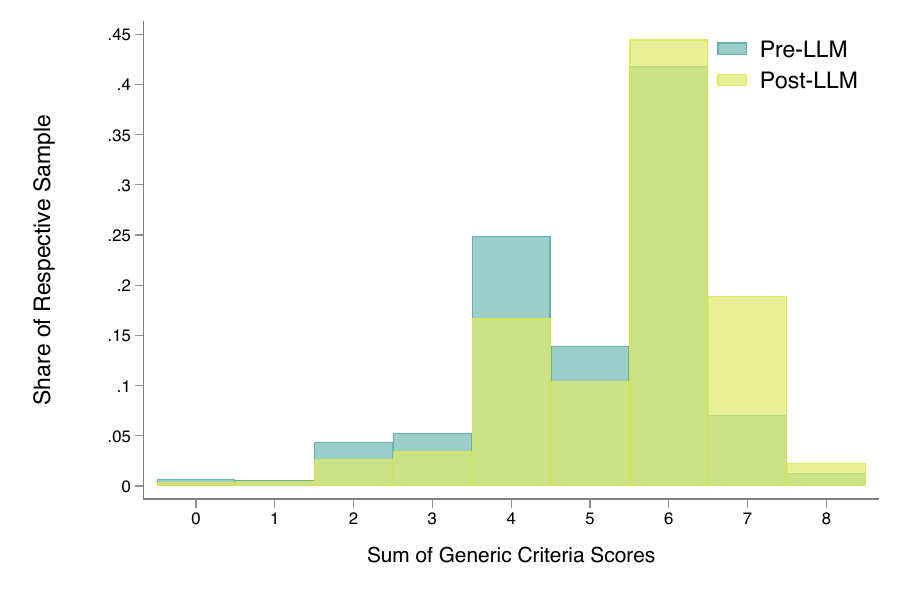}
    \caption{Distribution of Sum of Generic Criteria Scores: Pre-LLM vs. Post-LLM}
    \label{fig:generic_signals_prepost}
     \floatfoot{\footnotesize \textit{Notes:} This figure plots histogram estimates of the distributions of the sum of generic criteria scores in both the pre-LLM and post-LLM samples. The post-LLM sample corresponds to the subsample of post-LLM job posts that were posted after March 26, 2024.}
\end{figure}

\begin{figure}[H]
    \centering
\includegraphics[width=.74\linewidth]{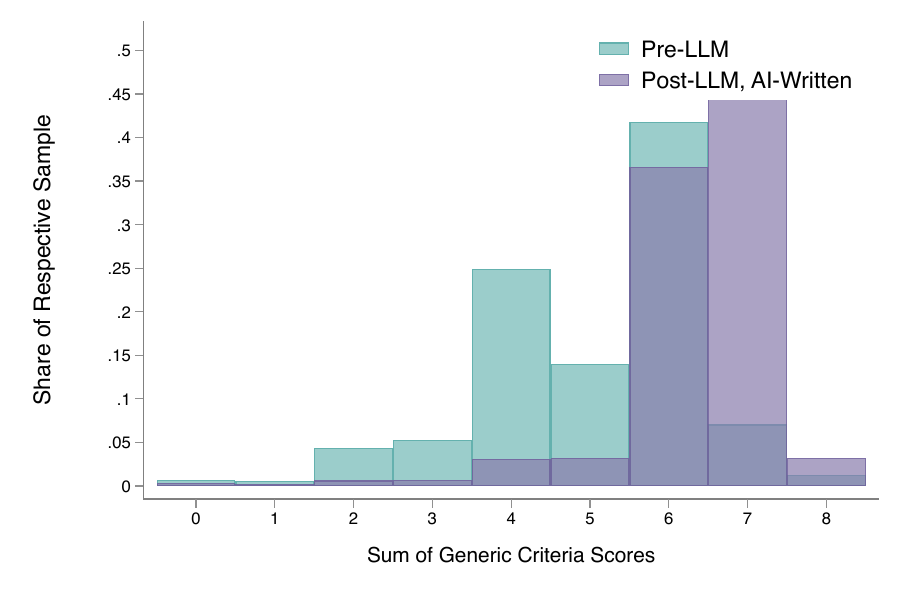}
    \caption{Distribution of Sum of Generic Criteria Scores: Pre-LLM vs. AI-Written Post-LLM}
    \label{fig:generic_signals_pre_vs_ai}
     \floatfoot{\footnotesize \textit{Notes:} This figure plots histogram estimates of the distributions of the sum of generic criteria scores in both the pre-LLM and post-LLM samples, where the latter is subsetted to only include the applications using the on-platform AI-writing tool. The post-LLM sample corresponds to the subsample of post-LLM job posts that were posted after March 26, 2024.}
\end{figure}

Figures~\ref{fig:custom_signals_prepost}, \ref{fig:custom_signals_pre_vs_ai}, \ref{fig:generic_signals_prepost}, and \ref{fig:generic_signals_pre_vs_ai} plot histogram estimates of the distributions of the sum of generic and copy-paste adjusted custom criteria scores in both the pre-LLM and post-LLM samples, as well as the post-LLM sample subsetted to only include the applications using the on-platform AI-writing tool.

\newpage

\newpage

\newpage

\FloatBarrier
\subsubsection{Signal Measurement LLM Prompt}\label{appendix_measurement_signal_prompt}

In this subsection, we provide the exact prompt we used, broken up by sections for readability, only missing some technical lines that request the LLM to output in JSON format and the scored examples we provided to the model along with each application job post pair.

\begin{tcolorbox}[colback=gray!5, colframe=gray!40, title=Prompt Setup, coltitle=black]
{%
\scriptsize
\begingroup
\setstretch{1}            
\parskip=0pt plus 0pt     
\parindent=0pt            
\frenchspacing             

You are evaluating a freelancer application against a job posting. You will assess the application across multiple criteria using a two-step process for each: first identify evidence, then assign scores.

\textbf{Process Overview}

1. For each evaluation criterion, systematically identify relevant evidence from both documents. Note both supporting evidence (what the applicant did demonstrate) and missing evidence (what they should have but didn't demonstrate based on the job posting and the rubric).

2. Based on the evidence, assign a score of 0, 1, or 2 according to the provided rubric.

3. Evaluate each criterion separately, without taking the other criteria into account.

\endgroup
}
\end{tcolorbox}

\begin{tcolorbox}[colback=gray!5, colframe=gray!40, title=Instructions, coltitle=black]
{%
\scriptsize
\begingroup
\setstretch{1}            
\parskip=0pt plus 0pt     
\parindent=0pt            
\frenchspacing             

You are evaluating a freelancer application against a job posting. You will assess the application across multiple criteria using a two‐step process for each: first identify evidence, then assign scores.

\textbf{Instructions}

For each criterion:
\begin{enumerate}[leftmargin=*, itemsep=0pt, topsep=0pt, parsep=0pt, partopsep=0pt]
  \item \textbf{Evidence Analysis:} Identify specific supporting evidence, missing evidence, and any ambiguous areas from the documents.
  \item \textbf{Scoring:} Apply the rubric to assign a score (0, 1, or 2).
\end{enumerate}

Do not let the score for one question inform your score for another question! Each criterion should be scored on its own merits, not in the context of the rest of the assessment.

\endgroup
}
\end{tcolorbox}

\begin{tcolorbox}[colback=gray!5, colframe=gray!40, title = Evaluation Criteria and Rubrics: Relevant Skills, coltitle=black]
{%
\scriptsize
\begingroup
\setstretch{1}
\parskip=0pt plus 0pt
\parindent=0pt
\frenchspacing

\textbf{Evaluation Criteria}

Does the application explicitly mention skills that are relevant and necessary for successfully completing the task described in the job post, even if these skills are not explicitly listed in the job post itself?
Keep in mind that the job posts are tagged with skill tags like "JavaScript", or "Web Scraping". Simply copy and pasting those tags into the application should not contribute to a higher score.
A telltale sign of copy-and-pasting these tags is when applicants list multiple skills in a way that mirrors the job post's skill tags (e.g., "I have tons of experience in Web Scraping, Java, Python, Web Crawling, Website Design"), particularly when the skills maintain the title casing or capitalization from the original tags and are mentioned in the application but were not mentioned in the job post description itself.
Simply sharing portfolio links or links to previous projects with no further context on how these examples demonstrate the relevant skills should not contribute the highest score.
Natural mentions of relevant skills with basic context (e.g., "I have 5 years of experience coding in Python") should contribute to the score when relevant to the job, even without detailed evidence. However, obvious copy-pasting of skill tag lists with no genuine integration into the applicant's experience should not contribute to a higher score.

\textbf{Rubric}

\begin{itemize}[leftmargin=*, itemsep=0pt, topsep=0pt, parsep=0pt, partopsep=0pt]
  \item \textbf{Score 2:} Application includes multiple specific and clearly relevant skills, commensurate with the complexity of the job post.
  \item \textbf{Score 1:} At least one relevant skill, but vague or incomplete or missing context.
  \item \textbf{Score 0:} No relevant skills, or a list of relevant skills that is lifted verbatim from the job posting.
\end{itemize}
\endgroup
}
\end{tcolorbox}

\begin{tcolorbox}[colback=gray!5, colframe=gray!40, title = Evaluation Criteria and Rubrics: Relevant Experience, coltitle=black]
{%
\scriptsize
\begingroup
\setstretch{1}
\parskip=0pt plus 0pt
\parindent=0pt
\frenchspacing

\textbf{Evaluation Criteria}

Does the applicant provide concrete evidence of having done similar work to that of the job post? 
Example evidence are portfolio links, project examples, demonstrated years of work experience. 
Demonstrated previous work should relate directly to the skills the poster specifically asked for. 
Make note of whether the evidence is provided with no further context on how it relates to the job post.

\textbf{Rubric}

\begin{itemize}[leftmargin=*, itemsep=0pt, topsep=0pt, parsep=0pt, partopsep=0pt]
  \item \textbf{Score 2:} Provides evidence (links or explicit descriptions of past projects) of previous work AND makes explicit that that work is relevant to job post.
  \item \textbf{Score 1:} Mentions years of experience in relevant work/skill but does not provide evidence AND/OR provides links with no context as to why they are relevant. 
  \item \textbf{Score 0:} No mention of relevant experience.
\end{itemize}
\endgroup
}
\end{tcolorbox}

\begin{tcolorbox}[colback=gray!5, colframe=gray!40, title = Evaluation Criteria and Rubrics: English Proficiency, coltitle=black]
{%
\scriptsize
\begingroup
\setstretch{1}
\parskip=0pt plus 0pt
\parindent=0pt
\frenchspacing

\textbf{Evaluation Criteria}

Does the application use well-written, natural English? Is the application written clearly using standard syntax suitable for a business environment? 
Focus on clarity and comprehensibility, accepting minor grammar issues if meaning is clear and structure is coherent.

\textbf{Rubric}

\begin{itemize}[leftmargin=*, itemsep=0pt, topsep=0pt, parsep=0pt, partopsep=0pt]
  \item \textbf{Score 2:} Clear, standard English. Grammatical errors do not go beyond missing punctuation or the occasional misspelling.
  \item \textbf{Score 1:} Mostly correct English. May include a few instances of non-standard phrasing or syntax or grammatical errors that do not impede understanding.
  \item \textbf{Score 0:} Uses non-standard syntax or phrasing heavily. May include many grammatical errors. 
\end{itemize}
\endgroup
}
\end{tcolorbox}

\begin{tcolorbox}[colback=gray!5, colframe=gray!40, title = Evaluation Criteria and Rubrics: Professional Tone, coltitle=black]
{%
\scriptsize
\begingroup
\setstretch{1}
\parskip=0pt plus 0pt
\parindent=0pt
\frenchspacing

\textbf{Evaluation Criteria}

Is the application respectful, business-like, well-organized, and credible, with realistic claims and no exaggerated guarantees? 
Minor grammar slips are fine if the message is clear, but language so poor that the meaning is hard to follow should lower the score.

\textbf{Rubric}

\begin{itemize}[leftmargin=*, itemsep=0pt, topsep=0pt, parsep=0pt, partopsep=0pt]
  \item \textbf{Score 2:} Highly professional and credible. Polite, organized, no slang or emojis. Demonstrates appropriate enthusiasm, makes only realistic claims, and errors (if any) do not impede comprehension.
  \item \textbf{Score 1:} Moderately professional. Generally respectful but may include mild slang, emojis, excessive punctuation, or minor exaggerated wording; message remains clear and mostly believable.
  \item \textbf{Score 0:} Unprofessional or not credible. Disrespectful, pushy, heavy use of slang or emojis. May make unrealistic promises of “perfect results” or use English so poor the core message is hard to understand.
\end{itemize}
\endgroup
}
\end{tcolorbox}

\begin{tcolorbox}[colback=gray!5, colframe=gray!40, title = Evaluation Criteria and Rubrics: Customization, coltitle=black]
{%
\scriptsize
\begingroup
\setstretch{1}
\parskip=0pt plus 0pt
\parindent=0pt
\frenchspacing

\textbf{Evaluation Criteria}

Is this application customized for this specific job post, or could it be submitted to a similar job without any changes? 
In other words, how much of the application's content and structure is purpose-built for this job vs. reusable boiler-plate? 
Evidence of customization can include job-specific greetings or introductions, bespoke plans or timelines, and/or skills rephrased around a client's needs (e.g. if a client says "I need to collect data from websites", the applicant might say "I have experience building web scrapers using Python")
Note that directly copy-pasted phrases and sentences from the job post do not count towards customization.

\textbf{Rubric}

\begin{itemize}[leftmargin=*, itemsep=0pt, topsep=0pt, parsep=0pt, partopsep=0pt]
  \item \textbf{Score 2:} Application is highly customized. May include some generic language, but includes substantial evidence of customization. 
  \item \textbf{Score 1:} Application shows modest evidence of customization, such as 1 or 2 sentences that mention things specific to this job post and not the type of task generally, but is otherwise generic.
  \item \textbf{Score 0:} Application is completely generic — application could be sent to any job of the same general type, no evidence of customizing or tailoring to the job post whatsoever. If the only customized part of the application is copy pasted, the score should be 0.
\end{itemize}
\endgroup
}
\end{tcolorbox}

\begin{tcolorbox}[colback=gray!5, colframe=gray!40, title = Evaluation Criteria and Rubrics: Read Through, coltitle=black]
{%
\scriptsize
\begingroup
\setstretch{1}
\parskip=0pt plus 0pt
\parindent=0pt
\frenchspacing

\textbf{Evaluation Criteria}

Does the applicant specifically reference requirements or core details from the job post (not just the title)? Saying that they have gone through/read the requirements does not count. Content copied directly from the job description or job name should not be considered as evidence of having read the post.
If an attention check is present (e.g. "if you read this through, please include the word X in your response"), and the bid fails it, then the score cannot be 2, but could be 1 if there are other indications of the applicant having read the job post.
If an attention check is present and the bid passes it, then the score should not be 0, but could still be 1 if there are other signs the applicant did not read the job post in detail.
A telltale sign of copy-pasting is identical phrasing and capitalization, potentially even including typos.
A higher score requires specific mentions or callouts of what is involved in the task (e.g., “Integrate a payment system onto our website using PHP Laravel, “Build a website for our online marketing company, using the existing logo”, or “Deliverables should include a zipped package of all your code with a detailed readme”) with no generic language that could apply broadly to all projects of the same general nature
Score this with consideration of the amount of details in the job post: If the job post contains few details to begin with, the applicant cannot reference as many details and should not be penalized because of it.

\textbf{Rubric}

\begin{itemize}[leftmargin=*, itemsep=0pt, topsep=0pt, parsep=0pt, partopsep=0pt]
  \item \textbf{Score 2:} Applicant specifically and clearly addresses the project goals, including the majority of the unique specifications (e.g. deliverables, technology, budget, or similar) of the job posting.
  \item \textbf{Score 1:} Applicant obliquely or by inference addresses the project goals. Some of the unique specifications (e.g. deliverables, technology, budget, or similar) of the job posting are omitted.
  \item \textbf{Score 0:} Applicant does not address the stated project goals at all. No major details or required features are included in the application.
\end{itemize}
\endgroup
}
\end{tcolorbox}

\begin{tcolorbox}[colback=gray!5, colframe=gray!40, title = Evaluation Criteria and Rubrics: Understood Goal, coltitle=black]
{%
\scriptsize
\begingroup
\setstretch{1}
\parskip=0pt plus 0pt
\parindent=0pt
\frenchspacing

\textbf{Evaluation Criteria}

Does the application restate or rephrase the job post’s main goal accurately? 
Text copied verbatim from the job post is not evidence of understanding.
When evaluating goal restatement, do not award points (1s or 2s) for generic acknowledgments that merely identify the job type (e.g., 'I can build websites'), but do award points when applicants demonstrate understanding of specific project details or context (e.g., 'I can build the payment page for your car company website').

\textbf{Rubric}

\begin{itemize}[leftmargin=*, itemsep=0pt, topsep=0pt, parsep=0pt, partopsep=0pt]
  \item \textbf{Score 2:} Clearly and accurately restates or paraphrases the main goal of the job.
  \item \textbf{Score 1:} Mentions or implies the goal, but does so at a very high level, vaguely, incompletely, or with partial misunderstanding.
  \item \textbf{Score 0:} Does not mention the goal at all, or clearly misunderstands it.
\end{itemize}
\endgroup
}
\end{tcolorbox}

\begin{tcolorbox}[colback=gray!5, colframe=gray!40, title = Evaluation Criteria and Rubrics: Understood Complexity, coltitle=black]
{%
\scriptsize
\begingroup
\setstretch{1}
\parskip=0pt plus 0pt
\parindent=0pt
\frenchspacing

\textbf{Evaluation Criteria}

Does the application indicate awareness of the complexity or difficulty level of the described task in the job post?
When evaluating complexity understanding, distinguish between basic task comprehension and awareness of implementation challenges—for example, an applicant saying 'I can create a payment page that accepts credit cards' shows they understand the goal but may receive a lower complexity comprehension score if they don't acknowledge factors like PCI compliance, system integration, or error handling requirements.

\textbf{Rubric}

\begin{itemize}[leftmargin=*, itemsep=0pt, topsep=0pt, parsep=0pt, partopsep=0pt]
  \item \textbf{Score 2:} Clearly demonstrates understanding of complexity.
  \item \textbf{Score 1:} Vague or implied understanding of complexity.
  \item \textbf{Score 0:} No indication of understanding or clear misunderstanding of complexity.
\end{itemize}
\endgroup
}
\end{tcolorbox}

\begin{tcolorbox}[colback=gray!5, colframe=gray!40, title = Evaluation Criteria and Rubrics: Initiative, coltitle=black]
{%
\scriptsize
\begingroup
\setstretch{1}
\parskip=0pt plus 0pt
\parindent=0pt
\frenchspacing

\textbf{Evaluation Criteria}

Does the application demonstrate a clear initiative or plan to complete the task in the job post (e.g. concise plan, roadmap, or specific clarifying questions grounded in the job posting)? 
Any plan, roadmap, or question should only count towards this criterion if the applicant is directly referencing this specific job post. 
Many applications that are completely generic say something like “Can I have more details?” or “I will start work right away and can return the product to you in 24 hours” or “You can have unlimited revisions, so satisfaction guaranteed”. 
None of these should count as evidence towards an application showing initiative, since they are all completely generic.
Clarifying questions may be evidence of initiative only if they are specific to the project and address genuine ambiguity, as opposed to a general request for more details.

\textbf{Rubric}

\begin{itemize}[leftmargin=*, itemsep=0pt, topsep=0pt, parsep=0pt, partopsep=0pt]
  \item \textbf{Score 2:} Features a clear concise plan, detailed roadmap, or key clarifying questions.
  \item \textbf{Score 1:} Poses some vague plan, roadmap, or asks specific clarifying questions.
  \item \textbf{Score 0:} Does not include any plan, roadmap, or clarifying question.
\end{itemize}
\endgroup
}
\end{tcolorbox}

\FloatBarrier
\subsubsection{Signal Measurement Examples}\label{appendix_measurement_signal_examples}

In this subsection, we provide four examples of application and job post pairs along with their measured efforts and signals. We include one example job post from the pre-LLM period along with two applications (one high-effort, one low-effort) and one example job post from the post-LLM period along with two applications (one high-effort and not using the on-platform AI-writing tool, one low-effort and using the on-platform AI-writing tool).

\begin{tcolorbox}[colback=gray!5, colframe=Plum4, title = Pre-LLM Job Post Example, coltitle=black]
{%
\small
\begingroup
\setstretch{1}
\parskip=0pt plus 0pt
\parindent=0pt
\frenchspacing

\textbf{Build me an optimized shopify store}

Already have the store need someone to build the back end which will allow customers to choose colors of blanket, text color of blanket, designs are step and repeat pattern as well as adding multiple names. Also need an option for customers to be able to upload their own photos  or logos for total customization  sample for what I am looking for: \_\_\_\_\_\_\_\_.com

\endgroup
}
\end{tcolorbox}

\begin{tcolorbox}[colback=gray!5, colframe=Aquamarine4, title = {Pre-LLM Application Example 1: Low Effort, Low Signal}, coltitle=black]
{%
\small
\begingroup
\setstretch{1}
\parskip=0pt plus 0pt
\parindent=0pt
\frenchspacing

Hi, I am clear about your details that you need to customize your Shopify store. I will build the website exactly with the functionality that you need. I am expert Web developer with command on HTML/CSS Wix, WordPress, Shopify, Magento and PHP/MySQL.  Here are some websites that I have made: \_\_\_\_\_\_\_\_ Looking forward to working with you.  \_\_\_\_.

\textbf{Effort} $\boldsymbol{= 0.55}$ \textbf{minutes}

\textbf{Signal} $\boldsymbol{= 6.00}$

\endgroup
}
\end{tcolorbox}

\begin{tcolorbox}[colback=gray!5, colframe=Aquamarine4, title = {Pre-LLM Application Example 2: High Effort, High Signal}, coltitle=black]
{%
\small
\begingroup
\setstretch{1}
\parskip=0pt plus 0pt
\parindent=0pt
\frenchspacing

Hello,  Yes I can do this customization work where customer can choose variants like color , choose blanket , add text , upload their own photos  or logos  etc.  I need to discuss this project in detail after that I will suggest you we do this help of theme or we need to create private app .  I am waiting for your response .  Thanks, \_\_\_\_

\textbf{Effort} $\boldsymbol{= 2.18}$ \textbf{minutes}

\textbf{Signal} $\boldsymbol{= 14.79}$

\endgroup
}
\end{tcolorbox}

\begin{tcolorbox}[colback=gray!5, colframe=Plum4, title = Post-LLM Job Post Example, coltitle=black]
{%
\small
\begingroup
\setstretch{1}
\parskip=0pt plus 0pt
\parindent=0pt
\frenchspacing

\textbf{Modern Car Finance Company Website}

I'm in need of a website for my Car Finance Company. This website should be designed with a modern and sleek aesthetic. Important functionalities I want to see implemented include a loan calculator for user ease and an online application form. This form should be carefully designed to collect personal, employment, and financial information from potential clients. I believe freelancers with creativity and experience in finance industry web design will be the perfect match for this project.

\endgroup
}
\end{tcolorbox}

\begin{tcolorbox}[colback=gray!5, colframe=Aquamarine4, title = {Post-LLM Application Example 1: AI-Written, Low Effort, High Signal}, coltitle=black]
{%
\small
\begingroup
\setstretch{1}
\parskip=0pt plus 0pt
\parindent=0pt
\frenchspacing

\_\_\_\_ here, ready to create your dream Car Finance Company website! With over 15 years of experience, my web development skills cut across various platforms, from PHP, Laravel, Livewire to Bootstrap. This means I'm well-equipped to incorporate important functionalities that you require like a sleek loan calculator or a user-friendly online application form that will effectively gather all the personal and financial information of your potential clients.   My background in managing and maintaining websites has given me a unique perspective on design and user experience. Your website will not just be a beautiful design but a functional money-spinning asset. I'm not only fluent in technical codes and frameworks but also knowledgeable in optimizing search engines which increases your chances to get found by more customers.  Lastly, my awareness of emerging trends is an added bonus for your project. My proficiency in Adobe Photoshop, Figma, Adobe XD ensures that I translate a modern and sleek aesthetic to your website. Let me convert this vision you have about a modern car finance company into a reality. You'll certainly appreciate the \_\_\_\_ touch on your project! Looking forward to collaborating with you!

\textbf{Effort} $\boldsymbol{= 0.52}$ \textbf{minutes}

\textbf{Signal} $\boldsymbol{= 14.79}$

\endgroup
}
\end{tcolorbox}

\begin{tcolorbox}[colback=gray!5, colframe=Aquamarine4, title = {Post-LLM Application Example 2: Not AI-Written, High Effort, High Signal}, coltitle=black]
{%
\small
\begingroup
\setstretch{1}
\parskip=0pt plus 0pt
\parindent=0pt
\frenchspacing

Dear \_\_\_\_, This is \_\_\_\_, Full stack developer for Car Finance Company. I can make modern and sleek websites include loan calculator for user ease and  Thanks for your post job . I'm confident that I can make best modern car finance company Thanks Best regards! \_\_\_\_

\textbf{Effort} $\boldsymbol{= 2.28}$ \textbf{minutes}

\textbf{Signal} $\boldsymbol{= 10.93}$

\endgroup
}
\end{tcolorbox}

\FloatBarrier
\subsection{Consideration Set Measurement Details}\label{appendix_measurement_consideration}

In this subsection, we describe our procedure for constructing consideration sets. Since we do not have direct data on which applications were considered by employers, we develop an algorithm that constructs a proxy for consideration based on click and timestamp data.

\medskip\noindent
\textbf{Step 1: Assign each application a consideration score}
An application is given one point for each of the following conditions:
\begin{enumerate}
    \item The applicant was hired, exchanged at least two messages with the employer, or the employer clicked on the application. 
    \item The application was among the first 8 and was submitted within 2 hours of the job being posted.
    \item The application appeared on the first page\footnote{Each page contains 8 applications.} by the end of bidding and was submitted within 2 hours of the job being posted.
    \item The application was submitted within 5 minutes of the job being posted, provided less than 30 applications were submitted within 5 minutes.
    \item The application was ranked higher on the bid list than the latest-submitted application with which the employer interacted at the time of that interaction.\footnote{Interactions correspond to the actions that award a point in condition (1).}
\end{enumerate}
At this stage, we mark applications with at least one point as considered, and we record the point total of each application as the ``consideration score.''

\medskip\noindent
\textbf{Step 2: Trim excessively large consideration sets.}
We count the number of considered applications per job according to the definition from Step 1. We record the 75th percentile of considered‑set sizes across all jobs for each of the pre‑ and post‑LLM samples separately. If a job's consideration set size exceeds the 75th percentile for its sample, we set consideration status to zero for all applications with a consideration score of one, provided their only point came from conditions (2)–(5) above.

\medskip\noindent
\textbf{Step 3a: Add to excessively small consideration sets: first pass.}
We recount the number of considered applications per job after Step 2. For jobs with fewer than 5 considered applications and more than 15 total applications, we set consideration status to one for all applications on the first two pages of the bid list at the end of bidding.

\medskip\noindent
\textbf{Step 3b: Add to excessively small consideration sets: second pass.}
We recount the number of considered applications per job after Step 3a. For jobs with fewer than 5 considered applications and fewer than 9 total applications, we set consideration status to one for all applications that were submitted within 12 hours of the job being posted.

\medskip\noindent
\textbf{Step 3c: Add to excessively small consideration sets: third pass.}
We recount the number of considered applications per job after Step 3b. For jobs with fewer than 3 considered applications, we set consideration status to one for all applications that were on the first page of the bid list by the end of bidding and were submitted within 12 hours of the job being posted.

\medskip\noindent
\textbf{Step 3d: Add to excessively small consideration sets: fourth pass.}
We recount the number of considered applications per job after Step 3c. For jobs with fewer than 3 considered applications and fewer than 8 total applications, we set consideration status to one for all applications.

\medskip\noindent
\textbf{Step 3e: Add to excessively small consideration sets: final pass.}
We recount the number of considered applications per job after Step 3d. For jobs with fewer than 3 considered applications, we set consideration status to one for all applications that were among the first 5 to be submitted.

\medskip\noindent
\textbf{Step 4a: Do final trim of excessively large consideration sets: first pass.}
We recount the number of considered applications per job after Step 3e. For jobs with more than 32 considered applications, we set consideration status to zero for all applications with a consideration score of one, provided their only point came from conditions (2)–(5) above.

\medskip\noindent
\textbf{Step 4b: Do final trim of excessively large consideration sets: second pass.}
We recount the number of considered applications per job after Step 4a. For jobs with more than 32 considered applications, we set consideration status to zero for all applications with a consideration score of two, provided neither of their points came from condition (1) above.

\medskip\noindent
\textbf{Step 4c: Do final trim of excessively large consideration sets: third pass.}
We recount the number of considered applications per job after Step 4b. For jobs with more than 32 considered applications, we set consideration status to zero for all applications with a consideration score of three, provided none of their points came from condition (1) above.

\medskip\noindent
\textbf{Step 4d: Do final trim of excessively large consideration sets: fourth pass.}
We recount the number of considered applications per job after Step 4c. For jobs with more than 32 considered applications, we set consideration status to zero for all applications with a consideration score of four, provided none of their points came from condition (1) above.

\medskip\noindent
\textbf{Step 4e: Do final trim of excessively large consideration sets: final pass.}
We recount the number of considered applications per job after Step 4d. For jobs with more than 32 considered applications, we set consideration status to zero for all applications with a consideration score of four, unless the applicant exchanged at least five messages with the employer.

\medskip\noindent
\textbf{Step 5: Drop Job Posts with Implausibly Large Consideration Sets.}
We recount the number of considered applications per job after Step 4e. We drop any jobs with more than 32 considered applications that did result in the employer hiring an applicant.\footnote{This last step only drops 16 job posts from the pre-LLM sample and 69 job posts from the post-LLM sample.}

\FloatBarrier
\renewcommand{\thefigure}{D\arabic{figure}} 
\setcounter{figure}{0}
\renewcommand{\thetable}{D\arabic{table}} 
\setcounter{table}{0}
\renewcommand{\thesubsection}{D.\arabic{subsection}}
\setcounter{subsection}{0}
\section{Estimation Details}\label{appendix_estimation}


This appendix provides various implementation details of our empirical analysis and then outlines our estimation procedure in detail. We begin by discussing how we stratify our data by observable characteristics and how we implement our effort correction. Then, we provide details on each step of our estimation procedure.

\subsection{Stratifying Data by Observables}\label{appendix_stratification}

To maintain nonparametric flexibility, we stratify our data by observable characteristics. We construct 56 groups using workers' country, arrival times, and reputations: country is defined as the worker's country of residence; arrival time is defined as time elapsed between the job being posted and the worker first clicking on the job post;  reputation is defined as the one-dimensional ``reputation score'' that the platform assigns to each worker in order to rank them in the bid list.

\begin{figure}[h!]
    \centering
    \caption{Distributions of Key Variables by Arrival Group}
    \label{fig:arrival_groups}

    \begin{minipage}[t]{0.5\linewidth}\centering
        \includegraphics[width=\linewidth]{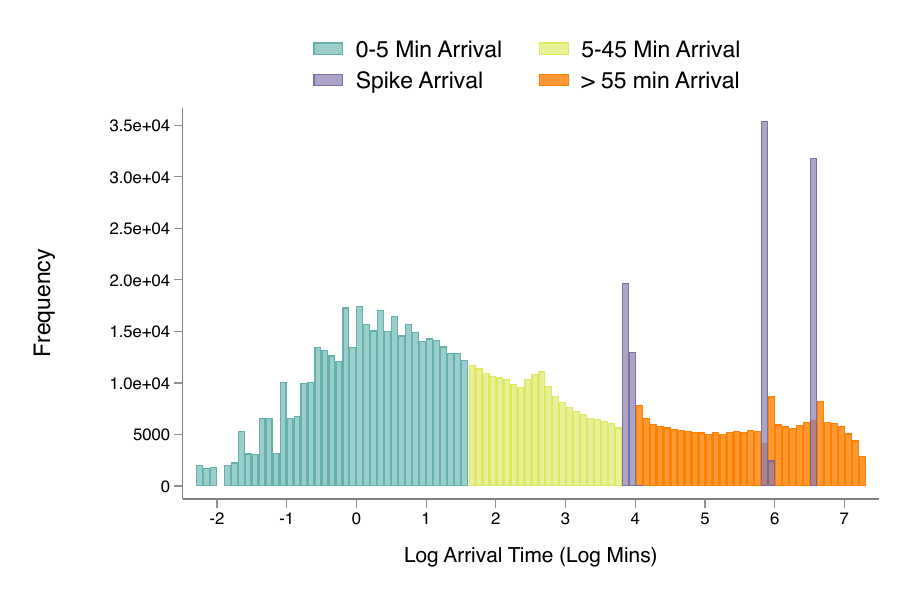}
    \end{minipage}\hfill
    \begin{minipage}[t]{0.5\linewidth}\centering
        \includegraphics[width=\linewidth]{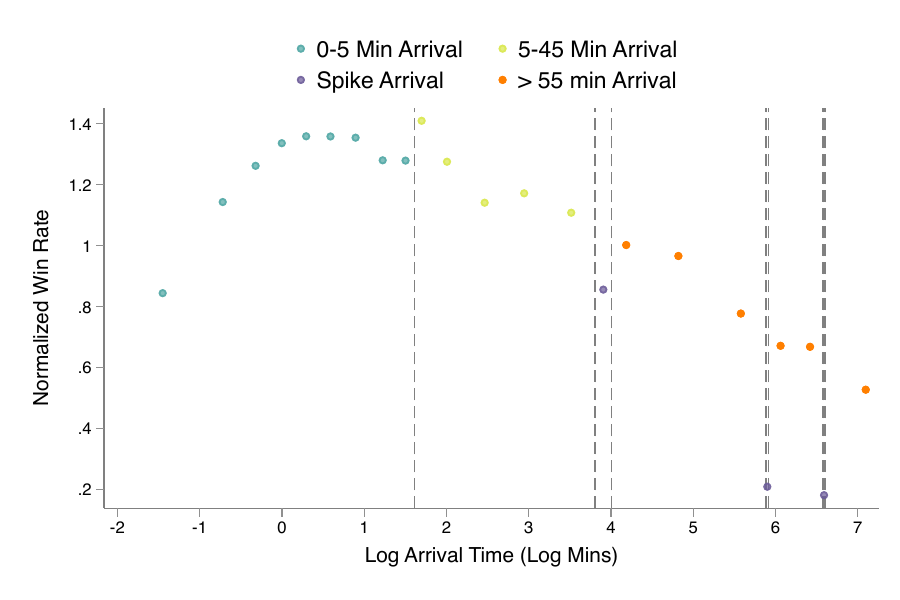}
    \end{minipage}

    \vspace{3mm}

    \begin{minipage}[t]{0.5\linewidth}\centering
        \includegraphics[width=\linewidth]{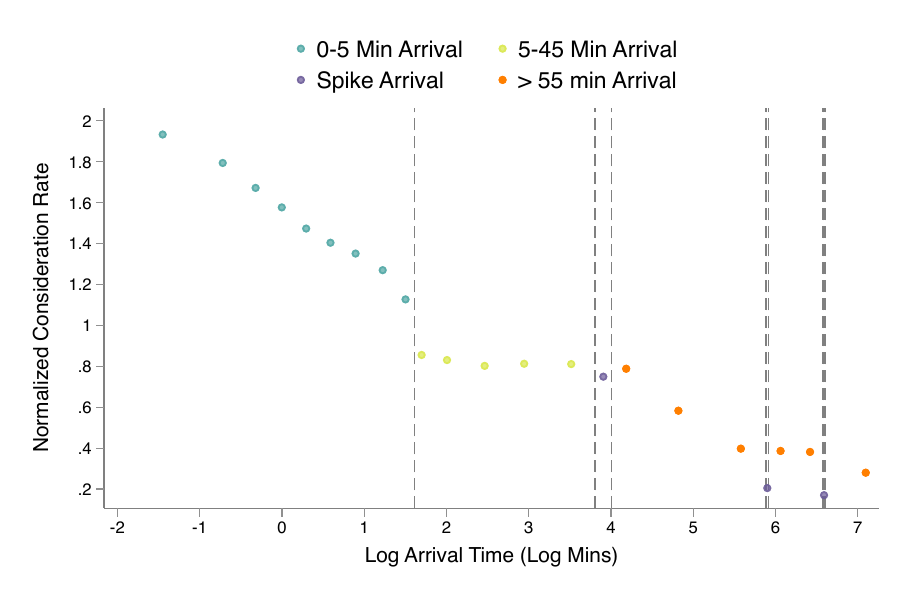}
    \end{minipage}\hfill
    \begin{minipage}[t]{0.5\linewidth}\centering
        \includegraphics[width=\linewidth]{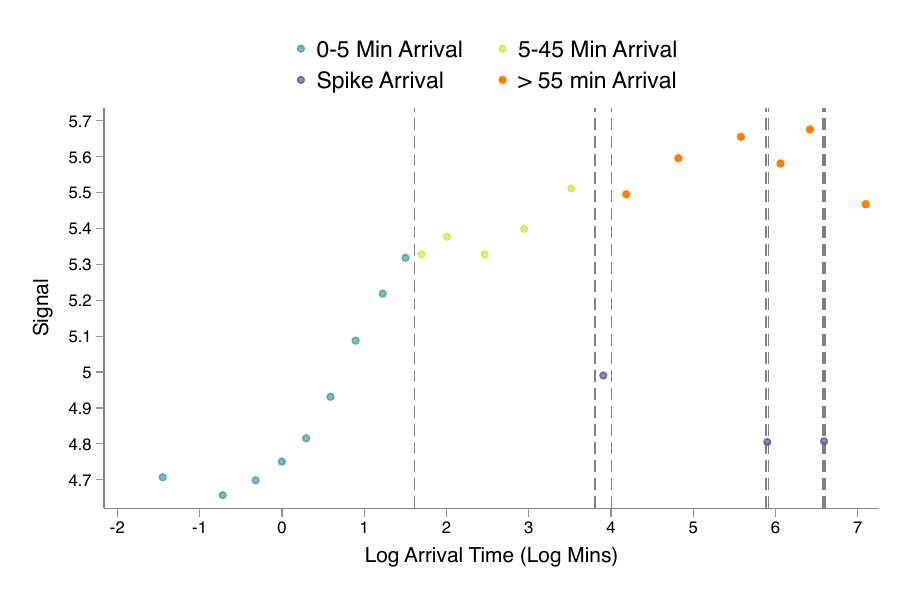}
    \end{minipage}

    \vspace{3mm}

    \begin{minipage}[t]{0.5\linewidth}\centering
        \includegraphics[width=\linewidth]{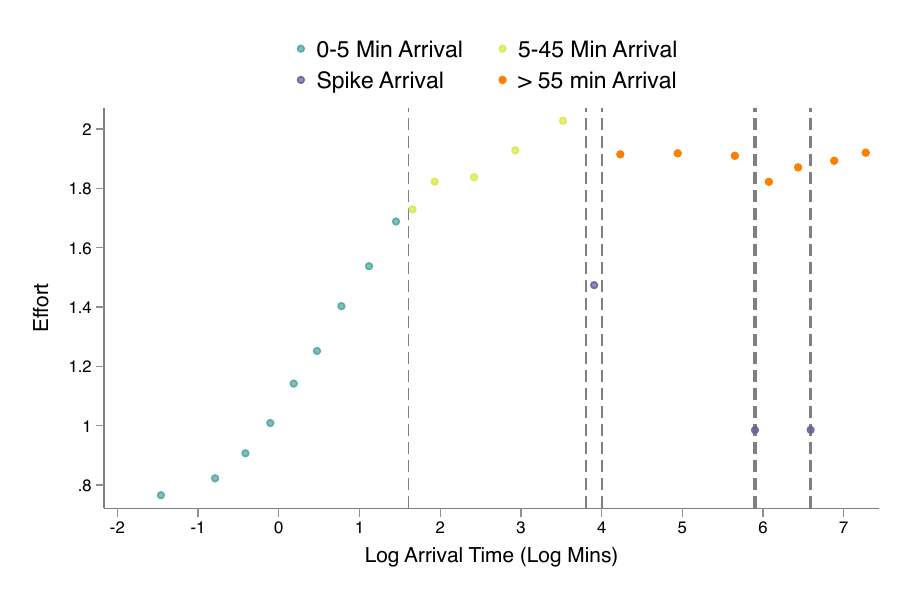}
    \end{minipage}\hfill
    \begin{minipage}[t]{0.5\linewidth}\centering
        \includegraphics[width=\linewidth]{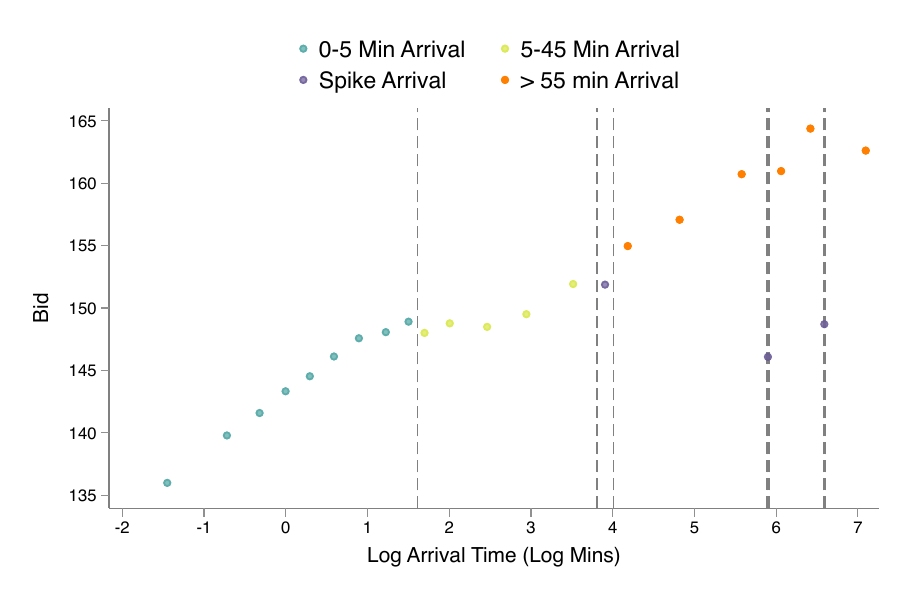}
    \end{minipage}
    \floatfoot{\footnotesize \textit{Notes:} This figure plots both a histogram estimate of the distribution of arrival times (top left) and binscatters of winning the job (top right), being considered (middle left), signal (middle right), effort (bottom left), and bid (bottom right) on arrival time, all of which separate the binscatters by our definitions of arrival group. Arrival groups are defined as follows: 0-5 Minutes, 5-45 Minutes, Spike Arrivals (45-55 minutes, 360-370 minutes, or 720-735 minutes), and Over 45 Minutes (no spikes).}
\end{figure}

To avoid estimating our model using bins that are too sparse, we partition countries into four coarse country groups: \textit{English Speaking}, \textit{Europe}, \textit{South Asia}, and \textit{Other}. Countries in the English Speaking group include Australia, Canada, Ireland, New Zealand, the United Kingdom, and the United States. Countries in the South Asia group include Bangladesh, Bhutan, India, Maldives, Nepal, Pakistan, and Sri Lanka. Countries in the Europe group includes all countries in the European continent, including Russia, Turkey, and Ukraine. All other countries are grouped into the Other category.

We partition arrival times into four groups: \textit{0-5 Minutes}, \textit{5-45 Minutes}, \textit{spike arrivals}, and \textit{Over 45 Minutes (no spikes)}. We observe that there are three major worker arrival spikes---45 minutes, 6 hours, and 12 hours---after the initial spike of attention when a job is posted. In particular, we find that workers that arrive within 10-15 minutes of these spikes behave differently when applying for jobs and are treated differently by employers. Thus, we define spike arrivals to be workers that arrive between 45-55 minutes, 360-370, or 720-735 after a job is posted. We can see how spike arrivals observably differ from other arrival groups in Figure~\ref{fig:arrival_groups}. Furthermore, we impute arrival time as the time elapsed between the job being posted and the worker's application being submitted for each worker whose first click is measured to be after their application was submitted.

Finally, we partition reputation scores and rookie-status into four groups: \textit{Rookies}, \textit{Low}, \textit{Middle}, and \textit{High}, where the latter three only contain workers who have received work on the platform. Note that these are dynamic variables that we measure at the time a worker applies to a job.

Due to some sparse bins, we combine all non-South Asian countries who are also spike arrivals into a single group. 

Summary statistics of each group is presented in Table~\ref{tab:x_groups_sum_stats}. 

\input{tables/x_groups_sum_stats.tex}
\FloatBarrier
\subsection{Effort Correction}\label{appendix_effort_correction}

Before using our raw effort measurements, we implement a correction that accounts for heterogeneity in workers' time efficiency. Some workers convert a given amount of time into higher-quality signals than others; thus two workers with the same measured time \(t_{ij}\) may not expend the same true effort. Our correction uses an empirical Bayes estimator to recover worker-specific efficiency multipliers and then re-scales measured time accordingly.

\paragraph{Measurement and microfoundation.}
Let \(t_{ij}>0\) denote the measured time worker $j$ spends reading and writing when applying to job post $i$. True effort is
\[
e_{ij} \;\equiv\; \eta_j \cdot t_{ij}, \qquad \phi_j \;\equiv\; \log (\eta_j),
\]
where \(\eta_j\) is a worker-\(j\) efficiency multiplier. Signals are produced according to
\begin{equation}
s_{ij} \;=\; K^s(x_{ij}) \;+\; \gamma^s(x_{ij}) \cdot \log (e_{ij}) \;+\; \varepsilon^s_{ij},
\label{eq:signal_production_effcorr}
\end{equation}
with \(\varepsilon^s_{ij}\) mean-zero noise. Substituting \(e_{ij}=\eta_j t_{ij}\) yields
\begin{equation}
s_{ij} \;=\; K^s(x_{ij}) \;+\; \gamma^s(x_{ij}) \cdot \log (t_{ij}) \;+\; \underbrace{\gamma^s(x_{ij}) \cdot \phi_j}_{\text{worker effect}} \;+\; \varepsilon^s_{ij}.
\label{eq:signal_production_effcorr_reduced}
\end{equation}

\paragraph{Assumptions.}
(i) \(\phi_j=\log\eta_j \stackrel{\text{i.i.d.}}{\sim} \mathcal{N}(0,V_\eta)\) and is independent of \((a_{ij},c_{ij},x_{ij})\) and of \(\varepsilon^s_{ij}\).  
(ii) Within each \(x\)-group, \(\varepsilon^s_{ij}\) is homoskedastic with variance \(V_{\varepsilon^s}(x)\) and mean zero conditional on \(t_{ij}\) and \(j\).

\medskip
\noindent The estimator proceeds in five steps.

\subsubsection*{Step 1: Estimate \(\widehat K^s(x)\), \(\widehat\gamma^s(x)\), and \(\widehat{V}_{\varepsilon^s}(x)\)}
Separately for each observable group \(x\), run a regression of \(s_{ij}\) on \(\log (t_{ij})\) with worker fixed effects:
\[
s_{ij} \;=\; K^s(x) \;+\; \gamma^s(x)\cdot\log (t_{ij}) \;+\; RE_j \;+\; u_{ij}.
\]
With this regression, we obtain estimates of \(\widehat{K}^s(x)\), \(\widehat{\gamma}^s(x)\), and \(\widehat{V}_{\varepsilon^s}(x)\). Define the residual
\[
r_{ij} \;\equiv\; s_{ij}-\widehat{K}^s(x_{ij})-\widehat{\gamma}^s(x_{ij}) \cdot \log (t_{ij})
\;\approx\; \gamma^s(x_{ij})\cdot \phi_j + \varepsilon^s_{ij}.
\]

\subsubsection*{Step 2: Normalization and sufficient statistics by worker}
Normalize by the estimated slope and attach precision weights:
\begin{equation}
y_{ij} \;\equiv\; \frac{r_{ij}}{\widehat\gamma^s(x_{ij})}
\;\approx\; \phi_j + \frac{\varepsilon^s_{ij}}{\gamma^s(x_{ij})},
\qquad
w_{ij} \;\equiv\; \frac{\big(\widehat\gamma^s(x_{ij})\big)^2}{\widehat{V}_{\varepsilon^s}(x_{ij})}.
\label{eq:normalized_y_w}
\end{equation}
For worker \(j\), define total precision \(S_j=\sum_i w_{ij}\) and the precision-weighted mean
\[
\tilde\phi_j \;\equiv\; \frac{\sum_i w_{ij}\cdot y_{ij}}{S_j}.
\]
Given \eqref{eq:normalized_y_w}, \(\tilde\phi_j \mid \phi_j \sim \mathcal{N}(\phi_j,\,1/S_j)\).

\subsubsection*{Step 3: Method-of-moments estimate of \(V_\eta\)}
Let \(\mathcal{J}_{2+}\) be workers with at least two observations. The precision-weighted variance across \(\{\tilde\phi_j\}_{j\in\mathcal{J}_{2+}}\) satisfies
\[
\mathbb{E}\big[\operatorname{Var}_w(\tilde\phi)\big] \;=\; V_\eta \;+\; \mathbb{E}_w\!\left[\frac{1}{S_j}\right],
\]
so we estimate
\begin{equation}
\widehat V_\eta \;\equiv\; \max\!\left\{\operatorname{Var}_w(\tilde\phi) - \frac{|\mathcal{J}_{2+}|}{\sum_{j\in\mathcal{J}_{2+}} S_j},\; 0\right\}.
\label{eq:Veta_hat}
\end{equation}

\subsubsection*{Step 4: Empirical Bayes posterior mean of \(\phi_j\)}
Under the normal prior \(\phi_j\sim\mathcal{N}(0,V_\eta)\) and sampling variance \(1/S_j\),
\begin{equation}
\widehat\phi^{\,EB}_j \;=\; \mathbb{E}[\phi_j \mid \tilde\phi_j,S_j;\widehat V_\eta]
\;=\; \left(\frac{\widehat V_\eta\cdot S_j}{1+\widehat V_\eta\cdot S_j}\right)\tilde\phi_j.
\label{eq:EB_phi}
\end{equation}
Low-precision workers (small \(S_j\)) are shrunk toward 0; high-precision workers are barely shrunk.

\subsubsection*{Step 5: Capping and corrected effort}
We translate posterior means into a worker-level shift in log effort and cap extremes:
\[
\Delta\log (e_j) \;\equiv\; \mathrm{clip}\!\big(\widehat\phi^{\,EB}_j,\,-\text{cap},+\text{cap}\big), \qquad \text{cap}=1.25.
\]
The corrected effort used in estimation is
\begin{equation}
\log (e^{\mathrm{corr}}_{ij}) \;=\; \log (t_{ij}) + \Delta\log (e_j),
\qquad
e^{\mathrm{corr}}_{ij} \;=\; t_{ij}\cdot\exp\!\big(\Delta\log (e_j)\big).
\label{eq:ecorr_main}
\end{equation}
Equivalently, defining a worker effect in signal units \(\widehat{RE}_{ij}\equiv \widehat\gamma^s(x_{ij})\cdot\widehat\phi^{\,EB}_j\),
\begin{equation}
e^{\mathrm{corr}}_{ij} \;=\; t_{ij}\,\exp\!\left(\mathrm{clip}\!\left(\frac{\widehat{RE}_{ij}}{\widehat\gamma^s(x_{ij})},-\text{cap},+\text{cap}\right)\right).
\label{eq:ecorr_RE}
\end{equation}

\paragraph{Discussion and use.}
By construction, \(x_{ij}\) is purely a vector of observables; we estimate \(\widehat K^s(x)\), \(\widehat\gamma^s(x)\), and \(\widehat{V}_{\varepsilon^s}(x)\) separately by observable group, but the latent efficiency \(\phi_j\) is worker-level and independent of \(x\). Because \(\phi_j\) is assumed independent of \((a_{ij},c_{ij},x_{ij})\), this correction does not affect our identification arguments. In all downstream steps, we replace the raw effort with \(e^{\mathrm{corr}}_{ij}\) from \eqref{eq:ecorr_main}.

\subsection{Estimation Procedure}\label{appendix_estimation_procedure}

In this subsection, we outline the details of our estimation procedure.

\subsubsection*{Step 1.1: Estimating Signal Production Functions}

For each observable group defined in Appendix~\ref{appendix_stratification}, we separately estimate the signal production function using OLS and a linear-log specification of the form:
\begin{align}
    s_{i j} = K^s(x_{i j}) + \gamma^s(x_{i j}) \cdot \log(e_{i j}) + \varepsilon^s_{i j},
\end{align}
where $e_{ij}$ is the effort measure corrected as in Appendix~\ref{appendix_effort_correction}. (For all future mentions of effort, we refer to this corrected measure.)

\subsubsection*{Step 1.2: Estimating Reduced-Form Demand}

In this step, we estimate the reduced-form demand for workers at the pre-LLM signaling equilbrium. The demand we are estimating in this step can be interpreted as the predicted equilibrium demand from the workers' perspective. 

First, we define the following notation:
\begin{align*}
    W_{i j} \equiv& \: \mathbbm{1}\{j \text{ wins } i\}; \\
    W_{i 0} \equiv & \: \mathbbm{1}\{i \text{ chooses outside option}\}; \\
    q_{i j} \equiv& \: \mathbbm{1}\{j \text{ is considered by } i\} \text{ (Note, we set } q_{i 0} = 1\text{.)}; \\
    h(s; x) \equiv& \: T(x) + \beta \cdot \mathbbm{E}[a | s; x];\footnote{} \\
    \theta \equiv& \: \left(\alpha, h, \pi\right); \\
    b_i \equiv& \: \left(b_{i j}\right)_{j = 1}^{N_i}; \\
    s_i \equiv& \: \left(s_{i j}\right)_{j = 1}^{N_i}; \\
    x_i \equiv& \: \left(x_{i j}\right)_{j = 1}^{N_i}; \\
    Q_i \equiv& \: \left\{j \in \{0, 1, ..., N_i\} \mid q_{i j} = 1\right\}; \\
    D_i \equiv& \: \left(N_i, b_i, s_i, x_i, Q_i\right) \text{ to be the non-outcome data for job } i; \\
    \delta_{i j}(D_i; \theta) \equiv&  \: h(s_{i j}; x_{i j}) + \alpha \cdot b_{i j}; \\
    \delta_{i 0} \equiv& \: 0.
\end{align*}
Note that we write that the bid component of utility enters positively instead of negatively. This choice is obviously without loss and is for notational convenience when coding our estimator. Thus, we expect to estimate negative values of $\alpha$.

We can now write down the probability that worker $j$ wins job $i$, conditional on the non-outcome data $D_i$ and parameters $\theta$ as:
\begin{align}
    \mathbbm{P}\big(W_{i j} = 1 \mid D_i; \theta\big) = \pi \cdot q_{i j} \cdot \frac{\exp\left(\delta_{ij}(D_i; \theta)\right)}{1 + \sum_{k \in Q_i \setminus \{0\}} \exp\left(\delta_{ik}(D_i; \theta)\right)}.
\end{align}

And we write down the probability that job poster $i$ chooses the outside option, conditional on the non-outcome data $D_i$ and the parameters $\theta$ as:
\begin{align}
    \mathbbm{P}\big(W_{i 0} = 1 \mid D_i; \theta\big) = \frac{\pi}{1 + \sum_{k \in Q_i \setminus \{0\}} \exp\left(\delta_{ik}(D_i; \theta)\right)} + (1-\pi).
\end{align}

Next define the linear function $\lambda(s; x, K^\lambda, \gamma^\lambda)$ used to approximate $h(s; x)$ in the reduced-form demand estimation step:
\begin{align}
    \lambda(s; x, K^\lambda, \gamma^\lambda) \equiv K^\lambda(x) + \gamma^\lambda(x)\cdot s,
\end{align}
where $K^\lambda \equiv (K^\lambda(x))_{x \in X}$, and $\gamma^\lambda \equiv (\gamma^\lambda(x))_{x \in X}$. 
Note that we estimate separate $\lambda$ functions, i.e., slopes and intercepts, for each value of $x$, and thus we index those parameters accordingly.

Using this $\lambda$ approximation, now define:
\begin{align*}
    \tilde{\theta} \equiv& \: (\alpha, K^\lambda, \gamma^\lambda, \pi); \\
    \tilde{\delta}_{i j}(D_i; \tilde{\theta}) \equiv& \: \lambda(s_{i j}; x_{i j}, K^\lambda, \gamma^\lambda) + \alpha \cdot b_{i j}.
\end{align*}

We can thus define the pseudo-probability that job poster $i$ chooses worker $j$ conditional on the non-outcome data $D_i$ and the parameters $\tilde{\theta}$ as:
\begin{align}
    \mathbbm{\tilde{P}}\big(W_{i j} = 1 \mid D_i; \tilde{\theta}\big) \equiv \pi \cdot q_{i j} \cdot \frac{\exp\left(\tilde{\delta}_{i j}(D_i; \tilde{\theta})\right)}{1 + \sum_{k \in Q_i \setminus \{0\}} \exp\left(\tilde{\delta}_{i k}(D_i; \tilde{\theta})\right)}.
\end{align}
Similarly, we can define the pseudo-probability that job poster $i$ chooses the outside option as:
\begin{align}
    \mathbbm{\tilde{P}}\big(W_{i 0} = 1 \mid D_i; \tilde{\theta}\big) \equiv \frac{\pi}{1 + \sum_{k \in Q_i \setminus \{0\}} \exp\left(\tilde{\delta}_{ik}(D_i; \tilde{\theta})\right)} + (1-\pi).
\end{align}
Define $D \equiv \left(D_i\right)_{i = 1}^I$, \: $W_i \equiv \left(W_{i j}\right)_{j = 0}^{N_i}$, and $W \equiv \left(W_i\right)_{i = 1}^I$, where $I$ is the number of jobs in the data.

In preparation for writing down the pseudo-likelihood function, we define 
$\tilde{P}_{i j}(\tilde{\theta} \mid D_i) \equiv \mathbbm{\tilde{P}}\big(W_{i j} = 1 \mid D_i; \tilde{\theta}\big)$ and $\tilde{P}_{i 0}(\tilde{\theta} \mid D_i) \equiv \mathbbm{\tilde{P}}\big(W_{i 0} = 1 \mid D_i; \tilde{\theta}\big)$.
We can then write down the pseudo-likelihood function as:
\begin{align}
\mathcal{\tilde{L}}\left(\tilde{\theta} \mid D, W\right) \equiv \prod_{i = 1}^I \prod_{j \in Q_i} \tilde{P}_{i j}(\tilde{\theta} \mid D_i)^{W_{i j}}.
\end{align}
Note that $W_{ij} = 1 \implies q_{ij} = 1$, and thus we can drop the $q_{i j}$ term from the pseudo-likelihood function, as long as we sum over considered workers for each job, allowing us to avoid a $0^0$ term in the pseudo-likelihood function.

Thus, the pseudo-log-likelihood function is:
\begin{align}
    \tilde{L}\left(\tilde{\theta} \mid D, W\right) = \sum_{i = 1}^I \sum_{j \in Q_i} W_{i j} \cdot \log\left(\tilde{P}_{i j}(\tilde{\theta} \mid D_i)\right),
\end{align}
where, for $j\neq 0$:
{\small
\begin{align}
    \log\left(\tilde{P}_{i j}(\tilde{\theta} \mid D_i)\right) =
    \tilde{\delta}_{i j}(D_i; \tilde{\theta})
     - \log\left(1 + \sum_{k \in Q_i \setminus \{0\}} \exp\left(\tilde{\delta}_{ik}(D_i; \tilde{\theta})\right)\right) + \log(\pi),
\end{align}}
and for $j = 0$:
\begin{align}
    \log\left(\tilde{P}_{i 0}(\tilde{\theta} \mid D_i)\right) =
    \log\left(1 + \left(\left(1 + \sum_{k \in Q_i \setminus \{0\}} \exp\left(\tilde{\delta}_{ik}(D_i; \tilde{\theta})\right)\right)^{-1} -  1\right)\cdot \pi\right).   
\end{align}

This step of the estimation procedure consists of maximizing the above pseudo-log-likelihood function with respect to the parameters $\tilde{\theta}$. 

\subsubsection*{Step 1.3: Simulating Job Posts}

To simulate the reduced-form win-probability function, $P^*(b, s; x)$, which is needed to invert workers' FOCs, we need to simulate many job posts and the bids and signals of the workers applying to those job posts. First, we simulate the observable types of workers applying to each job post and employer consideration sets by simply bootstrapping with replacement from the empirical distribution of job posts and consideration sets. 

Next, we need to estimate the joint distribution of bids and signals within each $x$-group, which we will use to simulate bids and signals in the next step. We do so by estimating the marginal distributions of bids and signals nonparametrically, and then we estimate a parametric copula to capture their joint distribution. This approach allows us to capture the non-standard shapes of the empirical marginal distributions and capture the main features of the dependence structure between bids and signals, while not falling prey to the curse of dimensionality of estimating a joint distribution nonparametrically.

Probably since workers do not only compete on bids, workers tend to choose discrete bid levels that are multiples of \$5, \$10, or \$50, with some workers choosing bids in a more continuous manner. To accurately capture this feature of the marginal bid distributions, we group bids into discrete bins, estimate a nonparametric discrete marginal distribution for each $x$-group, and then we build a second layer of the empirical model that explains the deviations from those discrete bid levels. First, we define these bid bins within $x$-group to be bid levels that either are an end point, i.e., \$30 or \$250, or have a mass of workers equal to at least the maximum of 20 and 0.5\% of the $x$-group's total workers. Second, once we have defined the modal set of discrete bid levels, we categorize every bid not exactly equal to one of those discrete bid levels into its nearest bid level, and then we compute the fraction of each bid bin that deviates from the exact value of the bin. One assumption that we make in this estimation step is that the deviation from the exact bid bin value is independent of the signal conditional on the bid bin. This assumption allows us to model the joint distribution of bids and signals with a copula between the marginal bid bin distribution and the marginal signal distribution. 

In particular, we fit a Student-$t$ copula via MLE separately for each $x$-group. We use the Student-$t$ copula rather than a Gaussian copula, for example, because we know that in the empirical joint distributions of bids and signals, correlated outliers in both variables are somewhat common.

Once we have estimated this empirical model of bids and signals, we draw from it via the following process: (1) we draw the pseudo-uniform scores of bids and signals from the estimated Student-$t$ copula; (2) we use the estimated marginal empirical CDFs of bids and signals to map those ranks to the actual values of discrete bid bins and signals; (3) we use the estimated frequency of deviations for each bid bin and $x$-group to simulate whether the bid deviates from the exact value of the bin; (4) If the bid deviates, we draw the deviation from $U(a,b)$, where  $a =$ lower-midpoint between this bid bin and the next lowest bid bin, and $b =$ upper-midpoint between this bid bin and the next highest bid bin (the \$30 bid bin uses $a = 30$; the \$250 bid bin uses $b = 250$).

\subsubsection*{Step 1.4: Computing Simulated Job Poster Mean Utilities}

Before continuing, we define some notation for our simulations. We index each simulated job post by $m$, with there being $M$ total simulated job posts. For each simulated job post $m$, we index workers by $j$, with there being $N_m$ total simulated workers applying to job post $m$. We also denote the $x$-group-specific count of workers applying to job post $m$ by $N_{m x}$, where $\sum_{x \in X} N_{m x} = N_m$, and we denote the set of considered workers for simulated job post $m$ by $Q_m$. We denote simulated bids by $b_{m j}$, simulated signals by $s_{m j}$, simulated observable types by $x_{m j}$, and simulated consideration status indicator $q_{m j}$. 

Once we have simulated bids $b_{m j}$ and signals $s_{m j}$ for each worker $j$ in each simulated job post $m$, we can compute the mean reduced-form expected utility of each simulated application. In particular, we plug the simulated bids and signals into the estimated reduced-form employer indirect expected utility function from Step 1.2:
\begin{align}
    \widehat{\delta}_{m j} \equiv \widehat{K}^\lambda(x_{m j}) + \widehat{\gamma}^\lambda(x_{m j}) \cdot s_{m j} + \widehat{\alpha} \cdot b_{m j}
\end{align}

\subsubsection*{Step 1.5: Computing Simulated Win-Probability Functions and Their Derivatives}
Having a simulated set of job posts with consideration sets and sets of indirect expected utilities $\widetilde{EU}_{m j}$, we can now compute the simulated ex-ante win-probability function for each worker observable type $x$.

First, we define some more notation:
\begin{align}
    \widehat{\lambda}_{m j} &\equiv \widehat{K}^\lambda(x_{m j}) + \widehat{\gamma}^\lambda(x_{m j}) \cdot s_{m j} \\
    \widehat{\omega}_{m j} &\equiv \exp\left(\widehat{\delta}_{m j}\right) \\
    \widehat{\Delta}_{m, -j} &\equiv \sum_{k \in Q_m \setminus \{0, j\}} \widehat{\omega}_{m k}
\end{align}

To compute the simulated ex-ante win-probability function for each worker of type $x$, we need to integrate over two distributions. First, we need to numerically integrate over the hypothetical worker's signaling noise. We draw the signaling noise $\varepsilon^s_{m j}$ for each worker $j$ applying to job post $m$ from the estimated $x$-group-specific distribution of signaling noise ($N(0, V_{\varepsilon^s}(x))$), which we estimated in Step 1.1 above. Note that this signaling noise does not affect any of the above defined objects, but instead only affects the signal that would be produced by the hypothetical worker from whose perspective we are computing the simulated win-probability function.

Second, we need to numerically integrate over all simulated workers of type $x$, even when multiple $x$-type workers apply to the same job post. We operationalize this sum by letting the hypothetical worker replace, one at a time, each simulated worker of type $x$, meaning they use that worker's consideration status, and they use that worker's simulated signaling noise to compute their signal, and we replace that worker's simulated expected mean utility with a newly computed one based on their choice of bid $b$ and effort $e$.

We now formalize this approach. First, we define the estimated signaling function:
\begin{align}
\widehat{s}(e; x, \varepsilon^s) = \widehat{K}^s(x) + \widehat{\gamma}^s(x) \cdot \log(e) + \varepsilon^s,
\end{align}
Next, we define three functions that take as input worker actions, worker observable type, and signaling noise, using previously estimated parameters:
\begin{align}
    \widehat{\lambda}(e; x, \varepsilon^s) &\equiv \widehat{K}^\lambda(x) + \widehat{\gamma}^\lambda(x) \cdot \widehat{s}(e; x, \varepsilon^s); \\
    \widehat{\delta}(b, e; x, \varepsilon^s) &\equiv \widehat{\lambda}(e; x, \varepsilon^s) + \widehat{\alpha} \cdot b; \\
    \widehat{\omega}(b, e; x, \varepsilon^s) &\equiv \exp\left(\widehat{\delta}(b, e; x, \varepsilon^s)\right).
\end{align}

Now, we define the ex-interim simulated win-probability function for each worker $j$ applying to job post $m$, conditional on $m$ not abandoning, as:
\begin{align}
    \widehat{P}^*_{m j}(b, e) \equiv q_{m j} \cdot \left(\frac{\widehat{\omega}\left(b, e; x_{m j}, \varepsilon^s_{m j}\right)}{1 + \widehat{\omega}\left(b, e; x_{m j}, \varepsilon^s_{m j}\right) + \widehat{\Delta}_{m, -j}}\right)
\end{align}
Now, we average over all simulated job posts $m$ to compute the ex-ante simulated win-probability function for a worker of type $x$:
\begin{align}
    \widehat{P}^*(b, e; x) = \pi \cdot \left(\sum_{m = 1}^M N_{m x} \right)^{-1} \sum_{m = 1}^M \sum_{j = 1}^{N_m} \widehat{P}^*_{m j}(b, e) \cdot \mathbbm{1}\{x_{m j} = x\}.
\end{align}
Note that the ex-interim simulated win-probability for worker $j$ applying to job post $m$ is only included in the average if that worker is considered by the job poster, i.e., if $q_{m j} = 1$, and if the worker's observable type matches $x_{m j} = x$. Also note that we do not need to simulate each job poster abandoning or not, since that rate is equally $\pi$ for all job posts, so we can simply multiply the entire function by $\pi$.

We can similarly write down the derivatives of the simulated ex-ante win-probability function as:
\begin{align}
    \frac{\partial}{\partial z} \widehat{P}^*(b, e; x) &= \pi \cdot \left(\sum_{m = 1}^M N_{m x} \right)^{-1} \sum_{m = 1}^M \sum_{j = 1}^{N_m} \frac{\partial}{\partial z} \widehat{P}^*_{m j}(b, e) \cdot \mathbbm{1}\{x_{m j} = x\},
\end{align}
where $z \in \{b, e\}$.

\subsubsection*{Step 1.6: Inverting Simulated FOCs} 

To estimate the joint distribution of cost $c$, ability $a$, and observable type $x$, we use the simulated ex-ante win-probability function $\widehat{P}^*(b, e; x)$ and its derivatives to invert simulated worker first order conditions (FOCs). This step mirrors the standard approach to estimating auction models, with the two main deviations being that we have two dimensions of the FOCs, and that we are simulating the win-probability function rather than computing it directly from the empirical distribution of bids.

Recall that a worker with cost $c$, ability $a$, and observable type $x$ choosing bid $b$ and effort $e$ solves the following problem:
\begin{align}
    \max_{b, e} \quad & P^*(b, e; x) \cdot (b - c) - 
    \frac{1}{2}\cdot \exp\left(-a\right)\cdot e^2
\end{align}
At each worker optimal choice of bid $b^*$ and effort $e^*$, the following FOCs hold:
\begin{align}
    \frac{\partial}{\partial b} P^*(b, e; x)\bigg|_{(b, e) = (b^*, e^*)} \cdot (b^* - c) + P^*(b^*, e^*; x) &= 0; \\
    \frac{\partial}{\partial e} P^*(b, e; x)\bigg|_{(b, e) = (b^*, e^*)} \cdot (b^* - c) - 
    \exp\left(-a\right) \cdot e^* &= 0.
\end{align}

We use these two equations to compute a pseudo-data set of
$\left(\left(\widehat{c}_{i j}, \widehat{a}_{i j}\right)_{j = 1}^{N_i}\right)_{i = 1}^I$:\footnote{We abuse notation in the indices, as we only can compute these pseudo-data points for workers for whom we observe effort $e$}
\begin{align}
    \widehat{c}_{i j} = b_{i j} + \widehat{P}^*\left(b_{i j}, e_{i j}; x_{i j}\right) \cdot \left(\frac{\partial}{\partial b} \widehat{P}^*\left(b, e; x_{i j}\right)\bigg|_{(b, e) = (b_{i j}, e_{i j})}\right)^{-1};
\end{align}
{\small
\begin{align}
    \widehat{a}_{i j} = \log \left(-e_{i j} \cdot \left(\frac{\partial}{\partial e_{i j}} \widehat{P}^*\left(b, e; x\right)\bigg|_{(b, e) = (b_{i j}, e_{i j})}\right)^{-1} \cdot \left(\frac{\partial}{\partial b} \widehat{P}^*\left(b, e; x_{i j}\right)\bigg|_{(b, e) = (b_{i j}, e_{i j})}\right) \cdot \left(\widehat{P}^*\left(b_{i j}, e_{i j} ; x_{i j}\right)\right)^{-1}\right).
\end{align}}
Note that $\pi$ falls out of the first FOC, but is be present in the second FOC, since effort costs are paid regardless of whether the employer abandons the job post.

\subsubsection*{Step 2: Estimating Employer Beliefs} 

Now that we have our pseudo-data set of $\left(\widehat{c}_{i j}, \widehat{a}_{i j}, s_{i j}, x_{i j}\right)$, we can estimate the employer belief function beyond our initial linear approximation. Recall that we grouped the taste-based shifters $\left(T(x)\right)_{x \in X}$, as well as the preference-weighted beliefs about ability $\beta \cdot \mathbb{E}[\widehat{a} \mid s ; x]$ into a single function, $h(s ; x)$, which we then approximated with a linear function $\lambda(s; x)$. To disentangle the pieces combined into $h$ in the structural demand estimation step, we need to get a separate estimate of the reduced-form employer belief function, which we denote $g(s; x) \equiv \mathbb{E}[\widehat{a} \mid s ; x]$, the estimate of which we denote $\widehat{g}(s; x)$.

We are aiming to estimate $g$ with more fidelity than our previous linear approximation, so we use a three-step nonparametric approach. All the following steps are done completely separately by $x$-group. First, we bin the data into 50 equally populated bins of the signal $s_{i j}$, and compute the mean estimated ability $\widehat{a}_{i j}$ for each bin. Second, we estimate an isotonic regression of the mean estimated ability $\widehat{a}_{i j}$ on the mean signal within each bin, which allows us to capture the monotonicity of the employer belief function, while also allowing for nonlinearity in the relationship between the signal and the employer belief. Third, we fit a Piecewise cubic Hermite Interpolating Polynomial (PCHIP) to the estimated isotonic regression, which allows us to maintain the data-driven nonlinearity of the previous step, while also ensuring that the estimated employer belief function is continuous and differentiable.

One important technical note is that the since the PCHIP interpolates from the mean of the lowest signal bin to the mean of the highest bin, it is undefined for the signals that fall below the lowest bin's mean or above the highest bin's mean. Since we are using 50 bins, and thus those tail bins are already small, and to reflect the nature of belief formation at those tails, instead of extrapolating the PCHIP to those tails, we simply set $\widehat{g}(s; x) = \widehat{g}(\text{mean of lowest bin}; x)$ for all $s < \text{mean of the lowest bin}$, and $\widehat{g}(s; x) = \widehat{g}(\text{mean of highest bin}; x)$ for all $s > \text{mean of the highest bin}$.

\subsubsection*{Step 3: Estimating Structural Demand} 

Now that we have estimated the employer belief function $\widehat{g}(s; x) \equiv \widehat{\mathbb{E}}[\widehat{a} \mid s ; x]$, we can plug our signal data $s_{i j}$ into it and generate a pseudo-data set of employer beliefs $\overline{a}_{i j} \equiv \widehat{g}(s_{i j}; x_{i j})$ for every observed $(i, j)$ pair. Now, we have everything we need to estimate the structural demand parameters that we excluded from the reduced-form demand estimation in Step 1.2. Specifically, we can estimate the structural demand parameters $\beta$ and $T \equiv \left(T(x)\right)_{x \in X}$, in addition to re-estimating the price coefficient $\alpha$ and the non-abandon rate $\pi$. 

Now define:
\begin{align}
\overline{\theta} &\equiv (\alpha, \beta, T, \pi); \\
\overline{D}_i &\equiv \left(N_i, b_i, x_i, Q_i, \overline{a}_i\right); \\
\overline{\delta}_{i j}(\overline{D}_i; \overline{\theta}) &\equiv T(x_{ij}) + \beta \cdot \overline{a}_{i j} + \alpha \cdot b_{i j}.
\end{align}

We can thus define the updated pseudo-probability that job poster $i$ chooses worker $j$ conditional on the non-outcome data $D_i$ and the parameters $\overline{\theta}$ as:
\begin{align}
    \mathbbm{\overline{P}}\big(W_{i j} = 1 \mid \overline{D}_i; \overline{\theta}\big) = \pi \cdot q_{i j} \cdot \frac{\exp\left(\overline{\delta_{ij}}(\overline{D}_i; \overline{\theta})\right)}{1 + \sum_{k \in Q_i \setminus \{0\}} \exp\left(\overline{\delta_{ik}}(\overline{D}_i; \overline{\theta})\right)}.
\end{align}
Similarly, we can define the updated pseudo-probability that job poster $i$ chooses the outside option as:
\begin{align}
    \mathbbm{\overline{P}}\big(W_{i 0} = 1 \mid \overline{D}_i; \overline{\theta}\big) = \frac{\pi}{1 + \sum_{k \in Q_i \setminus \{0\}} \exp\left(\overline{\delta_{ik}}(\overline{D}_i; \overline{\theta})\right)} + (1-\pi) .
\end{align}
Define $\overline{D} \equiv \left(\overline{D}_i\right)_{i = 1}^I$, and recall that $W_i \equiv \left(W_{i j}\right)_{j = 0}^{N_i}$ and $W \equiv \left(W_i\right)_{i = 1}^I$, where $I$ is the number of jobs in the data.

In preparation for writing down the updated pseudo-likelihood function, we define \\ $\overline{P}_{i j}(\overline{\theta} \mid \overline{D}_i) \equiv \mathbbm{\overline{P}}\big(W_{i j} = 1 \mid \overline{D}_i; \overline{\theta}\big)$ and $\overline{P}_{i 0}(\overline{\theta} \mid \overline{D}_i) \equiv \mathbbm{\overline{P}}\big(W_{i 0} = 1 \mid \overline{D}_i; \overline{\theta}\big)$.
We can then write down the updated pseudo-likelihood function as:
\begin{align}
\mathcal{\overline{L}}\left(\overline{\theta} \mid D, W\right) \equiv \prod_{i = 1}^I \prod_{j \in Q_i} \overline{P}_{i j}(\overline{\theta} \mid D_i)^{W_{i j}}.
\end{align}
Note that $W_{ij} = 1 \implies q_{ij} = 1$, and thus we can drop the $q_{i j}$ term from the updated pseudo-likelihood function, as long as we sum over considered workers for each job, allowing us to avoid a $0^0$ term in the updated pseudo-likelihood function.

Thus, the updated pseudo-log-likelihood function is:
\begin{align}
    \overline{L}\left(\overline{\theta} \mid \overline{D}, W\right) = \sum_{i = 1}^I \sum_{j \in Q_i} W_{i j} \cdot \log\left(\overline{P}_{i j}(\overline{\theta} \mid \overline{D}_i)\right),
\end{align}
where, for $j\neq 0$:
{\small
\begin{align}
    \log\left(\overline{P}_{i j}(\overline{\theta} \mid \overline{D}_i)\right) =
    \overline{\delta}_{i j}(\overline{D}_i; \overline{\theta})
     - \log\left(1 + \sum_{k \in Q_i \setminus \{0\}} \exp\left(\overline{\delta}_{ik}(\overline{D}_i; \overline{\theta})\right)\right) + \log(\pi),
\end{align}}
and for $j = 0$:
\begin{align}
    \log\left(\overline{P}_{i 0}(\overline{\theta} \mid \overline{D}_i)\right) =
    \log\left(1 + \left(\left(1 + \sum_{k \in Q_i \setminus \{0\}} \exp\left(\overline{\delta}_{ik}(\overline{D}_i; \overline{\theta})\right)\right)^{-1} -  1\right)\cdot \pi\right).   
\end{align}

This step of the estimation procedure consists of maximizing the above pseudo-log-likelihood function with respect to the parameters $\tilde{\theta}$. With this step, we complete the estimation procedure.

\FloatBarrier
\renewcommand{\thefigure}{E\arabic{figure}} 
\setcounter{figure}{0}
\renewcommand{\thetable}{E\arabic{table}} 
\setcounter{table}{0}
\renewcommand{\thesubsection}{E.\arabic{subsection}}
\setcounter{subsection}{0}
\section{Robustness with NLP Measures of Signal}\label{appendix_nlp}

In this section, we present a number of robustness checks of our descriptive findings using alternative Natural Language Processing (NLP) measures of textual similarity between job posts and proposals in lieu of our LLM-based measure of signal. We use three measures of textual similarity to quantify how closely the language of an application matches that of the corresponding job description, with the general idea that greater textual alignment reflects a higher degree of customization and effort. One major drawback of these approaches relative to our LLM-based measure is that simple copy-and-pasting of phrases or even entire sentences from the job post in a worker's proposal score highly on NLP similarity metrics, even though the proposals may be entirely non-customized and generic. In contrast, our LLM-based measure is prompted to disregard such blatant copy-pasting, much like a human reader would. Regardless, these NLP-based measures provide useful robustness checks for our main descriptive findings.

\subsection{NLP Measure Definitions} 

The three main measures of textual similarity we use---TF-IDF cosine similarity, average word embeddings, and Sentence-BERT embeddings---differ from each other primarily in the linguistic information they capture and their computational complexity.

\textbf{TF-IDF} quantifies similarity based on direct word overlap. Each document is represented as a high-dimensional sparse vector, where each entry corresponds to a term weighted by its frequency in the document and its rarity in the corpus. Cosine similarity between two such vectors measures the extent of shared vocabulary, giving more weight to uncommon but informative terms. TF-IDF thus captures lexical overlap, making it particularly informative when workers literally reuse text from job postings. However, it cannot recognize paraphrases or synonyms, as it treats for instance ``developer'' and ``programmer'' as distinct terms. Its main advantages are transparency, speed, and interpretability.

\textbf{Averaged word embeddings} (using GloVe as an embedding matrix) move beyond raw word counts by embedding each word in a continuous semantic space. Averaging these vectors for each document yields a dense, low-dimensional representation of overall semantic content. Cosine similarity between averaged embeddings can be high even when two texts use different but related words, allowing this measure to capture semantic proximity rather than literal word overlap. The drawback of this measure is that averaging discards word order and contextual nuance, effectively summarizing each document's ``topic'' rather than its detailed phrasing. Nevertheless, this approach often provides a more robust indicator of whether an application and job posting are discussing related ideas rather than merely sharing vocabulary.

\textbf{Sentence-BERT} extends this idea by producing contextualized sentence-level embeddings using a transformer model pre-trained on large text corpora. Unlike averaged embeddings, Sentence-BERT accounts for word order and syntactic relationships: the meaning of each word is conditioned on its surrounding context. This often yields more discriminating similarity scores, distinguishing, for instance, between a worker stating they ``led a design project'' and one who ``assisted on a design project.'' Empirically, Sentence-BERT embeddings have been shown to generally outperform both TF-IDF and simple averaged embeddings in capturing semantic similarity across a wide range of natural language tasks. Their main costs are computational—requiring the evaluation of a deep neural model—and interpretational, since the embedding dimensions are not human-readable. For the purpose of this paper, we use the ``all-MiniLM-L6-v2'' variant of Sentence-BERT (SentenceTransformer model), which balances performance and speed.

In summary, TF-IDF captures literal overlap, averaged embeddings capture broad semantic similarity, and Sentence-BERT captures contextual semantic alignment. In our setting, where applicants may paraphrase or adapt postings rather than quote them verbatim, embeddings-based measures (especially Sentence-BERT) provide a more informative measure of ``content fit.'' However, TF-IDF remains useful as a transparent benchmark emphasizing explicit textual reuse.

\paragraph{Cosine Similarity.}
Across all text-based representations used in this section---TF-IDF vectors, averaged word embeddings, and Sentence-BERT embeddings---we measure the alignment between a job post and a worker's proposal using the cosine similarity metric.  
Each document is represented as a vector in a high-dimensional space, where the specific construction of the vector depends on the representation method: sparse term-frequency weights for TF-IDF, averaged semantic vectors for word embeddings, or contextual sentence embeddings for Sentence-BERT.  
Cosine similarity between the job posting ($A_i$) and the application ($B_i$) is defined as
\[
\text{Similarity}_i = 
\frac{A_i \cdot B_i}{\|A_i\|\,\|B_i\|},
\]
where $A_i \cdot B_i$ denotes the dot product and $\|A_i\|$ and $\|B_i\|$ are the Euclidean norms of the two vectors.  

This metric measures the \emph{angle} between the two representations rather than their magnitude, making it scale-invariant to document length or overall word frequency.  
Intuitively, a higher cosine similarity indicates that the application and the job posting place weight on similar terms or occupy nearby regions in semantic space.  
When applied to TF-IDF vectors, it captures \textit{lexical overlap} in keywords; when applied to embedding-based representations, it captures \textit{semantic proximity} between the ideas expressed in each text.  

Cosine similarity is widely used in natural language processing because it provides an interpretable and computationally efficient measure of content alignment.  
In this context, it serves as a unified metric for comparing how closely a worker's proposal reflects the language, topics, or semantics of the corresponding job post.

\subsection{Distributions of NLP-Based Signal Measures} 

\begin{figure}[H]
  \centering
  \begin{subfigure}[t]{0.49\linewidth}
    \centering
    \subcaption{Weighted LLM score}
    \includegraphics[width=\linewidth]{hist_signal_s_wt_adj_04_2_pre_vs_post.pdf}
  \end{subfigure}\hfill
  \begin{subfigure}[t]{0.49\linewidth}
    \centering
    \subcaption{Sentence-BERT score}
    \includegraphics[width=\linewidth]{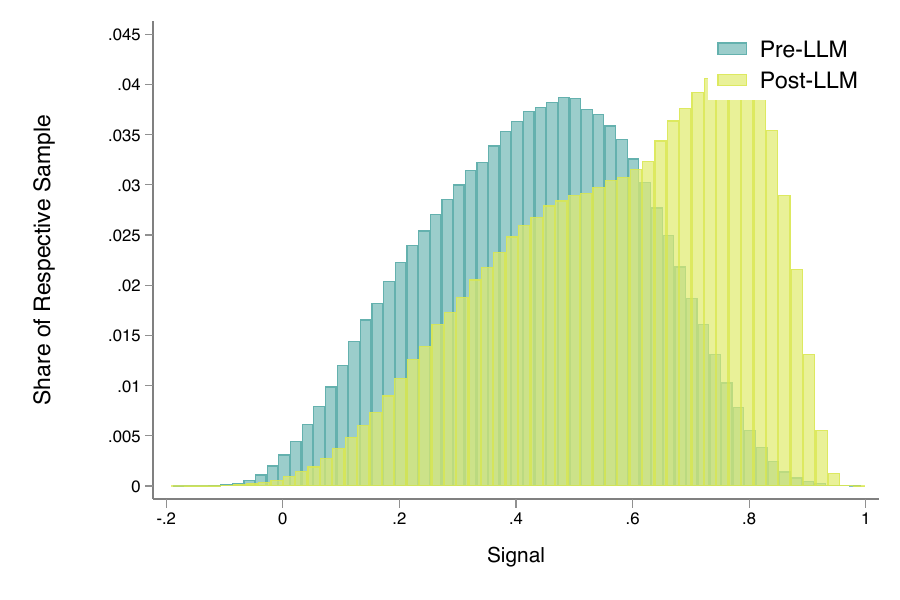}
  \end{subfigure}

  \vspace{0.6em}
  \begin{subfigure}[t]{0.49\linewidth}
    \centering
    \subcaption{Embedding similarity score}
    \includegraphics[width=\linewidth]{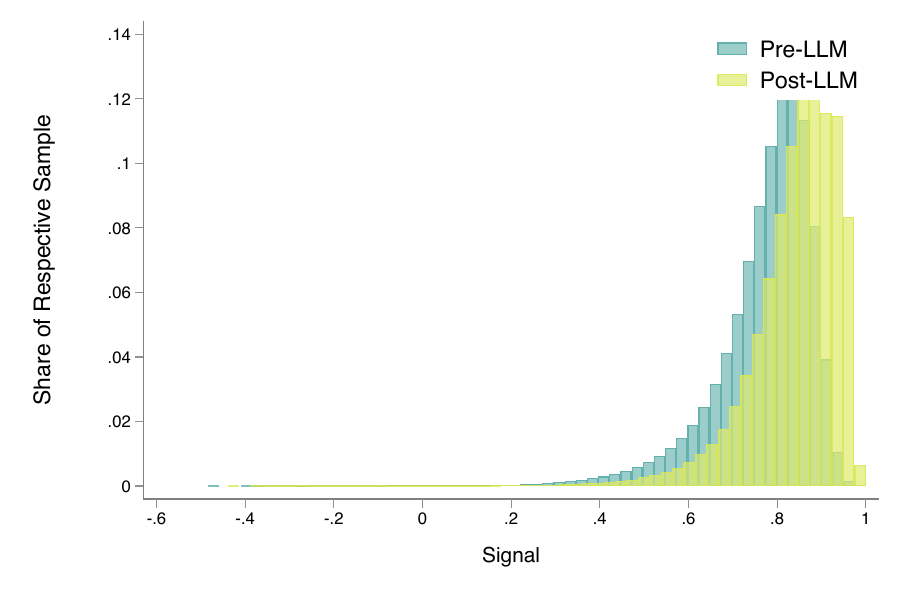}
  \end{subfigure}\hfill
  \begin{subfigure}[t]{0.49\linewidth}
    \centering
    \subcaption{TF-IDF similarity score}
    \includegraphics[width=\linewidth]{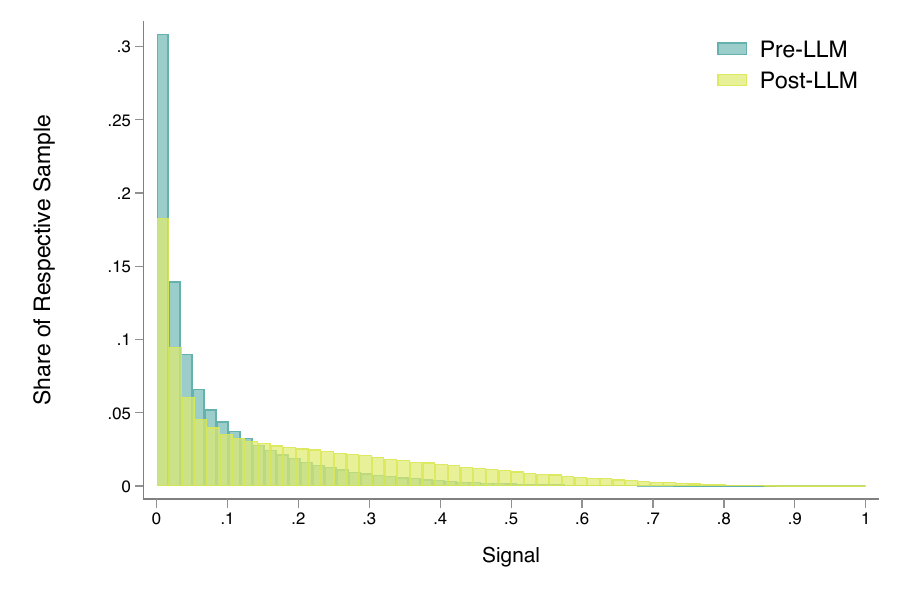}
  \end{subfigure}

  \caption{Distribution of Signals: Pre-LLM vs. Post-LLM}
  \begingroup
    \setstretch{1}
  \floatfoot{\footnotesize \textit{Notes:} Each panel plots histogram estimates of the distribution of the corresponding signal measure for the pre-LLM and post-LLM samples.  
  The post-LLM sample corresponds to job posts after March 26, 2024.}
  \endgroup
\end{figure}

\subsection{Win vs. Signal: NLP Measures} 

\begin{figure}[H]
  \centering
  \begin{subfigure}[t]{0.49\linewidth}
    \centering
    \subcaption{Weighted LLM score}
    \includegraphics[width=\linewidth]{norm_win_vs_signal_s_wt_adj_04_2_pre_vs_post.pdf}
  \end{subfigure}\hfill
  \begin{subfigure}[t]{0.49\linewidth}
    \centering
    \subcaption{Sentence-BERT score}
    \includegraphics[width=\linewidth]{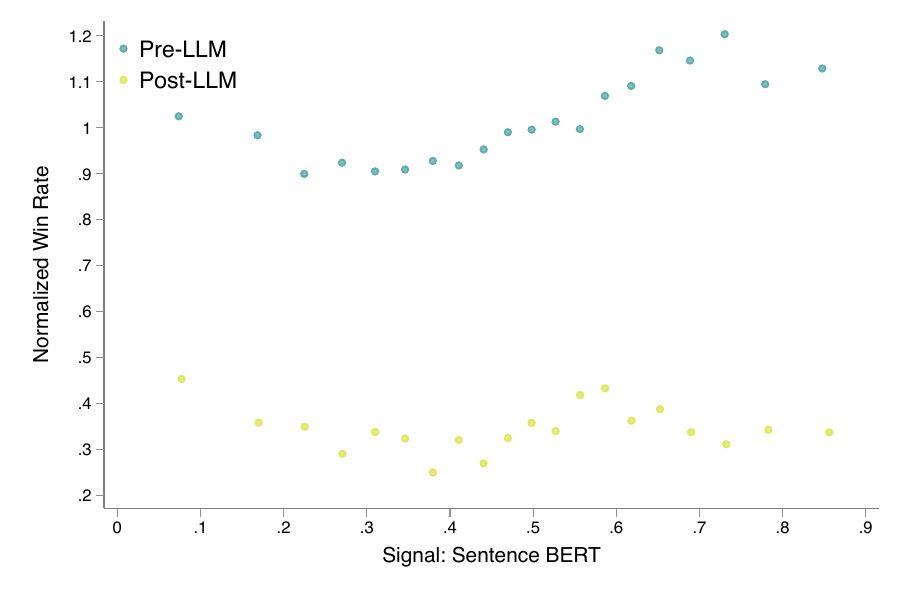}
  \end{subfigure}

  \vspace{0.6em}
  \begin{subfigure}[t]{0.49\linewidth}
    \centering
    \subcaption{Embedding similarity score}
    \includegraphics[width=\linewidth]{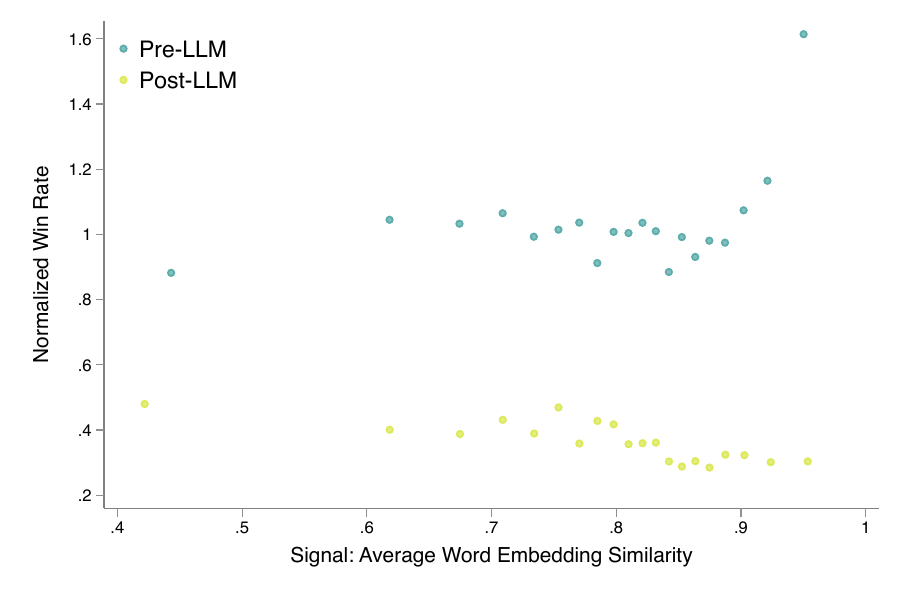}
  \end{subfigure}\hfill
  \begin{subfigure}[t]{0.49\linewidth}
    \centering
    \subcaption{TF-IDF similarity score}
    \includegraphics[width=\linewidth]{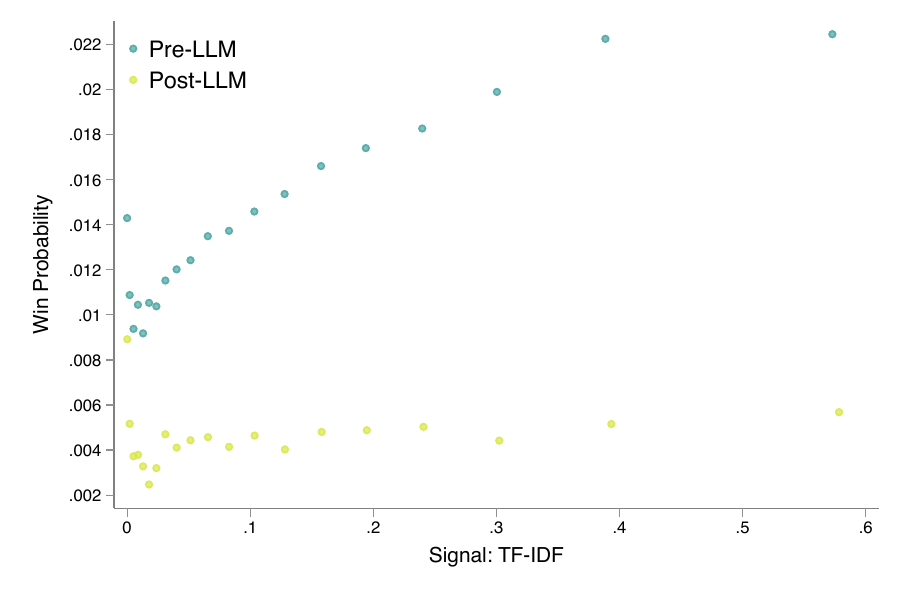}
  \end{subfigure}

  \captionsetup{justification=centering}
  \caption{Binscatter of Win on Signal\\[0.2em]
  \small Pre-LLM vs. Post-LLM}
  \begingroup
    \setstretch{1}
  \floatfoot{\footnotesize\raggedright
  \textit{Notes:} Each panel plots a binscatter of whether the applying worker is hired on the corresponding signal measure in both the pre-LLM and post-LLM samples. As per our research agreement with \texttt{Freelancer.com}, we do not disclose the level of hiring rates, so we normalize (i.e., divide) the $y$-axis by the unconditional pre-LLM overall win rate. The post-LLM sample corresponds to job posts after March 26, 2024}
    \endgroup
\end{figure}

\subsection{Signal vs. Effort: NLP Measures}

\begin{figure}[H]
  \centering
  \begin{subfigure}[t]{0.49\linewidth}
    \centering
    \subcaption{Weighted LLM score}
    \includegraphics[width=\linewidth]{signal_vs_effort_s_wt_adj_04_2_pre_vs_ai.pdf}
  \end{subfigure}\hfill
  \begin{subfigure}[t]{0.49\linewidth}
    \centering
    \subcaption{Sentence-BERT score}
    \includegraphics[width=\linewidth]{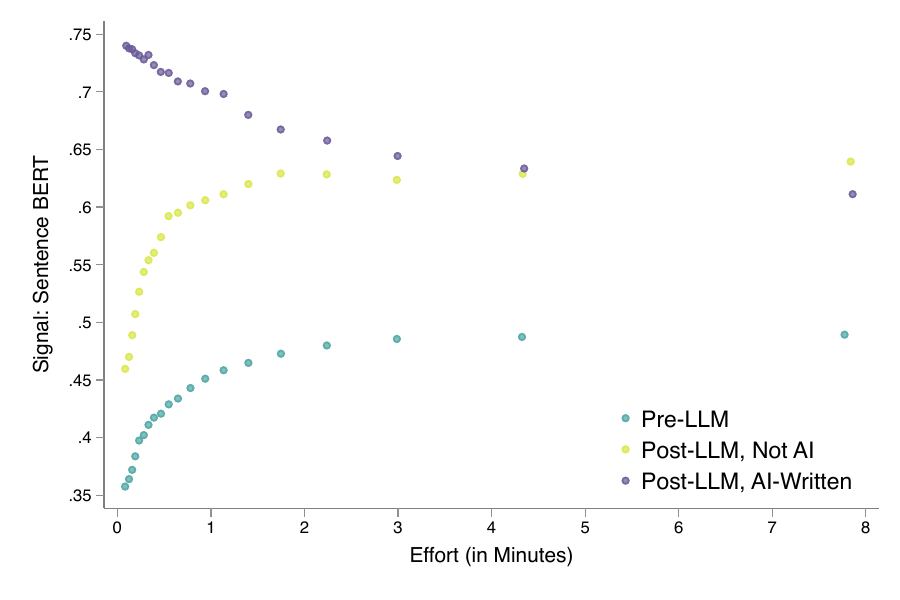}
  \end{subfigure}

  \vspace{0.6em}
  \begin{subfigure}[t]{0.49\linewidth}
    \centering
    \subcaption{Embedding similarity score}
    \includegraphics[width=\linewidth]{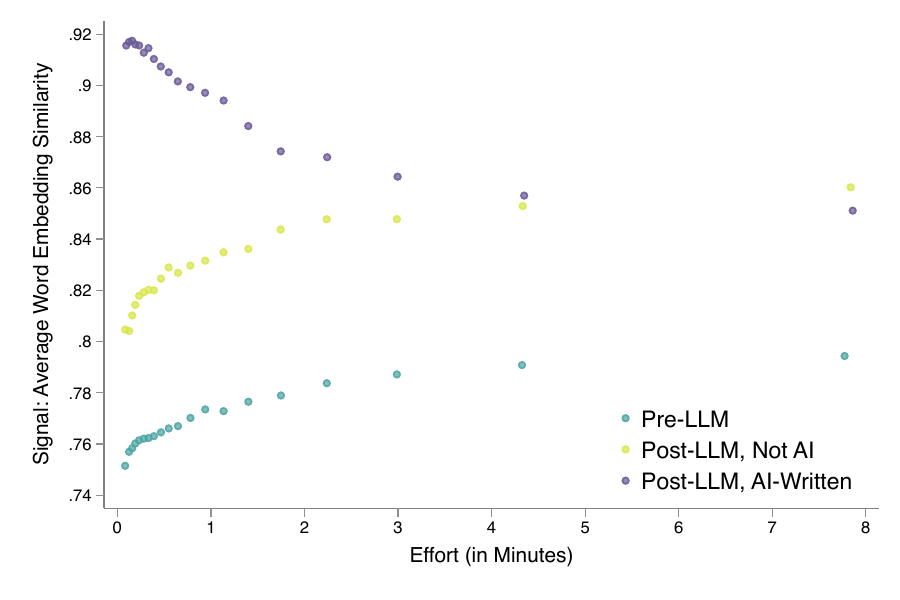}
  \end{subfigure}\hfill
  \begin{subfigure}[t]{0.49\linewidth}
    \centering
    \subcaption{TF-IDF similarity score}
    \includegraphics[width=\linewidth]{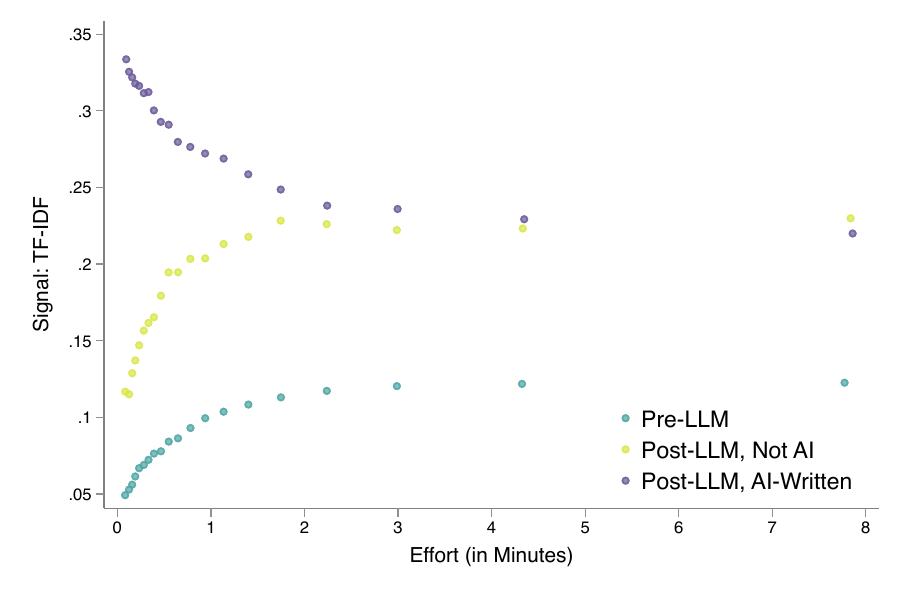}
  \end{subfigure}
  \vspace{0.6em}
  \captionsetup{justification=centering}
  \caption{Binscatter of Signal on Effort\\[0.2em]
  \small Pre-LLM vs. Post-LLM, Not-AI vs. Post-LLM, AI-Written}\label{NLP_effort_vs_signal}
  \begingroup
    \setstretch{1}
  \floatfoot{\footnotesize\raggedright
  \textit{Notes:} Each panel plots a binscatter of the corresponding signal measure on effort in both the pre-LLM and post-LLM samples, splitting the latter by on-platform AI usage.  
  The post-LLM sample corresponds to job posts after March 26, 2024.}
      \endgroup
\end{figure}
\newpage
\subsection{Job Completion vs. Signal: NLP Measures}

\begin{figure}[h!]
  \centering

  \begin{subfigure}[t]{0.49\linewidth}
    \centering
    \subcaption{Weighted LLM score}
    \includegraphics[width=\linewidth]{norm_complete_5_stars_vs_signal_s_wt_adj_04_2_pre_vs_post.pdf}
  \end{subfigure}\hfill
  \begin{subfigure}[t]{0.49\linewidth}
    \centering
    \subcaption{Sentence-BERT score}
    \includegraphics[width=\linewidth]{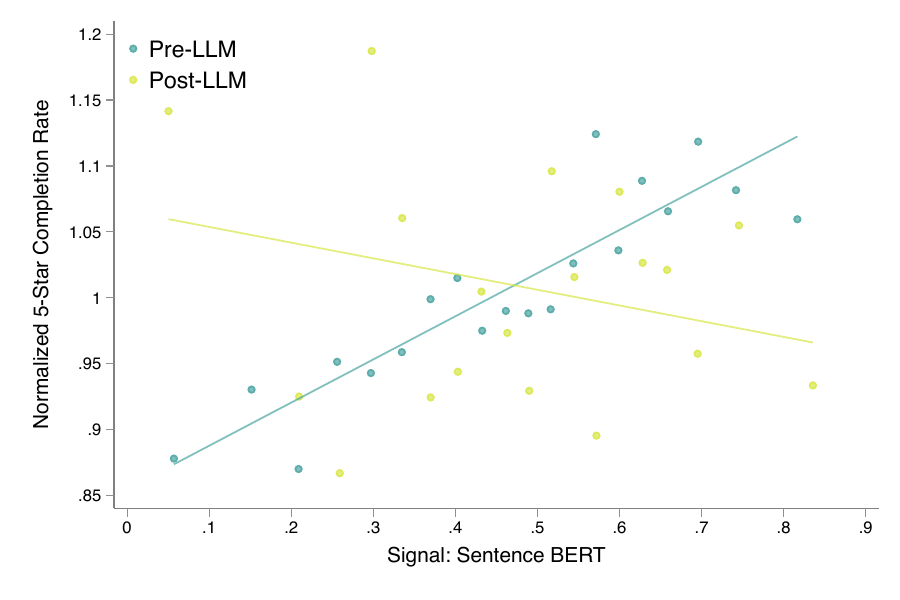}
  \end{subfigure}

  \vspace{0.6em}
  \begin{subfigure}[t]{0.49\linewidth}
    \centering
    \subcaption{Embedding similarity score}
    \includegraphics[width=\linewidth]{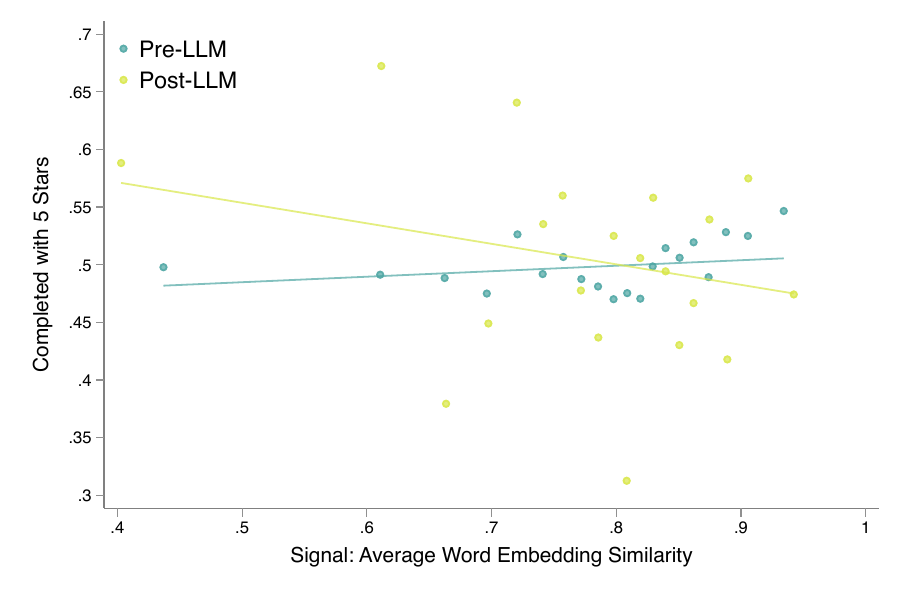}
  \end{subfigure}\hfill
  \begin{subfigure}[t]{0.49\linewidth}
    \centering
    \subcaption{TF-IDF similarity score}
    \includegraphics[width=\linewidth]{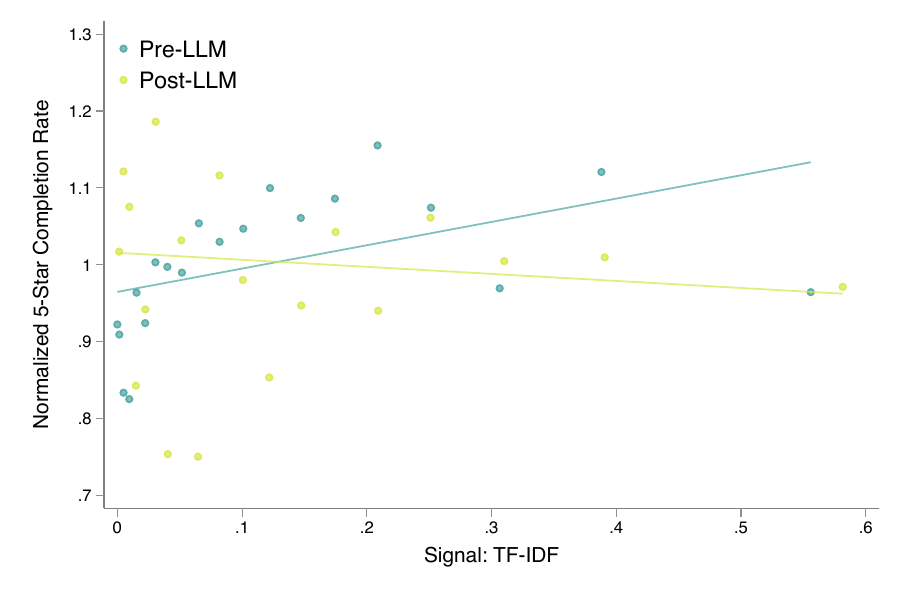}
  \end{subfigure}

  \captionsetup{justification=centering}
  \caption{Binscatter of 5-Star Completion on Signal\\[0.2em]
  \small Pre-LLM vs. Post-LLM}
  \begingroup
    \setstretch{1}
  \floatfoot{\footnotesize\raggedright
  \textit{Notes:} Each panel plots a binscatter of whether the hired worker completes the job with 5 stars on the corresponding signal measure, comparing the pre-LLM and post-LLM samples. As per our research agreement with \texttt{Freelancer.com}, we do not disclose the level of completion rates, so we normalize (i.e., divide) the $y$-axis by the unconditional pre-LLM overall 5-star completion rate. The post‑LLM sample corresponds to the subsample of job posts after March 26, 2024.}
    \endgroup
\end{figure}

\newpage
\subsection{Pre-LLM vs. AI-Written Signals: NLP Measures}

\begin{figure}[H]
  \centering
  \begin{subfigure}[t]{0.49\linewidth}
    \centering
    \subcaption{Weighted LLM score}
    \includegraphics[width=\linewidth]{hist_signal_s_wt_adj_04_2_pre_vs_ai.pdf}
  \end{subfigure}\hfill
  \begin{subfigure}[t]{0.49\linewidth}
    \centering
    \subcaption{Sentence-BERT score}
    \includegraphics[width=\linewidth]{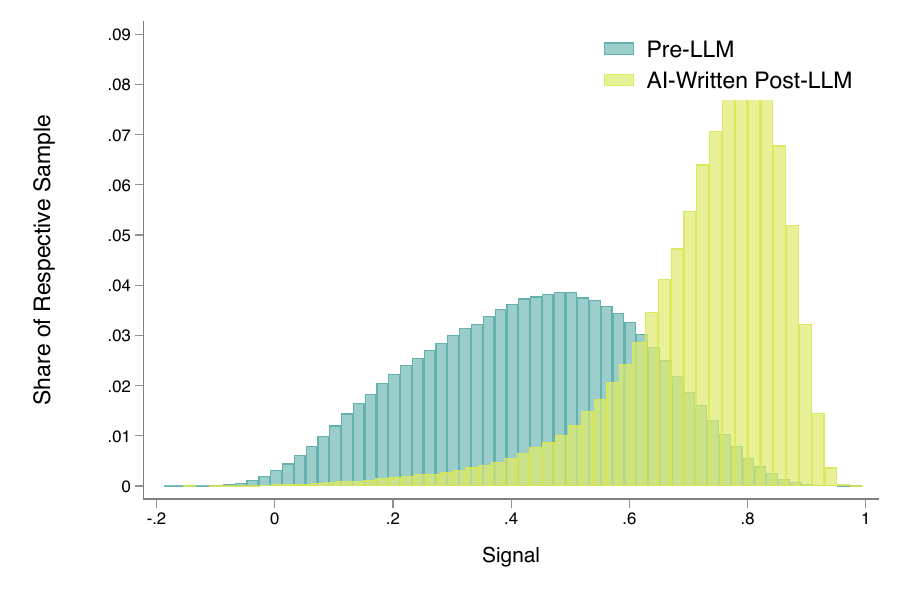}
  \end{subfigure}

  \vspace{0.6em}
  \begin{subfigure}[t]{0.49\linewidth}
    \centering
    \subcaption{Embedding similarity score}
    \includegraphics[width=\linewidth]{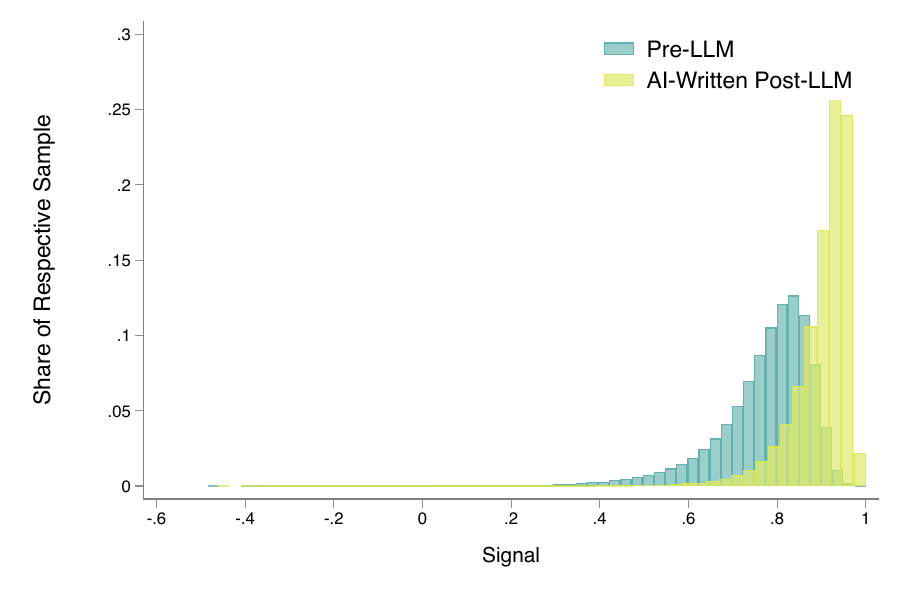}
  \end{subfigure}\hfill
  \begin{subfigure}[t]{0.49\linewidth}
    \centering
    \subcaption{TF-IDF similarity score}
    \includegraphics[width=\linewidth]{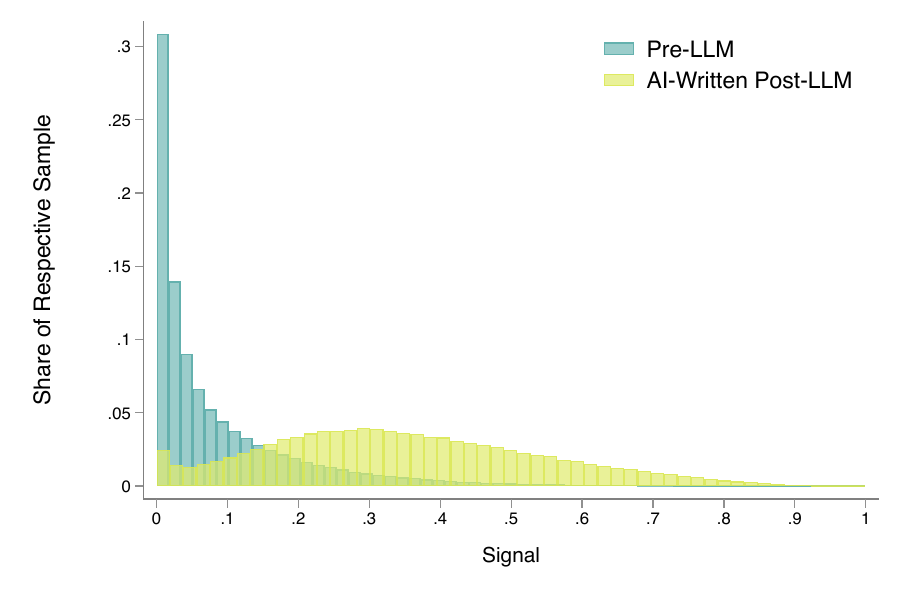}
  \end{subfigure}

  \caption{Distribution of Signals: Pre-LLM vs. Post-LLM}
  \begingroup
    \setstretch{1}
  \floatfoot{\footnotesize \textit{Notes:} Each panel plots histogram estimates of the distribution of the corresponding signal measure in both the pre-LLM and post-LLM samples, where the latter is subsetted to only include the applications using the on-platform AI-writing tool. The post-LLM sample corresponds to the subsample of post-LLM job posts that were posted after March 26, 2024.}
    \endgroup
\end{figure}
\newpage
\FloatBarrier
\renewcommand{\thefigure}{F\arabic{figure}} 
\setcounter{figure}{0}
\renewcommand{\thetable}{F\arabic{table}} 
\setcounter{table}{0}
\renewcommand{\thesubsection}{F.\arabic{subsection}}
\setcounter{subsection}{0}
\section{Platform Details}\label{appendix_market}

\begin{figure}[H]
\centering
\includegraphics[width=0.8\linewidth]{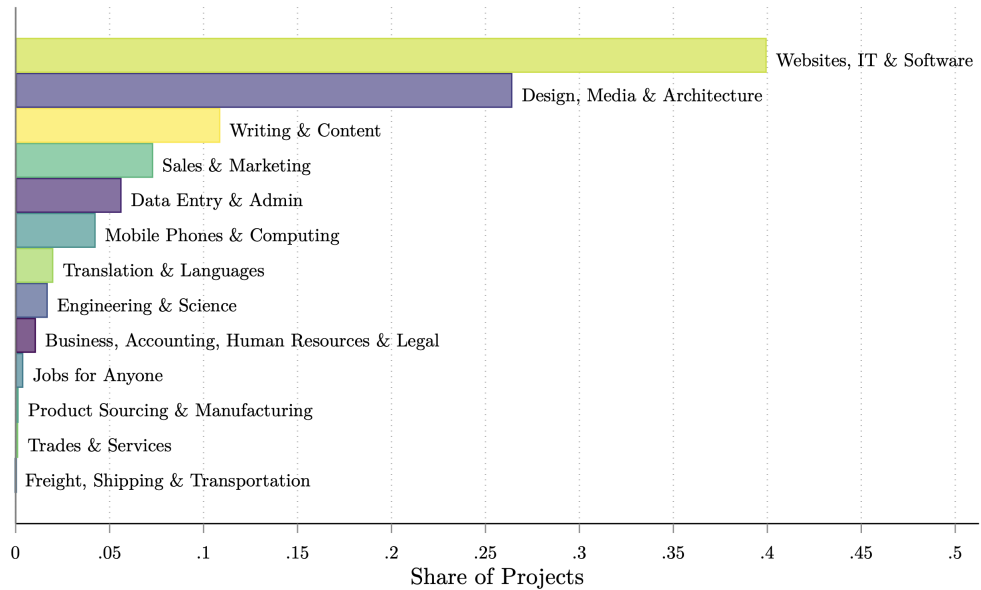}
\caption{What Types of Jobs Are Posted on the Platform?}
    \label{fig:projects_on_platform}
     \floatfoot{\footnotesize \textit{Notes:} This figure plots the aggregate job categories listed on the platform. Coding, design, and writing skills are most in demand on the platform. For a detailed list of all possible skills and job categories, see Appendix Table \ref{job_category}}
\end{figure}

On \texttt{Freelancer.com}, workers are primarily hired to complete tasks in areas such as software development, writing, data entry, and design (Figure~\ref{fig:projects_on_platform}). Each broad occupation can be further disaggregated into detailed categories and associated skills. A list of job categories and examples of their associated skills is provided in Appendix Table~\ref{job_category}.

\begin{figure}[H]
    \centering
  \includegraphics[width=0.7\linewidth]{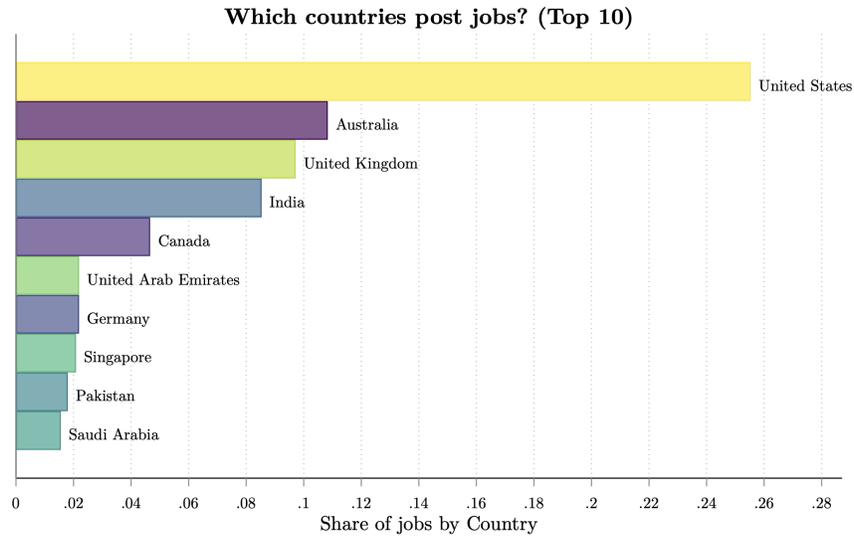} 
   \includegraphics[width=0.7\linewidth]{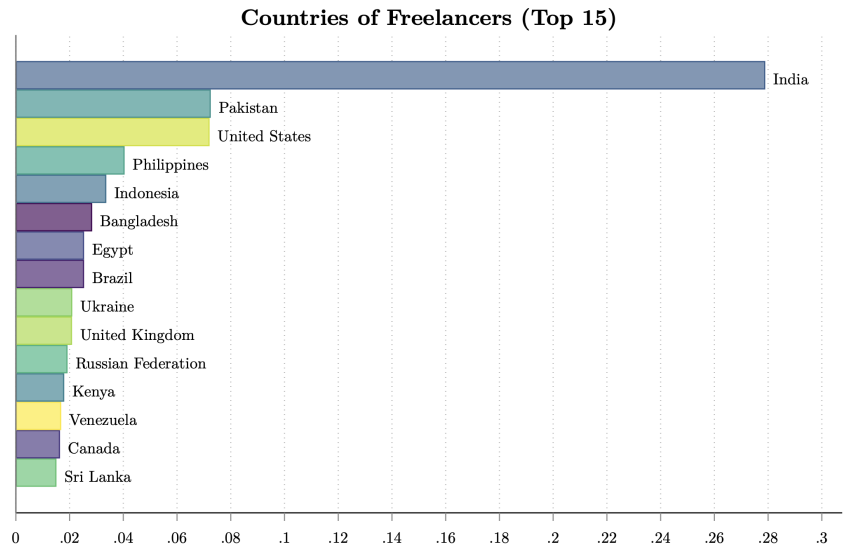}
    \caption{Who is on the platform?}
    \label{fig:who_is_on_platform}
    \floatfoot{\footnotesize \textit{Notes:} These figures plot the listed country of workers (``freelancers,'' at the bottom) applying to and employers (individuals or firms posting jobs, at the top) posting jobs in English in 2019 and soliciting bids in USD. Firms on the platform are largely from the Global North, while freelancers on the platform are mostly from the Global South}
\end{figure}

The outsourcing of tasks on \texttt{Freelancer.com} can be generally characterized by a North-South exchange, with many employers located in the Global North and most workers in the Global South. Figure \ref{fig:who_is_on_platform} compares the geographical distribution of workers on the platform applying to and employers on the platform posting jobs in English in 2019 and soliciting bids in USD. The top panel displays the top 10 countries of employers. 50\% of all jobs are posted from employers located in the United States (25.5\%), Australia (11\%), the United Kingdom (9.5\%) and Canada (5\%). While India also appears to be a large poster of jobs on the platform (8.5\%), it is above all the primary country of origin of workers: 28\% of all freelancers are from India. Pakistan (7\%), the Philippines (5\%), Indonesia (4\%) and Bangladesh (3\%) follow. 

\newpage

\newpage

\begin{table}[H]
\footnotesize
\begin{tabular} {@{} ll @{}} \\
\toprule
\addlinespace
\textbf{Job Category} & \textbf{Skill Examples} \\
\addlinespace
\midrule
\textbf{Websites, IT \& Software} & PHP, HTML, JavaScript, Software Architecture \\
\addlinespace
\textbf{Design, Media \& Architecture}  & Graphic Design, Website Design, Photoshop \\
\addlinespace
\textbf{Mobile Phones \& Computing}  & Mobile App Development, Android, iPhone \\
\addlinespace
\textbf{Data Entry \& Admin}  & Data Entry, Excel, Data Processing \\
\addlinespace
\textbf{Engineering \& Science}  & AutoCAD, CAD/CAM, Engineering \\
\addlinespace
\textbf{Freight Shipping \& Transportation}  & Dropshipping, Containerization, Import/Export \\
\addlinespace
\textbf{Business, Accounting, Human Resources \& Legal}  & Business Analysis, Accounting, Finance \\
\addlinespace
\textbf{Product Sourcing \& Manufacturing}  & Product Design, Product Sourcing, Supplier Sourcing \\
\addlinespace
\textbf{Sales \& Marketing}  & Internet Marketing, Social Media Marketing, Sales \\
\addlinespace
\textbf{Translation \& Languages}  & English (US) Translator, Spanish Translator, French Translator \\
\addlinespace
\textbf{Writing \& Content}  & Content Writing, Article Writing, Copywriting \\
\addlinespace
\textbf{Artificial Intelligence}  & Large Language Models (LLMs), Bot Development \\
\addlinespace
\textbf{Education \& Tutoring}  & English Tutoring, Language Tutoring, Chemistry Tutoring \\
\addlinespace
\textbf{Health \& Medicine}  & Food Safety, Counseling Psychology, Public Health \\
\addlinespace
\textbf{Telecommunications}  & Physical Site Survey, Active Site Survey, Pre-inspection visits \\
\addlinespace
\textbf{Jobs for Anyone}  & Virtual Assistant, Local Job, Travel Ready \\
\bottomrule
\end{tabular} 

 \caption{Job Categories and Skills}\label{job_category}
     \floatfoot{\footnotesize \textit{Notes:} This table summarizes the main job categories available on \texttt{Freelancer.com}. The second column provides examples of the type of skills workers can list on their profile and employers can link to projects. See \url{https://www.freelancer.com/job/} for a full list of skills.}
\end{table}

\newpage

\begin{figure}[H]
    \centering
   \includegraphics[width=\linewidth]{\detokenize{platform_overview.png}} 
    \vspace{0.4cm}
    \caption{Looking for Jobs - Platform Overview}\label{platform_overview}
    
\end{figure}

\newpage

\begin{figure}[H]
    \centering
        \includegraphics[width=.9\linewidth]{\detokenize{bid_list_screenshot.pdf}}
    \vspace{0.3cm}
    \caption{Bid List}\label{bid_list}
    \floatfoot{\footnotesize \textit{Notes:} This figure displays an example of how employers view applications to their job posts on \texttt{Freelancer.com}. This particular job post was posted by the authors of this paper, when seeking to interview workers about their application strategies on the platform.}
\end{figure}

\newpage

\begin{figure}[H]
    \centering
        \includegraphics[width=0.68\linewidth]{\detokenize{freelancer_profile.png}} \\ 
        \includegraphics[width=0.68\linewidth]{\detokenize{freelancer_profile_reviews.png}}   \\
    \includegraphics[width=0.68\linewidth]{\detokenize{freelancer_currentwork.png}} \\
    \vspace{0.5cm}
    \caption{Worker Profile}\label{worker_profile}
\end{figure}

\newpage

\begin{figure}[H]
    \centering
\includegraphics[width=0.45\linewidth]{\detokenize{confirmation_payment.png}} \\
   \includegraphics[width=0.75\linewidth]{\detokenize{milestone_payments.png}} 
    \vspace{0.4cm}
    \caption{Milestone Payment System}\label{milestone_payments}
\end{figure}

\newpage

\begin{figure}[H]
    \centering
\includegraphics[width=0.9\linewidth]{\detokenize{ai_bid_button.pdf}} \\
   \includegraphics[width=0.89\linewidth]{\detokenize{ai_written_bid.pdf}} 
    \vspace{0.4cm}
    \caption{AI-Written Proposal}\label{ai_bids}
    \floatfoot{\footnotesize \textit{Notes:} This figure presents both how the AI-writing tool is displayed on the platform and an example of an AI-written proposal. At the top of the figure we see an empty proposal text box with the ``Write my bid'' button at the top right corner above the text box. Workers wishing to use the AI-writing tool and with the requisite subscription can click this button to have their proposal generated automatically by an LLM within seconds. The bottom of the figure shows an example of a proposal generated by the AI-writing tool. The only information the AI-writing tool is given is the job title, job description, and information on the worker's profile. The job post for which this proposal was generated involved scraping data from public directories. The profile used to generate this proposal belongs to one of the authors of this paper.}
\end{figure}

\end{spacing}
\end{document}

%% file: packagesJMP.tex
\usepackage[english]{babel}
\usepackage[utf8]{inputenc}
\usepackage{lmodern}
\usepackage[sc]{mathpazo}
\usepackage{setspace}
\doublespacing

\usepackage[margin = 25mm]{geometry} 
\setlength{\parindent}{0.7cm}
\usepackage{changepage} 
\sloppy
\usepackage[x11names]{xcolor}

\usepackage{csquotes}
\usepackage{titling}
\PassOptionsToPackage{normalem}{ulem}
\usepackage{ulem}
\usepackage{comment}
\usepackage{soul}
\sethlcolor{Honeydew2}
\usepackage{enumitem}
\setlist{nosep} 

\usepackage{graphicx}
\usepackage{caption}
\usepackage{subcaption}
\usepackage{floatrow}
\floatsetup[table]{capposition=top}
\floatsetup[figure]{capposition=bottom, captionskip=-0.1cm}
\usepackage{tikz}
\usetikzlibrary{arrows,positioning}
\usepackage{adjustbox}
\usepackage{tcolorbox}

\usepackage{booktabs}
\usepackage{colortbl}
\usepackage{arydshln}

\usepackage{mathtools,amssymb,dsfont}
\usepackage{amsthm}
\usepackage{physics}
\usepackage{siunitx}
\usepackage{amsthm}
\usepackage{centernot}
\usepackage{gensymb}
\DeclareUnicodeCharacter{2212}{-}
\usepackage{bbm}
\usepackage{relsize}
\usepackage{threeparttable}
\usepackage{fancyvrb}
\usepackage{listings}

\usepackage{pdflscape}
\usepackage{afterpage}
\usepackage{longtable}
\usepackage{placeins} 

\usepackage{float}
\usepackage{etoolbox}
\setlength{\marginparwidth}{2cm}
\usepackage[colorinlistoftodos]{todonotes}

\usepackage{titlesec}
\titleformat*{\section}{\LARGE\bfseries\color{Black}}
\titleformat*{\subsection}{\Large\bfseries}
\titleformat*{\subsubsection}{\normalsize\bfseries}
\titlelabel{\thetitle.\quad}



\theoremstyle{remark}

\usepackage[title,titletoc]{appendix}

\usepackage{ragged2e}
\newlength\ubwidth

\newcommand{\lmat}{\left(\begin{matrix}}
\newcommand{\rmat}{\end{matrix}\right)}

\DeclareMathOperator*{\argmax}{arg\,max}

\usepackage[bottom]{footmisc}

\makeatletter
\def\blfootnote{\xdef\@thefnmark{}\@footnotetext}
\makeatother

\usepackage[colorlinks=true]{hyperref}
\hypersetup{
  linkcolor=DarkGoldenrod3,
  citecolor=Aquamarine4,
  urlcolor=Thistle3!140
}


\newcommand{\plum}[1]{\textcolor{Plum4}{\textbf{#1}}}

\usepackage[backend=biber,style=authoryear,natbib=true,%
  uniquelist=false,uniquename=false,backref=true]{biblatex}

\DefineBibliographyStrings{english}{%
  backrefpage  = {Cited on page },
  backrefpages = {Cited on pages},
}

\renewbibmacro*{volume+number+eid}{%
  \printfield{volume}%
  \setunit*{\addnbthinspace}%
  \printfield{number}%
  \printfield{eid}}
\DeclareFieldFormat[article]{number}{\mkbibparens{#1}}
\AtBeginBibliography{%
}
\renewbibmacro*{addendum+pubstate}{%
  \printfield{addendum}%
  \iffieldequalstr{labeldatesource}{pubstate}{}%
  {\newunit\newblock\printfield{pubstate}}}
\renewbibmacro*{url+urldate}{} 
\renewbibmacro*{eprint}{}      
\renewbibmacro*{doi+eprint+url}{%
  \iftoggle{bbx:doi}{}{%
    \printfield{doi}}%
  \newunit\newblock
  \iftoggle{bbx:eprint}{\usebibmacro{eprint}}{}%
  \newunit\newblock
  \iftoggle{bbx:url}{\usebibmacro{url+urldate}}{}}
\renewbibmacro*{finentry}{\finentrypunct}

\DeclareCiteCommand{\citet}
  {\usebibmacro{prenote}}
  {\usebibmacro{citeindex}\printtext[bibhyperref]{\printnames{labelname}\printtext{ (\printfield{year}\printfield{extrayear})}}}
  {\multicitedelim}
  {\usebibmacro{postnote}}
\DeclareCiteCommand{\citep}[\mkbibparens]
  {\usebibmacro{prenote}}
  {\usebibmacro{citeindex}\printtext[bibhyperref]{\printnames{labelname}\setunit{\nameyeardelim}\printlabeldateextra}}
  {\multicitedelim}
  {\usebibmacro{postnote}}
\DeclareCiteCommand{\citealt}
  {\usebibmacro{prenote}}
  {\usebibmacro{citeindex}\printtext[bibhyperref]{\printnames{labelname}\setunit{\space}\printfield{year}\printfield{extrayear}}}
  {\multicitedelim}
  {\usebibmacro{postnote}}

\addbibresource{\detokenize{JMP_bib.bib}}

%% file: tables/sample_sum_stats.tex
\begin{table}[!h]
\centering
\begin{threeparttable}
\notsosmallbutevenbigger
\caption{Sample Summary Statistics}\label{tab:sample_summary_stats}
\setlength{\tabcolsep}{8pt}
\begin{tabular}{@{}l ccc ccc@{}}
\toprule
\multicolumn{7}{c}{\textbf{Worker Summary Statistics}}\\
\midrule
& \multicolumn{3}{c}{\textbf{Pre-LLM}} & \multicolumn{3}{c}{\textbf{Post-LLM}} \\
\cmidrule(lr){2-4} \cmidrule(lr){5-7}
\multicolumn{1}{l}{\# Applications} & \multicolumn{3}{c}{957{,}442} & \multicolumn{3}{c}{1{,}723{,}462} \\
\multicolumn{1}{l}{\# Unique Workers} & \multicolumn{3}{c}{92{,}048} & \multicolumn{3}{c}{134{,}431} \\
\multicolumn{1}{l}{Mean Apps.\ per Worker} & \multicolumn{3}{c}{10.40} & \multicolumn{3}{c}{12.82} \\
\addlinespace[2pt]
\multicolumn{1}{l}{\% of Workers are Rookies} & \multicolumn{3}{c}{71.29\%} & \multicolumn{3}{c}{83.70\%} \\
\multicolumn{1}{l}{\% of Apps.\ from Rookies} & \multicolumn{3}{c}{12.52\%} & \multicolumn{3}{c}{20.69\%} \\
\multicolumn{1}{l}{\begin{tabular}[t]{@{}l@{}}Win Rate per Application \\ \hspace{0.8em}Divided by Pre-LLM Rate\end{tabular}} & \multicolumn{3}{c}{1} & \multicolumn{3}{c}{0.45} \\
\multicolumn{1}{l}{\% of Apps.\ using AI} & \multicolumn{3}{c}{---} & \multicolumn{3}{c}{14.03\%} \\
\multicolumn{1}{l}{\% of Workers ever using AI} & \multicolumn{3}{c}{---} & \multicolumn{3}{c}{16.89\%} \\
\addlinespace[2pt]
\cmidrule(lr){2-4} \cmidrule(lr){5-7}
& \textbf{Mean} & \textbf{Median} & \textbf{SD} & \textbf{Mean} & \textbf{Median} & \textbf{SD} \\
\multicolumn{1}{l}{Bid (\$)} & 149.43 & 140.00 & 65.91 & 142.92 & 140.00 & 64.43 \\
\addlinespace[2pt]
\multicolumn{1}{l}{Proposal Length (words)} & 89.03 & 79.00 & 49.60 & 110.67 & 104.00 & 55.64 \\
\addlinespace[2pt]
\multicolumn{1}{l}{Submit Time (mins)} & 204.59 & 10.58 & 388.09 & 210.43 & 8.83 & 370.22 \\
\addlinespace[2pt]
\cmidrule(lr){2-4} \cmidrule(lr){5-7}
\multicolumn{1}{l}{\emph{Modal 5 Worker Countries}} & \multicolumn{2}{l}{\textbf{Country}} & \multicolumn{1}{r}{\textbf{Share}} & \multicolumn{2}{l}{\textbf{Country}} & \multicolumn{1}{r}{\textbf{Share}} \\
& \multicolumn{2}{l}{India} & \multicolumn{1}{r}{36.48\%} & \multicolumn{2}{l}{India} & \multicolumn{1}{r}{30.38\%} \\
& \multicolumn{2}{l}{Pakistan} & \multicolumn{1}{r}{13.23\%} & \multicolumn{2}{l}{Pakistan} & \multicolumn{1}{r}{11.31\%} \\
& \multicolumn{2}{l}{Bangladesh} & \multicolumn{1}{r}{5.83\%} & \multicolumn{2}{l}{United States} & \multicolumn{1}{r}{10.84\%} \\
& \multicolumn{2}{l}{United States} & \multicolumn{1}{r}{3.63\%} & \multicolumn{2}{l}{Bangladesh} & \multicolumn{1}{r}{3.89\%} \\
& \multicolumn{2}{l}{Egypt} & \multicolumn{1}{r}{3.15\%} & \multicolumn{2}{l}{Germany} & \multicolumn{1}{r}{3.08\%} \\
\midrule
\multicolumn{7}{c}{\textbf{Job Post Summary Statistics}}\\
\midrule
& \multicolumn{3}{c}{\textbf{Pre-LLM}} & \multicolumn{3}{c}{\textbf{Post-LLM}} \\
\cmidrule(lr){2-4} \cmidrule(lr){5-7}
\multicolumn{1}{l}{\# Unique Job Posts} & \multicolumn{3}{c}{33{,}112} & \multicolumn{3}{c}{27{,}500} \\
\multicolumn{1}{l}{\# Unique Employers} & \multicolumn{3}{c}{19{,}573} & \multicolumn{3}{c}{16{,}131} \\
\addlinespace[2pt]
\multicolumn{1}{l}{\begin{tabular}[t]{@{}l@{}}Hiring Rate \\ \hspace{0.8em}Divided by Pre-LLM Rate\end{tabular}} & \multicolumn{3}{c}{1} & \multicolumn{3}{c}{0.97} \\
\multicolumn{1}{l}{\begin{tabular}[t]{@{}l@{}}Job Completion Rate \\ \hspace{0.8em}Divided by Pre-LLM Rate\end{tabular}} & \multicolumn{3}{c}{1} & \multicolumn{3}{c}{0.99} \\
\multicolumn{1}{l}{\begin{tabular}[t]{@{}l@{}}\% of Awarded Jobs \\ \hspace{0.8em}Where Lowest Bid Wins\end{tabular}} & \multicolumn{3}{c}{23.74\%} & \multicolumn{3}{c}{16.40\%} \\
\cmidrule(lr){2-4} \cmidrule(lr){5-7}
& \textbf{Mean} & \textbf{Median} & \textbf{SD} & \textbf{Mean} & \textbf{Median} & \textbf{SD} \\
\multicolumn{1}{l}{Job Post Length (words)} & 93.30 & 67.00 & 80.36 & 130.08 & 113.00 & 87.12 \\
\addlinespace[2pt]
\multicolumn{1}{l}{Apps.\ per Job Post} & 28.92 & 19.00 & 30.59 & 62.67 & 46.00 & 52.43 \\
\addlinespace[2pt]
\multicolumn{1}{l}{Winning Bid (\$)} & 113.24 & 100.00 & 63.78 & 107.98 & 100.00 & 62.85 \\
\addlinespace[2pt]
\cmidrule(lr){2-4} \cmidrule(lr){5-7}
\multicolumn{1}{l}{\emph{Modal 5 Employer Countries}} & \multicolumn{2}{l}{\textbf{Country}} & \multicolumn{1}{r}{\textbf{Share}} & \multicolumn{2}{l}{\textbf{Country}} & \multicolumn{1}{r}{\textbf{Share}} \\
& \multicolumn{2}{l}{United States} & \multicolumn{1}{r}{33.55\%} & \multicolumn{2}{l}{United States} & \multicolumn{1}{r}{33.55\%} \\
& \multicolumn{2}{l}{Saudi Arabia} & \multicolumn{1}{r}{4.14\%} & \multicolumn{2}{l}{Saudi Arabia} & \multicolumn{1}{r}{4.77\%} \\
& \multicolumn{2}{l}{United Kingdom} & \multicolumn{1}{r}{4.01\%} & \multicolumn{2}{l}{India} & \multicolumn{1}{r}{4.50\%} \\
& \multicolumn{2}{l}{India} & \multicolumn{1}{r}{3.94\%} & \multicolumn{2}{l}{United Kingdom} & \multicolumn{1}{r}{4.16\%} \\
& \multicolumn{2}{l}{United Arab Emirates} & \multicolumn{1}{r}{2.64\%} & \multicolumn{2}{l}{United Arab Emirates} & \multicolumn{1}{r}{3.42\%} \\
\bottomrule
\end{tabular}
\begin{tablenotes}[para,flushleft]
\scriptsize
\textit{Notes:} This table presents summary statistics for the pre-LLM and post-LLM samples. ``Rookies'' are workers on the platform who have not been hired for any jobs prior to the job post to which they are applying. The statistics on AI usage specifically refer to the use of the on-platform AI proposal-writing tool. We are not able to infer off-platform AI usage.
The variable ``Submit Time'' refers to the time between when a job post is created and when a worker submits her application. The statistic ``\% of Awarded Jobs Where Lowest Bid Wins'' refers to the share of job posts where a worker with the lowest bid was awarded the job, conditional on the job being awarded to any worker. The full distributions of bids and winning bids in each sample can be found in Figure~\ref{fig:bids_and_winning_bids_prepost}. As per our research agreement with \texttt{Freelancer.com}, we do not disclose win rates, hiring rates, or job completion rates in levels, so we present those rates normalized by their respective pre-LLM levels.
\end{tablenotes}
\end{threeparttable}
\end{table}

%% file: tables/logit_table_signal_edited.tex
\begin{table}[htbp]\centering
\begin{threeparttable}{
\notsosmallbutbigger
\def\sym#1{\ifmmode^{#1}\else\(^{#1}\)\fi}
{
\caption{Reduced-Form Demand for Signal: Pre-LLM vs. Post-LLM}\label{tab:logit_signal}
\begin{tabular}{l*{6}{c}}
\toprule
                &\multicolumn{2}{c}{Pre-LLM}          &\multicolumn{4}{c}{Post-LLM}                                               \\\cmidrule(lr){2-3}\cmidrule(lr){4-7}
                &\multicolumn{1}{c}{(1)}&\multicolumn{1}{c}{(2)}&\multicolumn{1}{c}{(3)}&\multicolumn{1}{c}{(4)}&\multicolumn{1}{c}{(5)}&\multicolumn{1}{c}{(6)}\\
                &\multicolumn{1}{c}{All}&\multicolumn{1}{c}{Considered}&\multicolumn{1}{c}{All}&\multicolumn{1}{c}{Considered}&\multicolumn{1}{c}{All}&\multicolumn{1}{c}{Considered}\\
\midrule
Signal          &   0.0881 &   0.0967 &   0.0250 &   0.0340 &   0.0138  &   0.0246 \\
                & [0.0030]         & [0.0031]         & [0.0051]         & [0.0050]         & [0.0061]         & [0.0062]         \\
\addlinespace
Bid Amount      &  -0.0103 &  -0.0105 &  -0.0101 &  -0.0103 &  -0.0100 &  -0.0102 \\
                & [0.0002]         & [0.0002]         & [0.0004]         & [0.0004]         & [0.0004]         & [0.0004]         \\
\addlinespace
Rep. Score      &   3.3521 &   2.0009 &   5.0119 &   2.9546 &   5.0424 &   2.9911 \\
                & [0.0732]         & [0.0732]         & [0.1960]         & [0.1954]         & [0.1956]         & [0.1955]         \\
\addlinespace
Log(Arrival Time)&  -0.0788 &   0.1673 &  -0.1144 &   0.2291 &  -0.1148 &   0.2285 \\
                & [0.0031]         & [0.0040]         & [0.0083]         & [0.0107]         & [0.0083]         & [0.0107]         \\
\addlinespace
AI-Written      &                  &                  &                  &                  &   0.2894 &   0.2365 \\
                &                  &                  &                  &                  & [0.0696]         & [0.0731]         \\
\midrule
Country Controls&\checkmark         &\checkmark         &\checkmark         &\checkmark         &\checkmark         &\checkmark         \\
WTP for $+$1 SD of Signal in \$ &  25.44         &  25.67         &  11.49         &  14.85         &   6.39         &  10.82         \\
WTP for $+$1 SD of Rep. Score in \$ &  48.71         &  27.96         &  78.04         &  46.24         &  79.28         &  47.22         \\
SD of Bid       &  65.91         &  65.68         &  64.25         &  65.31         &  64.25         &  65.31         \\
N Projects      &   33,105         &   33,109         &    5,476         &    5,344         &    5,476         &    5,344         \\
N Applications  &  990,001         &  434,756         &  438,736         &  102,117         &  438,736         &  102,117         \\
\bottomrule
\end{tabular}
\vspace{0.2cm}
    \begin{tablenotes}[para,flushleft] \footnotesize \textit{Notes:} This table presents a series of estimated multinomial logit models that predict reduced-form demand for signal in both the pre-LLM and post-LLM samples. The post-LLM sample corresponds to the subsample of post-LLM job posts that were posted after March 26, 2024. Each column represents a different specification using various controls and sample definitions. All specifications include controls for a worker's country of residence, grouped into seven broad categories. Columns (1), (3), and (5) use all applications, while columns (2), (4), and (6) only use applications that were measured to be in the employer's consideration set. Columns (1) and (2) are estimated using the pre-LLM sample, while columns (3)-(6) are estimated using the post-LLM sample. All standard errors are clustered at the job-post-level and displayed in brackets next to the estimates. In lieu of elasticities, we compute the increase in an application's bid, all else equal, needed to achieve the same predicted increase in demand as a standard deviation increase in signal and reputation score according to the following: WTP for $+$1 SD of variable in \$ $= \text{SD}(\text{variable}) \times \frac{\beta(\text{variable})}{|\beta(\text{Bid Amount})|}$.
\end{tablenotes}
}} 
\end{threeparttable}
\end{table}

%% file: tables/reg_s_on_log_e.tex
\begin{table}[htbp]\centering
\begin{threeparttable}
\small
\caption{Signal on Effort Regressions: Pre‑LLM vs. Post‑LLM}\label{tab:reg_s_on_log_e}
\begin{tabular}{l*{6}{c}}
\toprule
\multicolumn{7}{c}{Dependent Variable: Signal}\\ 
\midrule
          &\multicolumn{2}{c}{Pre‑LLM}&\multicolumn{2}{c}{Post‑LLM, Not AI}&\multicolumn{2}{c}{Post-LLM, AI-Written}\\\cmidrule(lr){2-3}\cmidrule(lr){4-5}\cmidrule(lr){6-7}
          &\multicolumn{1}{c}{(1)}&\multicolumn{1}{c}{(2)}&\multicolumn{1}{c}{(3)}&\multicolumn{1}{c}{(4)}&\multicolumn{1}{c}{(5)}&\multicolumn{1}{c}{(6)}\\

\midrule\addlinespace
Log(Effort)&   0.8683&   0.6246&   1.1339&   0.5044&  -0.9731&  -0.3483\\
          & [0.0031]& [0.0039]& [0.0075]& [0.0085]& [0.0139]& [0.0183]\\
\addlinespace Constant&   5.5062&   5.5011&   7.6710&   7.2718&  11.1838&  12.0064\\
          & [0.0041]& [0.0038]& [0.0104]& [0.0084]& [0.0196]& [0.0211]\\
\addlinespace\midrule
\addlinespace Worker FEs&         &$\checkmark$&         &$\checkmark$&         &$\checkmark$\\
\addlinespace Adj. $R^{2}$&    0.129&    0.391&    0.087&    0.643&    0.093&    0.168\\
\addlinespace N&  571,952&  531,607&  244,472&  229,971&   63,908&   58,221\\
\addlinespace
\bottomrule
\end{tabular}
\vspace{0.2cm}
\begin{tablenotes}[para,flushleft]
\footnotesize \textit{Notes:} This table presents estimates from a regression of our signal measure on log effort. Columns (1) and (2) are estimated on the pre‑LLM sample, columns (3) and (4) are estimated on the applications from the post‑LLM sample that did not employ the on-platform AI writing tool, and columns (5) and (6) are estimated on the applications from the post‑LLM sample that did employ the AI writing tool. Columns (1), (3), and (5) use only log effort as a regressor. Columns (2), (4), and (6) include worker fixed effects and are estimated only on the subset of applications from workers with multiple applications that have non-missing effort and signal measurements within the respective sample. The post-LLM sample corresponds to the subsample of post-LLM job posts that were posted after March 26, 2024. Heteroskedasticity-robust standard errors are in brackets.
\end{tablenotes}
\end{threeparttable}
\end{table}

%% file: tables/reg_outcomes_on_signal.tex
\begin{table}[!h]\centering
\begin{threeparttable}
\small
\caption{5‑Star Completion on Signal and Controls: Pre‑LLM vs. Post‑LLM}\label{tab:reg_5stars_on_signal}
\begin{tabular}{l*{6}{c}}
\toprule
\multicolumn{7}{c}{Dependent Variable: Completed with 5 Stars}\\ 
\midrule
          &\multicolumn{3}{c}{Pre‑LLM}  &\multicolumn{3}{c}{Post‑LLM} \\\cmidrule(lr){2-4}\cmidrule(lr){5-7}
          &\multicolumn{1}{c}{(1)}&\multicolumn{1}{c}{(2)}&\multicolumn{1}{c}{(3)}&\multicolumn{1}{c}{(4)}&\multicolumn{1}{c}{(5)}&\multicolumn{1}{c}{(6)}\\

\midrule\addlinespace
Signal    &   0.0174&   0.0158&   0.0093&  -0.0001&   0.0025&   0.0026\\
          & [0.0016]& [0.0017]& [0.0018]& [0.0026]& [0.0026]& [0.0026]\\
\addlinespace Reputation Score&         &   0.5336&   0.6040&         &   0.4335&   0.4743\\
          &         & [0.0365]& [0.0367]&         & [0.0824]& [0.0819]\\
\addlinespace Bid (\$)&         &  -0.0008&  -0.0008&         &  -0.0008&  -0.0008\\
          &         & [0.0001]& [0.0001]&         & [0.0002]& [0.0002]\\
\addlinespace Log(Arrival Time)&         &   0.0022&  -0.0034&         &  -0.0039&  -0.0087\\
          &         & [0.0024]& [0.0025]&         & [0.0053]& [0.0054]\\
\addlinespace Log(Effort)&         &         &   0.0502&         &         &   0.0462\\
          &         &         & [0.0053]&         &         & [0.0109]\\
\addlinespace\midrule
\addlinespace Country Controls&         &$\checkmark$&$\checkmark$&         &$\checkmark$&$\checkmark$\\
\addlinespace Adj. $R^{2}$&    0.013&    0.048&    0.058&   -0.001&    0.029&    0.039\\
\addlinespace N&    8,185&    8,185&    8,185&    1,659&    1,659&    1,659\\
\addlinespace
\bottomrule
\end{tabular}
\vspace{0.2cm}
\begin{tablenotes}[para,flushleft]
\footnotesize \textit{Notes:} This table presents estimates from a regression of whether the hired worker completes the job with 5 stars on signal and various controls. Columns (1)–(3) are estimated on winning applications from the pre‑LLM sample, and columns (4)–(6) are estimated on the winning applications from the post‑LLM sample. Columns (1) and (4) use only signal as a regressor. Columns (2) and (5) include various controls that were likely to be predictive of demand, including reputation score, bid, log arrival time, and country fixed effects. Columns (3) and (6) include the same regressors as columns (2) and (5), and they additionally include log effort as a regressor. Since columns (3) and (6) could only be estimated on those applications with non-missing effort, we apply this sampling restriction equally to all columns. The post-LLM sample corresponds to the subsample of post-LLM job posts that were posted after March 26, 2024. Heteroskedasticity-robust standard errors are in brackets. As per our agreement with \texttt{Freelancer.com}, we omit the constant term so that we do not disclose completion rates.
\end{tablenotes}
\end{threeparttable}
\end{table}

%% file: tables/supply_estimates_sum_stats.tex
\begin{table}[htbp]
\centering
\begin{threeparttable}
\caption{Supply Estimates}\label{tab:supply_estimates}
\begin{tabular}{lcccccc}
\toprule
& \textbf{Mean} & \textbf{SD} & \textbf{P20} & \textbf{P40} & \textbf{P60} & \textbf{P80} \\
\midrule
Worker Cost   & \$46.26  & \$73.81  & -\$13.79  & \$38.28  & \$50.18 & \$107.85 \\
\vspace{0.5mm} \\
Ability in \$  & \$31.83  & \$52.16  & -\$16.60  & \$13.82  & \$44.96  & \$80.85 \\
\bottomrule
\end{tabular}
\begin{tablenotes}[para,flushleft] 
\footnotesize
Notes: This table shows summary statistics of the marginal distributions of estimated worker costs and abilities. Ability is shown in dollars of employers' willingness to pay (WTP) computed as $\frac{\widehat{\beta}}{|\widehat{\alpha}|}\cdot\widehat{a}$, where $\widehat{\beta}$ is the estimated coefficient on ability and $\widehat{\alpha}$ is the estimated coefficient on bid, both from estimated employer demand. We only obtain estimates of worker cost and ability for those workers who we observe a valid effort measurement.
\end{tablenotes}
\end{threeparttable}
\end{table}

%% file: tables/demand_estimates_sum_stats.tex
\begin{table}[htbp]
\centering
\begin{threeparttable}
\caption{Demand Estimates}\label{tab:demand_estimates}
\begin{tabular}{l@{\hspace{6mm}}lc}
\toprule
\multicolumn{3}{c}{\textbf{Main Demand Parameters}} \\
\midrule
\multicolumn{1}{c}{\textbf{Parameter}} & \textbf{Value} & \textbf{Description} \\
\midrule
\addlinespace[0.75em]
\multicolumn{1}{c}{$|\alpha|$}   & 0.0110 & Price Coefficient \\
\addlinespace[0.5em]
\multicolumn{1}{c}{$\beta$}      & 0.1644 & Ability Coefficient \\
\addlinespace[0.5em]
\multicolumn{1}{c}{$1-\pi$}      & 42.51\% & Abandon Rate \\
\addlinespace[0.5em]
\midrule
\multicolumn{3}{c}{$\mathbf{T(x)}$ \textbf{by Group Summary Statistics}} \\
\midrule
\textbf{Group} & \textbf{Mean} & \textbf{Standard Deviation} \\
\midrule
\addlinespace[0.5em]
Overall              & \$-7.03  & \$62.58 \\
\addlinespace[0.5em]
\multicolumn{3}{l}{\textit{\underline{Reputation Groups}}} \\
\addlinespace[0.5em]
Rookie               & \$-33.58 & \$59.84 \\
Low                  & \$-28.88 & \$53.46 \\
Middle              & \$-7.65  & \$45.11 \\
High                 & \$53.69  & \$43.04 \\
\addlinespace[0.5em]
\multicolumn{3}{l}{\textit{\underline{Arrival Groups}}} \\
0--5 mins.           & \$-52.61 & \$56.88 \\
5--45 mins.          & \$-5.83  & \$48.15 \\
Spike Arrivals       & \$-1.14  & \$54.82 \\
$>$55 mins.          & \$43.17  & \$44.92 \\
\addlinespace[0.5em]
\multicolumn{3}{l}{\textit{\underline{Country Groups}}} \\
South Asia           & \$-48.78 & \$57.21 \\
English Speaking     & \$0.67   & \$54.11 \\
Europe               & \$26.26  & \$54.82 \\
Other                & \$5.44   & \$55.17 \\
\addlinespace[0.5em]
\bottomrule
\end{tabular}
\begin{tablenotes}[para,flushleft]
\footnotesize
Notes: This table presents the estimated demand parameters from structural estimation described in Section \ref{estimation_demand}. The main parameters of interest are the price coefficient $\alpha$, and the ability coefficient $\beta$. Also included is the estimate of the abandon rate $(1-\pi)$, as well as summary statistics of the distribution of estimated group-level ``fixed effects'' $T(x)$, which we report in dollars of employers' willingness to pay (WTP) computed as $\widehat{T(x)} / |\widehat{\alpha}|$. The summary statistics of $T(x)$ are aggregated by three different groupings corresponding to the three dimensions on which we group workers into $x$-groups: reputation, arrival time, and country of residence. These groups aggregate across the 56 raw estimates of $T(x)$ by the Cartesian product of reputation group, arrival group, and country group.
\end{tablenotes}
\end{threeparttable}
\end{table}

%% file: tables/coding_skill_tags.tex
\begin{table}[h!]
\centering
\begin{threeparttable}
\notsosmallbutbigger
\caption{Tagged Coding Skills in Sample (Top 15)}\label{tab:coding_skills}
\begin{tabular}{l c | l c}
\noalign{\hrule height 0.7pt}
\addlinespace[3pt]
\multicolumn{2}{c|}{\textbf{Pre-LLM Sample}} & \multicolumn{2}{c}{\textbf{Post-LLM Sample}} \\
\noalign{\hrule height 0.5pt}
\addlinespace[3pt]
\textbf{Skill Tag} & \textbf{Fraction of Jobs with Tag} & \textbf{Skill Tag} & \textbf{Fraction of Jobs with Tag} \\
\addlinespace[3pt]
\noalign{\hrule height 0.5pt}
\addlinespace[3pt]
PHP & 49.61\% & PHP & 51.11\% \\
Website Design & 14.03\% & Website Design & 14.33\% \\
JavaScript & 11.57\% & JavaScript & 10.60\% \\
Python & 9.17\% & Python & 8.91\% \\
Mobile App Development & 7.60\% & Mobile App Development & 6.72\% \\
Java & 4.49\% & C Programming & 4.03\% \\ 
C Programming & 4.22\% & Java & 3.94\% \\
Software Architecture & 3.09\% & Software Architecture & 2.48\% \\
C\# Programming & 1.94\% & C\# Programming & 1.67\% \\
.NET & 1.65\% & .NET & 1.37\% \\
Web Scraping & 1.29\% & Node.js & 1.09\% \\
Node.js & 1.21\% & Web Scraping & 1.05\% \\
C++ Programming & 0.79\% & C++ Programming & 0.66\% \\
React.js & 0.78\% & React.js & 0.76\% \\
Amazon Web Services & 0.56\% & Amazon Web Services & 0.57\% \\
\noalign{\hrule height 0.7pt}
\end{tabular}
\begin{tablenotes}[para,flushleft]
\footnotesize
\textit{Notes:} This table shows the top 15 most common skill tags associated with job posts in our pre-LLM and post-LLM samples. The first column lists the skill tag as they appear on the platform. The second column displays the fraction of projects in our samples with each skill tag. Note that as projects can have multiple skill tags, these fractions, if extended to the full list of skill tags, would sum to over 100\%. We first classified all tags that appear in the pre-LLM data as being coding-related or not. We then kept any job post in our sample that had at least one of these pre-LLM coding-related skill tags. In total, there were 147 tags that merged to the job posts that satisfied all of our other sample restrictions. Note to make the job posts comparable across the pre-LLM and post-LLM samples, we do not consider coding-related skill tags that only appear in the post-LLM sample, e.g., anything related to generative AI or LLMs.
\end{tablenotes}
\end{threeparttable}
\end{table}

%% file: tables/considered_sum_stats.tex
\begin{table}[h!]
\centering
\begin{threeparttable}
\notsosmallbutbigger
\caption{Consideration Set Summary Statistics}\label{tab:consideration_sum_stats}
\setlength{\tabcolsep}{8pt}
\begin{tabular}{@{}l ccc ccc@{}}
\toprule
\multicolumn{7}{c}{\textbf{Bid-Level Consideration Set Statistics}}\\
\midrule
& \multicolumn{3}{c}{\textbf{Pre-LLM}} & \multicolumn{3}{c}{\textbf{Post-LLM}} \\
\cmidrule(lr){2-4} \cmidrule(lr){5-7}
\multicolumn{1}{l}{\begin{tabular}[t]{@{}l@{}}Consideration Rate \\ \hspace{0.8em}Divided by Pre-LLM Rate\end{tabular}} & \multicolumn{3}{c}{1} & \multicolumn{3}{c}{0.66} \\
\addlinespace[2pt]
\multicolumn{1}{l}{\begin{tabular}[t]{@{}l@{}}Win Rate $\mid$ Considered \\ \hspace{0.8em}Divided by Pre-LLM Rate\end{tabular}} & \multicolumn{3}{c}{1} & \multicolumn{3}{c}{0.68} \\
\addlinespace[4pt]
\midrule
\multicolumn{7}{c}{\textbf{Job-Level Consideration Set Statistics}}\\
\midrule
& \multicolumn{3}{c}{\textbf{Pre-LLM}} & \multicolumn{3}{c}{\textbf{Post-LLM}} \\
\cmidrule(lr){2-4} \cmidrule(lr){5-7}
\multicolumn{1}{l}{} & \textbf{Mean} & \textbf{Median} & \textbf{SD} & \textbf{Mean} & \textbf{Median} & \textbf{SD} \\
\multicolumn{1}{l}{\# Apps. Considered per Job} & 12.13 & 11.00 & 6.63 & 17.38 & 17.00 & 8.86 \\
\addlinespace[2pt]
\multicolumn{1}{l}{\begin{tabular}[t]{@{}l@{}}Consideration Rate per Job \\ \hspace{0.8em}Divided by Pre-LLM Mean Rate\end{tabular}} & \multicolumn{1}{c}{1} & \multicolumn{1}{c}{1.03} & \multicolumn{1}{c}{0.43} & \multicolumn{1}{c}{0.70} & \multicolumn{1}{c}{0.64} & \multicolumn{1}{c}{0.45} \\
\addlinespace[4pt]
\midrule
\multicolumn{7}{c}{\textbf{Consideration Set Composition}}\\
\midrule
& \multicolumn{3}{c}{\textbf{Pre-LLM}} & \multicolumn{3}{c}{\textbf{Post-LLM}} \\
\cmidrule(lr){2-4} \cmidrule(lr){5-7}
\multicolumn{1}{l}{\% Rookies $\mid$ Considered} & \multicolumn{3}{c}{8.01\%} & \multicolumn{3}{c}{14.97\%} \\
\multicolumn{1}{l}{\% AI-Written $\mid$ Considered} & \multicolumn{3}{c}{---} & \multicolumn{3}{c}{12.53\%} \\
\addlinespace[2pt]
\cmidrule(lr){2-4} \cmidrule(lr){5-7}
\multicolumn{1}{l}{} & \textbf{Mean} & \textbf{Median} & \textbf{SD} & \textbf{Mean} & \textbf{Median} & \textbf{SD} \\
\multicolumn{1}{l}{Bid (\$) $\mid$ Considered} & 148.36 & 140.00 & 65.68 & 139.80 & 140.00 & 65.01 \\
\addlinespace[2pt]
\multicolumn{1}{l}{Proposal Length (words) $\mid$ Considered} & 86.54 & 77.00 & 48.63 & 109.54 & 101.00 & 56.97 \\
\addlinespace[2pt]
\multicolumn{1}{l}{Submit Time (mins) $\mid$ Considered} & 63.01 & 2.57 & 219.77 & 53.28 & 1.42 & 199.47 \\
\addlinespace[4pt]
\cmidrule(lr){2-4} \cmidrule(lr){5-7}
\multicolumn{1}{l}{\emph{Modal 5 Worker \ Countries $\mid$ Considered}} & \multicolumn{2}{l}{\textbf{Country}} & \multicolumn{1}{r}{\textbf{Share}} & \multicolumn{2}{l}{\textbf{Country}} & \multicolumn{1}{r}{\textbf{Share}} \\
& \multicolumn{2}{l}{India} & \multicolumn{1}{r}{36.51\%} & \multicolumn{2}{l}{Pakistan} & \multicolumn{1}{r}{32.93\%} \\
& \multicolumn{2}{l}{Pakistan} & \multicolumn{1}{r}{25.36\%} & \multicolumn{2}{l}{India} & \multicolumn{1}{r}{29.02\%} \\
& \multicolumn{2}{l}{Ukraine} & \multicolumn{1}{r}{5.06\%} & \multicolumn{2}{l}{United States} & \multicolumn{1}{r}{6.85\%} \\
& \multicolumn{2}{l}{Russia} & \multicolumn{1}{r}{4.73\%} & \multicolumn{2}{l}{Ukraine} & \multicolumn{1}{r}{5.65\%} \\
& \multicolumn{2}{l}{United States} & \multicolumn{1}{r}{3.39\%} & \multicolumn{2}{l}{Bangladesh} & \multicolumn{1}{r}{3.11\%} \\
\bottomrule
\end{tabular}
\begin{tablenotes}[para,flushleft]
\footnotesize
\textit{Notes:} This table presents summary statistics on our measured consideration sets for the pre-LLM and post-LLM samples. “Considered” means the application appears in the employer’s consideration set. The statistic ``Consideration Rate per Job'' is the share of applications to each job that are considered. All statistics under ``Consideration Set Composition'' are calculated across all considered applications, and are not calculated at the job post level, so for example, the standard deviation of bid across all considered applications in the pre-LLM sample is \$65.68. As per our research agreement with \texttt{Freelancer.com}, we do not disclose consideration rates or win rates conditional on consideration in levels, so we present those rates normalized by their respective pre-LLM levels. Moreover, they have requested that we do not disclose statistics on consideration rates per job in levels, so we present the mean, median, and standard deviation of consideration rates per job normalized by the pre-LLM mean consideration rate per job.
\end{tablenotes}
\end{threeparttable}
\end{table}

%% file: tables/logit_table_custom_edited.tex
\begin{table}[htbp]\centering
\begin{threeparttable}{
\notsosmallbutbigger
{
\caption{Reduced-Form Demand for Signal (Custom Criteria Only): Pre-LLM vs. Post-LLM}\label{tab:logit_signal_custom}
\begin{tabular}{l*{6}{c}}
\toprule
                &\multicolumn{2}{c}{Pre-LLM}          &\multicolumn{4}{c}{Post-LLM}                                               \\\cmidrule(lr){2-3}\cmidrule(lr){4-7}
                &\multicolumn{1}{c}{(1)}&\multicolumn{1}{c}{(2)}&\multicolumn{1}{c}{(3)}&\multicolumn{1}{c}{(4)}&\multicolumn{1}{c}{(5)}&\multicolumn{1}{c}{(6)}\\
                &\multicolumn{1}{c}{All}&\multicolumn{1}{c}{Considered}&\multicolumn{1}{c}{All}&\multicolumn{1}{c}{Considered}&\multicolumn{1}{c}{All}&\multicolumn{1}{c}{Considered}\\
\midrule
Sum of Custom Criteria &   0.1388&   0.1490&   0.0414&   0.0534&   0.0267&   0.0410\\\
                & [0.0043]         & [0.0044]         & [0.0072]         & [0.0072]         & [0.0086]         & [0.0087]         \\
\addlinespace
Bid Amount      &  -0.0102&  -0.0104&  -0.0101&  -0.0103&  -0.0100&  -0.0102\\\
                & [0.0002]         & [0.0002]         & [0.0004]         & [0.0004]         & [0.0004]         & [0.0004]         \\
\addlinespace
Rep. Score      &   3.3690&   2.0227&   5.0257&   2.9687&   5.0524&   3.0011\\\
                & [0.0728]         & [0.0728]         & [0.1960]         & [0.1953]         & [0.1956]         & [0.1955]         \\
\addlinespace
Log(Arrival Time)&  -0.0797&   0.1660&  -0.1148&   0.2284&  -0.1152&   0.2278\\\
                & [0.0032]         & [0.0040]         & [0.0083]         & [0.0107]         & [0.0083]         & [0.0107]         \\
\addlinespace
AI-Written      &                  &                  &                  &                  &   0.2655&   0.2187\\\
                &                  &                  &                  &                  & [0.0685]         & [0.0720]         \\
\midrule
Country Controls&\checkmark         &\checkmark         &\checkmark         &\checkmark         &\checkmark         &\checkmark         \\
WTP for $+$1 SD of Signal in \$ &  27.4648         &  26.9244         &  13.0157         &  15.9261         &   8.4912         &  12.3177         \\
WTP for $+$1 SD of Rep. Score in \$ &  49.3732         &  28.5236         &  78.1078         &  46.3952         &  79.2080         &  47.2766         \\
SD of Bid       &  65.9050         &  65.6786         &  64.2486         &  65.3095         &  64.2486         &  65.3095         \\
N Projects      &   33,105         &   33,109         &    5,476         &    5,344         &    5,476         &    5,344         \\
N Applications  &  990,001         &  434,756         &  438,736         &  102,117         &  438,736         &  102,117         \\
\bottomrule
\end{tabular}
\vspace{0.2cm}
    \begin{tablenotes}[para,flushleft] \footnotesize \textit{Notes:} This table presents a series of estimated multinomial logit models that predict reduced-form demand for the sum of custom criteria in both the pre-LLM and post-LLM samples. The post-LLM sample corresponds to the subsample of post-LLM job posts that were posted after March 26, 2024. Each column represents a different specification using various controls and sample definitions. All specifications include controls for a worker's country of residence, grouped into seven broad categories. Columns (1), (3), and (5) use all applications, while columns (2), (4), and (6) only use applications that were measured to be in the employer's consideration set. Columns (1) and (2) are estimated using the pre-LLM sample, while columns (3)-(6) are estimated using the post-LLM sample. All standard errors are clustered at the job-post-level and displayed in brackets next to the estimates. In lieu of elasticities, we compute the increase in an application's bid, all else equal, needed to achieve the same predicted increase in demand as a standard deviation increase in the sum of custom criteria and reputation score according to the following: WTP for $+$1 SD of variable in \$ util $= \text{SD}(\text{variable}) \times \frac{\beta(\text{variable})}{|\beta(\text{Bid Amount})|}$.
\end{tablenotes}
}} 
\end{threeparttable}
\end{table}

%% file: tables/logit_table_generic_edited.tex
\begin{table}[htbp]\centering
\begin{threeparttable}{
\notsosmallbutbigger
{
\caption{Reduced-Form Demand for Signal (Generic Criteria Only): Pre-LLM vs. Post-LLM}\label{tab:logit_signal_generic}
\begin{tabular}{l*{6}{c}}
\toprule
                &\multicolumn{2}{c}{Pre-LLM}          &\multicolumn{4}{c}{Post-LLM}                                               \\\cmidrule(lr){2-3}\cmidrule(lr){4-7}
                &\multicolumn{1}{c}{(1)}&\multicolumn{1}{c}{(2)}&\multicolumn{1}{c}{(3)}&\multicolumn{1}{c}{(4)}&\multicolumn{1}{c}{(5)}&\multicolumn{1}{c}{(6)}\\
                &\multicolumn{1}{c}{All}&\multicolumn{1}{c}{Considered}&\multicolumn{1}{c}{All}&\multicolumn{1}{c}{Considered}&\multicolumn{1}{c}{All}&\multicolumn{1}{c}{Considered}\\
\midrule
Generic Criteria&   0.0275&   0.0503&  -0.0035         &   0.0283         &  -0.0401 &  -0.0081         \\
                & [0.0070]         & [0.0071]         & [0.0185]         & [0.0189]         & [0.0194]         & [0.0201]         \\
\addlinespace
Bid Amount      &  -0.0102&  -0.0105&  -0.0099&  -0.0101&  -0.0098&  -0.0100\\
                & [0.0002]         & [0.0002]         & [0.0004]         & [0.0004]         & [0.0004]         & [0.0004]         \\
\addlinespace
Rep. Score      &   3.5310&   2.2230&   4.9475&   2.8846&   5.0288&   2.9772\\
                & [0.0732]         & [0.0730]         & [0.1955]         & [0.1952]         & [0.1950]         & [0.1956]         \\
\addlinespace
Log(Arrival Time)&  -0.0689&   0.1809&  -0.1114&   0.2346&  -0.1131&   0.2314\\
                & [0.0031]         & [0.0039]         & [0.0082]         & [0.0107]         & [0.0082]         & [0.0107]         \\
\addlinespace
AI-Written      &                  &                  &                  &                  &   0.4152&   0.3927\\
                &                  &                  &                  &                  & [0.0633]         & [0.0661]         \\
\midrule
Country Controls&\checkmark         &\checkmark         &\checkmark         &\checkmark         &\checkmark         &\checkmark         \\
WTP for $+$1 SD of Signal in \$ &   3.7382         &   6.5842         &  -0.4823         &   3.8457         &  -5.5679         &  -1.1032         \\
WTP for $+$1 SD of Rep. Score in \$ &  51.4391         &  31.0209         &  78.5630         &  46.0948         &  80.3884         &  47.9101         \\
SD of Bid       &  65.9050         &  65.6786         &  64.2486         &  65.3095         &  64.2486         &  65.3095         \\
N Projects      &   33,105         &   33,109         &    5,476         &    5,344         &    5,476         &    5,344         \\
N Applications  &  990,001         &  434,756         &  438,736         &  102,117         &  438,736         &  102,117         \\
\bottomrule
\end{tabular}
\vspace{0.2cm}
    \begin{tablenotes}[para,flushleft] \footnotesize \textit{Notes:} This table presents a series of estimated multinomial logit models that predict reduced-form demand for the sum of generic criteria in both the pre-LLM and post-LLM samples. The post-LLM sample corresponds to the subsample of post-LLM job posts that were posted after March 26, 2024. Each column represents a different specification using various controls and sample definitions. All specifications include controls for a worker's country of residence, grouped into seven broad categories. Columns (1), (3), and (5) use all applications, while columns (2), (4), and (6) only use applications that were measured to be in the employer's consideration set. Columns (1) and (2) are estimated using the pre-LLM sample, while columns (3)-(6) are estimated using the post-LLM sample. All standard errors are clustered at the job-post-level and displayed in brackets next to the estimates. In lieu of elasticities, we compute the increase in an application's bid, all else equal, needed to achieve the same predicted increase in demand as a standard deviation increase in the sum of generic criteria and reputation score according to the following: WTP for $+$1 SD of variable in \$ util $= \text{SD}(\text{variable}) \times \frac{\beta(\text{variable})}{|\beta(\text{Bid Amount})|}$.
\end{tablenotes}
}} 
\end{threeparttable}
\end{table}

%% file: tables/reg_outcomes_on_signal_effcorr.tex
\begin{table}[!h]\centering
\begin{threeparttable}
\small
\caption{5‑Star Completion on Signal and Controls Using Corrected Effort}\label{tab:reg_5stars_on_signal_effcorr}
\begin{tabular}{l*{4}{c}}
\toprule
\multicolumn{5}{c}{Dependent Variable: Completed with 5 Stars}\\ 
\midrule
          &\multicolumn{1}{c}{(1)}&\multicolumn{1}{c}{(2)}&\multicolumn{1}{c}{(3)}&\multicolumn{1}{c}{(4)}\\

\midrule\addlinespace
Signal    &   0.0174&   0.0158&   0.0039&   0.0047\\
          & [0.0016]& [0.0017]& [0.0020]& [0.0020]\\
\addlinespace Reputation Score&         &   0.5336&   0.5818&   0.6112\\
          &         & [0.0365]& [0.0363]& [0.0368]\\
\addlinespace Bid (\$)&         &  -0.0008&  -0.0008&  -0.0010\\
          &         & [0.0001]& [0.0001]& [0.0001]\\
\addlinespace Log(Arrival Time)&         &   0.0022&  -0.0024&  -0.0006\\
          &         & [0.0024]& [0.0025]& [0.0024]\\
\addlinespace Log(Corrected Effort)&         &         &   0.0380&         \\
          &         &         & [0.0038]&         \\
\addlinespace Estimated Ability&         &         &         &   0.0188\\
          &         &         &         & [0.0020]\\
\addlinespace\midrule
\addlinespace Country Controls&         &$\checkmark$&$\checkmark$&$\checkmark$\\
\addlinespace Adj. $R^{2}$&    0.013&    0.048&    0.059&    0.058\\
\addlinespace N&    8,185&    8,185&    8,185&    8,185\\
\addlinespace
\bottomrule
\end{tabular}
\vspace{0.2cm}
\begin{tablenotes}[para,flushleft]
\footnotesize \textit{Notes:} This table presents estimates from a regression of whether the hired worker completes the job with 5 stars on signal and various controls. All columns are estimated on winning applications from the pre‑LLM sample. Column (1) uses only signal as a regressor. Column (2) includes various controls that were likely to be predictive of demand, including reputation score, bid, log arrival time, and country fixed effects. Column (3) includes the same regressors as column (2), and it additionally includes log corrected effort as a regressor. Column (4) includes the same regressors as column (2), and it additionally includes estimated ability as a regressor. This table is identical to Table~\ref{tab:reg_5stars_on_signal}, except that it uses the corrected effort measure in column (3) instead of the original effort measure and it includes estimated ability in column (4). The corrected effort measurement accounts for worker fixed effects in signal production and is described in Sections\ref{estimation_signal_production} and Appendix~\ref{appendix_effort_correction}. Estimated ability comes from our structural supply estimates, the results of which are reported in Section~\ref{est_results_supply}. Since columns (3) and (4) could only be estimated on those applications with non-missing effort, we apply this sampling restriction equally to all columns. Heteroskedasticity-robust standard errors are in brackets. As per our agreement with \texttt{Freelancer.com}, we omit the constant term so that we do not disclose completion rates.
\end{tablenotes}
\end{threeparttable}
\end{table}

%% file: tables/cost_ability_by_x.tex
\begin{table}[!ht]
\centering
\begin{threeparttable}
\scriptsize
\setlength{\tabcolsep}{3pt}
\begin{tabular}{lllcccc}
\caption{Marginal Supply Estimates by Observable Group}\label{tab:supply_estimates_by_x_group}\\
\toprule
\textbf{Country Group} & \textbf{Arrival Group} & \textbf{Reputation Group} & \textbf{Mean(Ability)} & \textbf{SD(Ability)} & \textbf{Mean(Cost)} & \textbf{SD(Cost)} \\
\midrule
English Speaking & $0$--$5$ mins. & Rookie  & 30.35 & 47.26 & 38.96 & 66.47 \\
English Speaking & $0$--$5$ mins. & Low     & 53.28 & 48.61 & 49.66 & 62.63 \\
English Speaking & $0$--$5$ mins. & Middle & 27.22 & 49.42 & 55.38 & 65.31 \\
English Speaking & $0$--$5$ mins. & High    & 10.52 & 41.82 & 27.99 & 92.79 \\
English Speaking & $5$--$45$ mins. & Rookie  & 39.81 & 43.55 & 35.17 & 73.61 \\
English Speaking & $5$--$45$ mins. & Low     & 24.31 & 50.45 & 38.75 & 73.31 \\
English Speaking & $5$--$45$ mins. & Middle & 28.86 & 53.30 & 54.46 & 66.67 \\
English Speaking & $5$--$45$ mins. & High    & 37.75 & 45.25 & 58.15 & 79.23 \\
English Speaking & $>45$ mins. & Rookie  & 46.77 & 49.81 & 28.41 & 83.85 \\
English Speaking & $>45$ mins. & Low     & 24.45 & 56.74 & 30.22 & 73.93 \\
English Speaking & $>45$ mins. & Middle & 23.56 & 59.25 & 54.69 & 72.96 \\
English Speaking & $>45$ mins. & High    & 40.41 & 55.16 & 53.78 & 84.86 \\
South Asia       & Spike Arrivals & Rookie  & 46.28 & 47.67 & 41.48 & 61.59 \\
South Asia       & Spike Arrivals & Low     & 24.16 & 51.98 & 45.95 & 68.96 \\
South Asia       & Spike Arrivals & Middle & 31.78 & 50.48 & 50.75 & 72.82 \\
South Asia       & Spike Arrivals & High    & 24.21 & 50.36 & 45.91 & 82.97 \\
Europe           & $0$--$5$ mins. & Rookie  & 26.02 & 45.92 & 20.18 & 64.69 \\
Europe           & $0$--$5$ mins. & Low     & 29.58 & 46.74 & 37.89 & 56.05 \\
Europe           & $0$--$5$ mins. & Middle & 21.25 & 45.79 & 47.37 & 56.94 \\
Europe           & $0$--$5$ mins. & High    &  1.79 & 42.45 & 33.99 & 62.97 \\
Europe           & $5$--$45$ mins. & Rookie  & 43.35 & 45.69 & 21.17 & 70.35 \\
Europe           & $5$--$45$ mins. & Low     & 29.19 & 48.97 & 26.91 & 69.73 \\
Europe           & $5$--$45$ mins. & Middle & 29.69 & 48.62 & 46.06 & 63.55 \\
Europe           & $5$--$45$ mins. & High    & 33.65 & 51.01 & 16.03 & 88.70 \\
Europe           & $>45$ mins. & Rookie  & 41.30 & 46.65 & 15.86 & 80.96 \\
Europe           & $>45$ mins. & Low     & 23.63 & 52.40 & 26.56 & 77.24 \\
Europe           & $>45$ mins. & Middle & 27.60 & 51.69 & 44.01 & 68.44 \\
Europe           & $>45$ mins. & High    & -11.85 & 56.35 & -41.34 & 152.09 \\
Not South Asia   & Spike Arrivals & Rookie  & 34.80 & 47.07 & 26.95 & 67.25 \\
Not South Asia   & Spike Arrivals & Low     & 23.53 & 48.27 & 44.24 & 65.69 \\
Not South Asia   & Spike Arrivals & Middle & 27.25 & 48.67 & 45.18 & 64.53 \\
Not South Asia   & Spike Arrivals & High    &  9.60 & 45.56 & 39.58 & 84.01 \\
Other            & $0$--$5$ mins. & Rookie  & 40.91 & 42.49 & 29.14 & 64.54 \\
Other            & $0$--$5$ mins. & Low     & 20.16 & 49.40 & 21.75 & 65.03 \\
Other            & $0$--$5$ mins. & Middle & 15.47 & 50.46 & 28.66 & 64.79 \\
Other            & $0$--$5$ mins. & High    & 27.05 & 46.81 & 26.66 & 72.63 \\
Other            & $5$--$45$ mins. & Rookie  & 66.07 & 41.55 & 35.08 & 62.94 \\
Other            & $5$--$45$ mins. & Low     & 27.51 & 49.35 & 20.18 & 74.65 \\
Other            & $5$--$45$ mins. & Middle & 23.64 & 52.32 & 25.94 & 75.17 \\
Other            & $5$--$45$ mins. & High    & 11.34 & 52.25 & 21.34 & 79.68 \\
Other            & $>45$ mins. & Rookie  & 49.92 & 45.01 & 31.22 & 71.24 \\
Other            & $>45$ mins. & Low     & 30.49 & 50.54 & 36.30 & 81.15 \\
Other            & $>45$ mins. & Middle & 24.88 & 54.66 & 33.08 & 77.41 \\
Other            & $>45$ mins. & High    & 19.77 & 56.09 & 40.46 & 78.56 \\
South Asia       & $0$--$5$ mins. & Rookie  & 42.22 & 44.84 & 38.33 & 63.19 \\
South Asia       & $0$--$5$ mins. & Low     & 40.86 & 52.88 & 48.16 & 65.63 \\
South Asia       & $0$--$5$ mins. & Middle & 33.16 & 50.79 & 52.12 & 71.44 \\
South Asia       & $0$--$5$ mins. & High    & 23.11 & 49.91 & 36.01 & 81.57 \\
South Asia       & $5$--$45$ mins. & Rookie  & 48.29 & 47.14 & 37.09 & 62.22 \\
South Asia       & $5$--$45$ mins. & Low     & 42.35 & 56.12 & 51.14 & 68.55 \\
South Asia       & $5$--$45$ mins. & Middle & 38.59 & 55.09 & 51.87 & 74.17 \\
South Asia       & $5$--$45$ mins. & High    & 34.16 & 53.35 & 43.60 & 81.54 \\
South Asia       & $>45$ mins. & Rookie  & 48.88 & 47.93 & 39.55 & 66.11 \\
South Asia       & $>45$ mins. & Low     & 43.95 & 57.66 & 55.86 & 70.47 \\
South Asia       & $>45$ mins. & Middle & 40.13 & 56.76 & 65.88 & 74.53 \\
South Asia       & $>45$ mins. & High    & 32.45 & 56.42 & 52.28 & 83.81 \\
\bottomrule
\end{tabular}
\vspace{0.2cm}
\begin{tablenotes}[para,flushleft]
\footnotesize \textit{Notes:} This table presents summary statistics (mean and standard deviation) of the estimated cost and ability distributions by observable worker groups. Mean and standard deviation of ability are reported in dollars of employers' WTP, i.e., $\frac{\widehat{\beta}}{|\widehat{\alpha}|}\cdot\widehat{a}$.
\end{tablenotes}
\end{threeparttable}
\end{table}

%% file: tables/signal_production_estimates.tex
\begin{table}[!htbp]
\centering
\begin{threeparttable}
\scriptsize
\setlength{\tabcolsep}{3pt}
\begin{tabular}{lllccc}
\caption{Signal-Production Estimates}\label{tab:signal_production_estimates}\\
\toprule
\textbf{Country Group} & \textbf{Arrival Group} & \textbf{Reputation Group} & $\boldsymbol{\widehat{\gamma^s}}$ & $\boldsymbol{\widehat{K^s}}$ & $\boldsymbol{V_{\varepsilon^s}}$ \\
\midrule
English Speaking & $0-5$ mins. & Rookie & 1.1717 & 5.2111 & 4.9692 \\
English Speaking & $0-5$ mins. & Low & 0.9535 & 5.1805 & 5.1008 \\
English Speaking & $0-5$ mins. & Middle & 0.9063 & 5.5999 & 5.4669 \\
English Speaking & $0-5$ mins. & High & 0.9294 & 6.2544 & 6.7678 \\
English Speaking & $5-45$ mins. & Rookie & 1.4194 & 5.1873 & 5.0741 \\
English Speaking & $5-45$ mins. & Low & 0.9935 & 5.3489 & 5.4480 \\
English Speaking & $5-45$ mins. & Middle & 1.0186 & 5.7269 & 6.2796 \\
English Speaking & $5-45$ mins. & High & 0.8755 & 6.1702 & 7.2451 \\
English Speaking & $> 45$ mins. & Rookie & 1.4663 & 5.4206 & 5.4354 \\
English Speaking & $> 45$ mins. & Low & 1.1272 & 5.5169 & 6.3270 \\
English Speaking & $> 45$ mins. & Middle & 1.1176 & 6.0147 & 5.6003 \\
English Speaking & $> 45$ mins. & High & 1.0451 & 6.3107 & 5.8462 \\
South Asia & Spike Arrivals & Rookie & 0.9846 & 4.7802 & 5.2454 \\
South Asia & Spike Arrivals & Low & 0.9931 & 5.3942 & 4.4702 \\
South Asia & Spike Arrivals & Middle & 0.9086 & 5.7699 & 5.3539 \\
South Asia & Spike Arrivals & High & 0.9307 & 6.2295 & 6.1613 \\
Europe & $0-5$ mins. & Rookie & 0.8225 & 4.5927 & 5.2863 \\
Europe & $0-5$ mins. & Low & 0.7020 & 4.6897 & 4.4453 \\
Europe & $0-5$ mins. & Middle & 0.7308 & 5.0208 & 4.6880 \\
Europe & $0-5$ mins. & High & 1.0247 & 6.0213 & 5.1050 \\
Europe & $5-45$ mins. & Rookie & 1.1618 & 4.5472 & 5.6753 \\
Europe & $5-45$ mins. & Low & 0.8154 & 4.9199 & 5.4259 \\
Europe & $5-45$ mins. & Middle & 0.8330 & 5.3074 & 6.1675 \\
Europe & $5-45$ mins. & High & 1.1756 & 6.3297 & 6.9104 \\
Europe & $> 45$ mins. & Rookie & 1.1846 & 4.6517 & 6.0705 \\
Europe & $> 45$ mins. & Low & 1.0470 & 5.0596 & 5.6789 \\
Europe & $> 45$ mins. & Middle & 0.9554 & 5.5739 & 6.1728 \\
Europe & $> 45$ mins. & High & 1.1701 & 6.6537 & 6.9725 \\
Not South Asia & Spike Arrivals & Rookie & 1.0477 & 4.6156 & 5.0425 \\
Not South Asia & Spike Arrivals & Low & 0.8346 & 4.9764 & 4.2831 \\
Not South Asia & Spike Arrivals & Middle & 0.8055 & 5.2227 & 5.0128 \\
Not South Asia & Spike Arrivals & High & 1.0559 & 6.0927 & 6.4536 \\
Other & $0-5$ mins. & Rookie & 1.1871 & 4.0601 & 4.1066 \\
Other & $0-5$ mins. & Low & 0.9304 & 4.7517 & 4.1335 \\
Other & $0-5$ mins. & Middle & 0.9032 & 5.2746 & 4.8462 \\
Other & $0-5$ mins. & High & 0.9233 & 5.5763 & 5.3316 \\
Other & $5-45$ mins. & Rookie & 1.1601 & 3.7847 & 5.6961 \\
Other & $5-45$ mins. & Low & 1.0965 & 4.9349 & 4.8759 \\
Other & $5-45$ mins. & Middle & 1.0151 & 5.4756 & 6.0390 \\
Other & $5-45$ mins. & High & 1.0250 & 5.8741 & 6.1553 \\
Other & $> 45$ mins. & Rookie & 1.3188 & 4.3500 & 4.7674 \\
Other & $> 45$ mins. & Low & 1.2021 & 5.3846 & 5.6008 \\
Other & $> 45$ mins. & Middle & 1.0618 & 5.7417 & 6.0590 \\
Other & $> 45$ mins. & High & 0.9856 & 5.8750 & 6.5123 \\
South Asia & $0-5$ mins. & Rookie & 1.0540 & 4.5308 & 4.7143 \\
South Asia & $0-5$ mins. & Low & 0.9807 & 5.2677 & 4.7612 \\
South Asia & $0-5$ mins. & Middle & 0.9227 & 5.7227 & 5.3834 \\
South Asia & $0-5$ mins. & High & 0.9298 & 6.2374 & 6.1692 \\
South Asia & $5-45$ mins. & Rookie & 1.0218 & 4.5044 & 5.3035 \\
South Asia & $5-45$ mins. & Low & 1.0102 & 5.2615 & 5.4920 \\
South Asia & $5-45$ mins. & Middle & 0.9759 & 5.7368 & 6.1759 \\
South Asia & $5-45$ mins. & High & 0.9683 & 6.2108 & 6.8218 \\
South Asia & $> 45$ mins. & Rookie & 1.1185 & 4.7418 & 5.4906 \\
South Asia & $> 45$ mins. & Low & 1.0629 & 5.4996 & 5.4358 \\
South Asia & $> 45$ mins. & Middle & 1.0001 & 5.8716 & 6.0273 \\
South Asia & $> 45$ mins. & High & 1.0000 & 6.3032 & 7.1680 \\
\bottomrule
\end{tabular}
\vspace{0.2cm}
\begin{tablenotes}[para,flushleft]
\footnotesize \textit{Notes:} presents the estimates of the constant $K^s(x)$, slope $\gamma^s(x)$, and variance of the error term $V_{\varepsilon^s}(x)$ for each observable group. All estimates across all groups are statistically significant at the 0.1\% level.
\end{tablenotes}
\end{threeparttable}
\end{table}

%% file: tables/tx_by_x.tex
\begin{table}[!ht]
\centering
\begin{threeparttable}
\scriptsize
\setlength{\tabcolsep}{3pt}
\begin{tabular}{lll r}
\caption{Estimated $T(X)$ by Observable Group}\label{tab:t_by_x_group}\\
\toprule
\textbf{Country Group} & \textbf{Arrival Group} & \textbf{Reputation Group} & $\boldsymbol{T(X)}$ \\
\midrule
English Speaking & $0$--$5$ mins.  & Rookie & \$-73.99 \\
English Speaking & $0$--$5$ mins.  & Low & \$-87.66 \\
English Speaking & $0$--$5$ mins.  & Middle & \$-73.06 \\
English Speaking & $0$--$5$ mins.  & High & \$-19.41 \\
English Speaking & $5$--$45$ mins. & Rookie & \$-23.21 \\
English Speaking & $5$--$45$ mins. & Low & \$-24.24 \\
English Speaking & $5$--$45$ mins. & Middle & \$2.82 \\
English Speaking & $5$--$45$ mins. & High & \$-1.06 \\
English Speaking & $>45$ mins.     & Rookie & \$58.77 \\
English Speaking & $>45$ mins.     & Low & \$33.54 \\
English Speaking & $>45$ mins.     & Middle & \$68.97 \\
English Speaking & $>45$ mins.     & High & \$83.59 \\
South Asia       & Spike Arrivals  & Rookie & \$-126.09 \\
South Asia       & Spike Arrivals  & Low & \$-60.88 \\
South Asia       & Spike Arrivals  & Middle & \$-48.29 \\
South Asia       & Spike Arrivals  & High & \$20.15 \\
Europe           & $0$--$5$ mins.  & Rookie & \$-14.62 \\
Europe           & $0$--$5$ mins.  & Low & \$-75.50 \\
Europe           & $0$--$5$ mins.  & Middle & \$-36.90 \\
Europe           & $0$--$5$ mins.  & High & \$55.94 \\
Europe           & $5$--$45$ mins. & Rookie & \$-5.62 \\
Europe           & $5$--$45$ mins. & Low & \$21.38 \\
Europe           & $5$--$45$ mins. & Middle & \$10.27 \\
Europe           & $5$--$45$ mins. & High & \$81.96 \\
Europe           & $>45$ mins.     & Rookie & \$62.89 \\
Europe           & $>45$ mins.     & Low & \$61.66 \\
Europe           & $>45$ mins.     & Middle & \$49.60 \\
Europe           & $>45$ mins.     & High & \$143.48 \\
Not South Asia   & Spike Arrivals  & Rookie & \$6.32 \\
Not South Asia   & Spike Arrivals  & Low & \$-19.90 \\
Not South Asia   & Spike Arrivals  & Middle & \$-5.19 \\
Not South Asia   & Spike Arrivals  & High & \$84.40 \\
Other            & $0$--$5$ mins.  & Rookie & \$-88.45 \\
Other            & $0$--$5$ mins.  & Low & \$-73.61 \\
Other            & $0$--$5$ mins.  & Middle & \$-28.97 \\
Other            & $0$--$5$ mins.  & High & \$53.60 \\
Other            & $5$--$45$ mins. & Rookie & \$-91.33 \\
Other            & $5$--$45$ mins. & Low & \$4.41 \\
Other            & $5$--$45$ mins. & Middle & \$22.95 \\
Other            & $5$--$45$ mins. & High & \$68.22 \\
Other            & $>45$ mins.     & Rookie & \$3.21 \\
Other            & $>45$ mins.     & Low & \$37.16 \\
Other            & $>45$ mins.     & Middle & \$48.95 \\
Other            & $>45$ mins.     & High & \$65.25 \\
South Asia       & $0$--$5$ mins.  & Rookie & \$-125.28 \\
South Asia       & $0$--$5$ mins.  & Low & \$-140.86 \\
South Asia       & $0$--$5$ mins.  & Middle & \$-104.72 \\
South Asia       & $0$--$5$ mins.  & High & \$-8.22 \\
South Asia       & $5$--$45$ mins. & Rookie & \$-92.83 \\
South Asia       & $5$--$45$ mins. & Low & \$-64.94 \\
South Asia       & $5$--$45$ mins. & Middle & \$-16.79 \\
South Asia       & $5$--$45$ mins. & High & \$14.69 \\
South Asia       & $>45$ mins.     & Rookie & \$-39.64 \\
South Asia       & $>45$ mins.     & Low & \$-32.78 \\
South Asia       & $>45$ mins.     & Middle & \$-1.58 \\
South Asia       & $>45$ mins.     & High & \$47.60 \\
\bottomrule
\end{tabular}
\vspace{0.2cm}
\begin{tablenotes}[para,flushleft]
\footnotesize \textit{Notes:} This table presents estimates of the demand preference for each observable worker group, i.e., $T(x)$. The estimates are reported in dollars of employers' WTP, i.e., $\widehat{T(x)} /|\widehat{\alpha}|$.
\end{tablenotes}
\end{threeparttable}
\end{table}

%% file: tables/pct_diff_sq_vs_no_by_obs.tex
\begin{table}[!htbp]
\centering
\begin{threeparttable}
\captionsetup{justification=centering}
\caption{Percent Changes in Hiring Rates by Reputation, Arrival, and Country: \\ No-Signaling vs. Status-Quo}\label{tab:percent_changes_hiring}
\setlength{\tabcolsep}{8pt}
\begin{tabular}{l r}
\toprule
 & \textbf{Percentage Difference} \\
\midrule
\multicolumn{2}{l}{\textit{Reputation Groups}} \\
\cmidrule(lr){1-2}
Rookie & $-6.66\%$ \\
Low & $-7.46\%$ \\
Middle & $-2.58\%$ \\
High & $+2.16\%$ \\
\addlinespace[0.8ex]
\multicolumn{2}{l}{\textit{Arrival Groups}} \\
\cmidrule(lr){1-2}
$0$--$5$ mins. & $-0.98\%$ \\
$5$--$45$ mins. & $-3.67\%$ \\
Spike Arrivals & $-0.06\%$ \\
$> 55$ mins. & $-0.73\%$ \\
\addlinespace[0.8ex]
\multicolumn{2}{l}{\textit{Country Groups}} \\
\cmidrule(lr){1-2}
South Asia & $-2.50\%$ \\
English Speaking & $+4.50\%$ \\
Europe & $+0.30\%$ \\
Other & $-3.07\%$ \\
\bottomrule
\end{tabular}
\vspace{0.2cm}
\begin{tablenotes}[para,flushleft]
\footnotesize \raggedright \textit{Notes:} This table presents percent changes in hiring rates by reputation, arrival, and country groups going from the status-quo equilibrium to the no-signaling equilibrium.
\end{tablenotes}
\end{threeparttable}
\end{table}

%% file: tables/x_groups_sum_stats.tex
\begin{table}[!htbp]
\centering
\scriptsize
\setlength{\tabcolsep}{3pt}
\begin{tabular}{lllcccccc}
\caption{Summary Statistics by Observable Group in Estimation Sample}\label{tab:x_groups_sum_stats}\\
\toprule
\textbf{Country Group} & \textbf{Arrival Group} & \textbf{Reputation Group} & \textbf{\# Apps.} & \textbf{\# with Valid Effort} & \textbf{Mean Signal} & \textbf{Mean Effort} & \textbf{Mean Bid} \\
\midrule
English Speaking & $0-5$ mins. & Rookie & 1956 & 1227 & 4.61 & 1.80 & 139.86 \\
English Speaking & $0-5$ mins. & Low  & 6378 & 3647 & 4.38 & 1.07 & 150.21 \\
English Speaking & $0-5$ mins. & Middle  & 20423 & 11961 & 4.91 & 1.09 & 145.88 \\
English Speaking & $0-5$ mins. & High  & 6383 & 2794 & 5.73 & 0.95 & 127.65 \\
English Speaking & $5-45$ mins. & Rookie & 1562 & 854 & 5.60 & 2.99 & 141.35 \\
English Speaking & $5-45$ mins. & Low  & 2877 & 1347 & 5.05 & 1.93 & 144.87 \\
English Speaking & $5-45$ mins. & Middle  & 8028 & 4558 & 5.70 & 1.81 & 158.30 \\
English Speaking & $5-45$ mins. & High  & 1719 & 913 & 6.26 & 1.60 & 155.18 \\
English Speaking & $> 45$ mins.  & Rookie & 2265 & 1285 & 6.25 & 2.99 & 148.61 \\
English Speaking & $> 45$ mins.  & Low  & 2817 & 1485 & 5.43 & 1.92 & 148.37 \\
English Speaking & $> 45$ mins.  & Middle  & 7932 & 5065 & 5.78 & 1.88 & 165.04 \\
English Speaking & $> 45$ mins.  & High  & 1553 & 934 & 6.17 & 1.81 & 167.41 \\
South Asia & Spike Arrivals & Rookie & 4522 & 3103 & 4.30 & 1.95 & 137.85 \\
South Asia & Spike Arrivals & Low  & 5580 & 3605 & 4.43 & 1.14 & 144.21 \\
South Asia & Spike Arrivals & Middle  & 38681 & 25826 & 4.91 & 0.99 & 150.74 \\
South Asia & Spike Arrivals & High  & 20931 & 14083 & 5.50 & 0.96 & 151.02 \\
Europe & $0-5$ mins. & Rookie & 4006 & 2789 & 4.40 & 1.86 & 128.44 \\
Europe & $0-5$ mins. & Low  & 8502 & 6254 & 4.20 & 1.20 & 136.77 \\
Europe & $0-5$ mins. & Middle  & 47506 & 32269 & 4.49 & 1.15 & 147.76 \\
Europe & $0-5$ mins. & High  & 5212 & 3682 & 5.04 & 0.86 & 146.54 \\
Europe & $5-45$ mins. & Rookie & 3071 & 1643 & 4.96 & 2.73 & 129.24 \\
Europe & $5-45$ mins. & Low  & 3567 & 2240 & 4.86 & 1.96 & 138.40 \\
Europe & $5-45$ mins. & Middle  & 21226 & 13088 & 5.12 & 1.64 & 152.10 \\
Europe & $5-45$ mins. & High  & 2146 & 1315 & 5.90 & 1.36 & 145.12 \\
Europe & $> 45$ mins.  & Rookie & 3948 & 2085 & 5.36 & 2.94 & 140.84 \\
Europe & $> 45$ mins.  & Low  & 3352 & 2139 & 5.19 & 2.04 & 143.54 \\
Europe & $> 45$ mins.  & Middle  & 17134 & 9722 & 5.46 & 1.67 & 159.11 \\
Europe & $> 45$ mins.  & High  & 2666 & 1665 & 5.97 & 1.45 & 157.34 \\
Not South Asia & Spike Arrivals & Rookie & 3176 & 2141 & 4.50 & 2.25 & 136.29 \\
Not South Asia & Spike Arrivals & Low  & 4875 & 3136 & 4.17 & 1.17 & 148.29 \\
Not South Asia & Spike Arrivals & Middle  & 22291 & 15231 & 4.54 & 1.15 & 147.76 \\
Not South Asia & Spike Arrivals & High  & 2537 & 1720 & 5.52 & 1.12 & 154.20 \\
Other & $0-5$ mins. & Rookie & 3813 & 2660 & 4.13 & 2.63 & 131.56 \\
Other & $0-5$ mins. & Low  & 3547 & 2305 & 4.13 & 1.45 & 128.86 \\
Other & $0-5$ mins. & Middle  & 16899 & 10486 & 4.63 & 1.16 & 129.77 \\
Other & $0-5$ mins. & High  & 4736 & 3089 & 5.00 & 1.24 & 144.33 \\
Other & $5-45$ mins. & Rookie & 5617 & 3024 & 4.59 & 3.38 & 135.45 \\
Other & $5-45$ mins. & Low  & 2529 & 1379 & 5.11 & 2.20 & 134.34 \\
Other & $5-45$ mins. & Middle  & 9723 & 5585 & 5.40 & 1.84 & 140.38 \\
Other & $5-45$ mins. & High  & 2570 & 1456 & 5.60 & 1.45 & 146.10 \\
Other & $> 45$ mins.  & Rookie & 7808 & 4354 & 5.15 & 3.22 & 142.09 \\
Other & $> 45$ mins.  & Low  & 3419 & 1885 & 5.58 & 2.51 & 150.06 \\
Other & $> 45$ mins.  & Middle  & 12539 & 6674 & 5.56 & 1.94 & 148.14 \\
Other & $> 45$ mins.  & High  & 3904 & 2015 & 5.35 & 1.45 & 160.83 \\
South Asia & $0-5$ mins. & Rookie & 13835 & 9271 & 4.20 & 2.09 & 135.12 \\
South Asia & $0-5$ mins. & Low  & 22222 & 12995 & 4.40 & 1.23 & 141.55 \\
South Asia & $0-5$ mins. & Middle  & 161046 & 92375 & 4.89 & 1.08 & 148.09 \\
South Asia & $0-5$ mins. & High  & 93314 & 56324 & 5.54 & 0.99 & 139.48 \\
South Asia & $5-45$ mins. & Rookie & 16293 & 8953 & 4.49 & 2.70 & 136.49 \\
South Asia & $5-45$ mins. & Low  & 12267 & 6572 & 5.05 & 2.04 & 150.19 \\
South Asia & $5-45$ mins. & Middle  & 66189 & 36421 & 5.50 & 1.79 & 154.40 \\
South Asia & $5-45$ mins. & High  & 38167 & 21619 & 5.93 & 1.55 & 151.30 \\
South Asia & $> 45$ mins.  & Rookie & 18007 & 10321 & 4.88 & 2.85 & 142.37 \\
South Asia & $> 45$ mins.  & Low  & 14783 & 8392 & 5.34 & 2.13 & 157.23 \\
South Asia & $> 45$ mins.  & Middle  & 83127 & 49072 & 5.55 & 1.74 & 167.88 \\
South Asia & $> 45$ mins.  & High  & 51865 & 30886 & 5.88 & 1.55 & 165.46 \\
\bottomrule
\end{tabular}
\end{table}